\documentclass[twocolumn]{aastex631}

\newcommand \msun{\mathrm{M}_{\odot}}
\newcommand \rcore{r_{\mathrm{core}}}
\newcommand \rcorecorr{r_{\mathrm{core,corr}}}

\usepackage{comment}
\usepackage{booktabs}
\usepackage{hhline}
\usepackage{multirow}
\usepackage[shortlabels]{enumitem}

\shorttitle{AASTeX v6.3.1 Sample article}
\shortauthors{GASP team}

\begin{document}
\def\linefilter{F$680$N$_{\mathrm{line}}$\,}

\title{\textit{HST} imaging of star-forming clumps in 6 GASP ram-pressure stripped galaxies}

\author[0000-0002-3818-1746]{Eric Giunchi}
\affiliation{INAF-Osservatorio astronomico di Padova, Vicolo Osservatorio 5, 35122 Padova, Italy}
\affiliation{Dipartimento di Fisica e Astronomia, Universit\`a di Padova, Vicolo Osservatorio 3, 35122 Padova, Italy}

\author[0000-0002-7296-9780]{Marco Gullieuszik}
\affiliation{INAF-Osservatorio astronomico di Padova, Vicolo Osservatorio 5, 35122 Padova, Italy}

\author[0000-0001-8751-8360]{Bianca M. Poggianti}
\affiliation{INAF-Osservatorio astronomico di Padova, Vicolo Osservatorio 5, 35122 Padova, Italy}

\author[0000-0002-1688-482X]{Alessia Moretti}
\affiliation{INAF-Osservatorio astronomico di Padova, Vicolo Osservatorio 5, 35122 Padova, Italy}

\author[0000-0002-4382-8081]{Ariel Werle}
\affiliation{INAF-Osservatorio astronomico di Padova, Vicolo Osservatorio 5, 35122 Padova, Italy}

\author[0000-0002-9136-8876]{Claudia Scarlata}
\affiliation{Minnesota Institute for Astrophysics, School of Physics and Astronomy, University of Minnesota, 316 Church Street SE, Minneapolis, MN 55455, USA}

\author[0000-0001-8600-7008]{Anita Zanella}
\affiliation{INAF-Osservatorio astronomico di Padova, Vicolo Osservatorio 5, 35122 Padova, Italy}

\author[0000-0003-0980-1499]{Benedetta Vulcani}
\affiliation{INAF-Osservatorio astronomico di Padova, Vicolo Osservatorio 5, 35122 Padova, Italy}

\author[0000-0002-5189-8004]{Daniela Calzetti}
\affiliation{Department of Astronomy, University of Massachusetts, 710 N. Pleasant Street, LGRT 619J, Amherst, MA 01002, USA}

\begin{abstract}
Exploiting broad- and narrow-band images of the \textit{Hubble Space Telescope} from near-UV to I-band restframe, we study the star-forming clumps of six galaxies of the GASP sample undergoing strong ram-pressure stripping (RPS).
Clumps are detected in H$\alpha$ and near-UV, tracing star formation on different timescales.
We consider clumps located in galaxy disks, in the stripped tails and those formed in stripped gas but still close to the disk, called extraplanar.
We detect 2406 H$\alpha$-selected clumps (1708 in disks, 375 in extraplanar regions, and 323 in tails) and 3745 UV-selected clumps (2021 disk clumps, 825 extraplanar clumps and 899 tail clumps).
Only $\sim15\%$ of star-forming clumps are spatially resolved, meaning that most are smaller than $\sim 140$ pc.
We study the luminosity and size distribution functions (LDFs and SDFs, respectively) and the luminosity-size relation.
The average LDF slope is $1.79\pm 0.09$, while the average SDF slope is $3.1\pm 0.5$.
Results suggest the star formation to be turbulence driven and scale-free, as in main-sequence galaxies.
All the clumps, whether they are in the disks or in the tails, have an enhanced H$\alpha$ luminosity at a given size, compared to the clumps in main-sequence galaxies. Indeed, their H$\alpha$ luminosity is closer to that of clumps in starburst galaxies, indicating that ram pressure is able to enhance the luminosity.
No striking differences are found among disk and tail clumps, suggesting that the different environments in which they are embedded play a minor role in influencing the star formation.
\end{abstract}


\section{Introduction} \label{sec:intro}
Star formation is the mechanism driving the condensation of atomic gas from galactic to sub-kpc scales down to the dense cores (on pc/sub-pc scales) in which stars eventually form (Section 4 in \citealt{Kennicutt2012}) and therefore studying which processes are capable of influencing it is fundamental to our understanding of galaxy formation and evolution.
The bridge between the galactic- and core-scale regimes is represented by $\gtrsim10$ pc-scaled star-forming clumps, with masses $\gtrsim10^4\,\msun$ \citep{Portegies2010}. Our knowledge about these structures has greatly improved in the last decade thanks to observational surveys of low redshift galaxies with the \textit{Hubble Space Telescope} (\textit{HST}), which is able to achieve the resolution necessary to study their morphology and size properties (LEGUS, \citealt{Calzetti2015}; DYNAMO, \citealt{Fisher2017}; LARS, \citealt{Messa2019}; PHANGS-HST; \citealt{Lee2022}).
Exploiting LEGUS data, many studies find hints that star formation is a turbulence-driven process fragmenting the gas following a scale-free hierarchy \citep{Elmegreen2014,Gouliermis2015,Gouliermis2017}. The hierarchical structure is then reflected also in the emerging spatial distribution of stars formed from such gas \citep{Elmegreen1996,Elmegreen2006,Grasha2017}.

Moreover, the star formation mechanism can be strongly influenced by the properties of the local medium in which the star forming clumps form, and this leaves an imprint that can be studied using different diagnostics.\par
Models describing the fragmentation of star-forming regions as a scale-free, turbulence-driven process predict the mass distribution function of these regions to be a power law with slope 2 \citep{Elmegreen2006}. The corresponding luminosity distribution function (LDF), if derived using a tracer of a narrow age range, is expected to have a similar slope, provided that 1) the Initial Mass Function (IMF) is well sampled and independent of the mass of the initial cloud from which the clumps are formed and 2) the star formation history (SFH) and therefore the stellar age distributions of all clouds are the same \citep{Elmegreen1996,Elmegreen2006}.\par
Indeed, from the observational point of view, LDFs of recently formed star-forming regions are known to be well described by a power law, independently of wavelength, with a drop at luminosities fainter than a peak luminosity $L_{\mathrm{peak}}$ due to incompleteness (e.g., H$\alpha$: \citealt{Kennicutt1989,Santoro2022}; UV \citealt{Cook2016,Messa2019}; V-band \citealt{Larsen2002,Bastian2007}, R-band \citealt{Whitmore2014}; IR and radio \citealt{Mascoop2021}). In general, observed slopes are found to be consistent with 2, even though in some cases there are hints of a slope slightly smaller than 2 (\citealt{Cook2016}: $1.76\pm0.3$; \citealt{Santoro2022}: $1.73\pm0.15$). Interestingly, some of these works find the value of the slope to be affected by the local environment. \cite{Cook2016} and \cite{Santoro2022} show that the LDF flattens in regions with high star-formation rate (SFR) surface density ($\Sigma_\mathrm{SFR}$). The same trend is then reflected in the LDF of clumps belonging to different intergalactic environments: \cite{Messa2018b} show that the LDF of UV clumps in the spiral arms of the LEGUS galaxy M51 is flatter than that of clumps in the inter-arm region (with likely lower $\Sigma_\mathrm{SFR}$).\par
The size distribution function (SDF) of star-forming clumps is known to be well described by a power law as well \citep{Kennicutt1980,Gusev2014}, with slopes between 2.5 and 4.5, and a flatter distribution for an increasing level of clustering in the clumps.\par
The luminosity ($\mathrm{L_{H\alpha}}$ typically)-size relation, which many works (\citealt{Wisnioski2012,Cosens2018} and references therein) have shown to be a linear relation in the logarithmic plane, is another proxy of the properties of star formation. The slope and the normalization of the correlation are thought to be related to the geometrical properties of the HII regions ionized by young stars and to the star-formation rate surface density $\Sigma_\mathrm{SFR}$. As shown in \cite{Cosens2018}, clumps in starburst-like environments, therefore with a high $\Sigma_\mathrm{SFR}$, are likely to have a higher H$\alpha$ luminosity at a given size and to follow a flatter distribution.\par
Jellyfish galaxies are a great laboratory to study star formation in peculiar regimes and environments. Jellyfish galaxies are cluster galaxies undergoing strong ram-pressure stripping (RPS, \citealt{Gunn1972}). The ram pressure exerted by the intracluster medium (ICM) is able to strip the gas from the galactic interstellar medium (ISM), eventually producing tails up to more than 100 kpc long, but leaving the stellar disk almost undisturbed \citep{Poggianti2017b}. The gas removal accelerates the quenching of the star formation in the galaxy \citep{Cortese2021}. Several previous works find RPS to be able to briefly enhance star formation during the first stages of the stripping process and to trigger in-situ star formation in compact knots of gas stripped out of the galactic disks \citep{Yoshida2008,Smith2010,Merluzzi2013,Fossati2016,Consolandi2017,Jachym2019}. The tails of these galaxies give the unprecedented opportunity to study the star-formation mechanism in a hotter and higher-pressure environment than the galactic ISM and without the influence of the underlying contamination of the old stellar populations in the disk.

\cite{Abramson2014} and \cite{Kenney2015} observed galaxies undergoing RPS in the Virgo cluster (NGC 4402 and NGC 4522) and in the Coma cluster (NGC 4921) with \textit{HST}, respectively.
These works show that RPS is able to decouple the high-density component of the ISM (and in particular the Giant Molecular Clouds) from the low-density one, which is more prone to stripping. Also, dust is characterized by elongated morphology and filaments aligned with the stripping direction. \cite{Cramer2019} studied the long and narrow H$\alpha$ tail of D100 in the Coma cluster with \textit{HST} and found unbound young UV sources with sizes $\sim50-100$ pc, which they consider likely to disperse with ageing.

One of the aims of the GASP (GAs Stripping Phenomena in galaxies with MUSE, \citealt{Poggianti2017b}) ESO Large Program is to study the properties of galaxies affected by different gas removal processes in the field, in groups and clusters. This includes cluster galaxies in different RPS stages, from pre-stripping (the control sample) to post-stripping \citep{Fritz2017}. Targets were chosen from the catalog in \cite{Poggianti2016} as galaxies with long unilateral debris in optical images, suggestive of gas-only removal. The final sample includes galaxies in the mass range $10^9-10^{11.5}\,\msun$ and at redshift $0.04<z<0.07$. Targets were observed with MUSE on the VLT (details about the MUSE data are given in Sec. \ref{sec:musedata}), in order to investigate the properties of both the ionized gas phase and the stellar component in the disks and in the stripped tails. Results of GASP confirmed the presence of clumps with in-situ star formation in the tails of individual jellyfish galaxies of the sample \citep{Bellhouse2017,Gullieuszik2017,Moretti2018b,Moretti2020,Poggianti2019b}. In particular, \cite{Poggianti2019a} analyzed star-forming clumps in the tails and in the disks of 16 GASP galaxies, finding that tail clumps are less massive, both in terms of stellar and gas mass, and with a smaller gas velocity dispersion than the clumps in the disks. However, the spatial resolution of MUSE at the typical redshift of the GASP galaxies ($\sim1$ kpc) did not allow us to study the morphology and the size of these clumps.

In order to better characterize the properties of the star-forming clumps detected in GASP galaxies with MUSE, six galaxies of the GASP sample have been observed with \textit{HST}, whose resolution is about a factor 14 better than MUSE one (details in Sec. \ref{sec:hstdata}). The broad-band filters which were adopted are the F275W, F336W, F606W and F814W, covering a spectral range going from UV- to I-band restframe. In addition, galaxies were observed also with the narrow-band filter F680N, collecting the H$\alpha$ emission.

This paper is structured as follows: in Sec. \ref{sec:data} we present the \textit{HST} and MUSE data; in Sec. \ref{tail:flag} we define the spatial categories; Sec. \ref{clu:sam} is focused on the steps followed for detecting and selecting star-forming clumps and complexes; in Sec. \ref{sec:results} we present the samples of clumps and complexes (\ref{sec:results}); in Sec. \ref{sec:df} we study the luminosity and size distribution functions of clumps and complexes (Secs. \ref{lum:dis} and \ref{siz:dis}, respectively); Sec. \ref{lum_size} is dedicated to the luminosity-size relation of the clumps; in Sec. \ref{sec:catalog} the catalogs which will be publicly available are described; in Sec. \ref{sec:summary} we summarize our results.

This paper adopts standard concordance cosmology parameters $H_0=70\,\mathrm{km\,s^{-1}\,Mpc^{-1}}$, $\Omega_M=0.3$ and $\Omega_\Lambda=0.7$ and a \cite{Chabrier2003} Initial Mass Function (IMF).

\section{Data}\label{sec:data}

\subsection{\textit{HST} data}\label{sec:hstdata}
In this work, we focus on the study of the luminosities and sizes of star-forming clumps and complexes in a sub-sample of 6 GASP galaxies, whose main properties are listed in Table \ref{gal:table}. The galaxies were selected from the GASP sample of ram-pressure stripped galaxies \citep{Poggianti2017a} for their extended H$\alpha$ emitting tails and, in particular, the large number of H$\alpha$ clumps detected with MUSE observations (\citealt{Poggianti2019a}, see also \cite{Vulcani2020}).

\begin{deluxetable*}{cccccccccc}
\tablecaption{Summary of the main properties of the galaxies studied in this paper and of their host clusters.
}
\tablewidth{0pt}
\tablehead{
\colhead{ID$_{P16}$} & \colhead{RA} & \colhead{Dec} & \colhead{$M_*$} & \colhead{$\frac{\mathrm{H\alpha}}{\mathrm{H\alpha+[NII]}}$} & \colhead{$z_{gal}$} & \colhead{cluster} & \colhead{$\sigma_{clus}$} & \colhead{$z_{clus}$} & \colhead{ref}\\
& \colhead{(J2000)} & \colhead{(J2000)} & \colhead{$\mathrm{10^{10}\,M_{\odot}}$} &&&&$\mathrm{km/s}$&&
}
\decimalcolnumbers
\startdata
JO$175$ & 20:51:17.593 & -52:49:22.34 & $3.2^{3.7}_{2.7}$ & 0.705 & 0.0468 & A3716 & $753^{+36}_{-38}$ & 0.0457 & (4,10,14)\\
\multirow{2}{*}{JO$201$} & \multirow{2}{*}{00:41:30.295} & \multirow{2}{*}{-09:15:45.98} & \multirow{2}{*}{$6.2^{7.0}_{2.4}$} & \multirow{2}{*}{0.660} & \multirow{2}{*}{0.0446} & \multirow{2}{*}{A85} & \multirow{2}{*}{$859^{+42}_{-44}$} & \multirow{2}{*}{0.0559} & (3,4,5,6,8,9,10,\\ 
&&&&&&&&& 13,14,15,16,18)\\
JO$204$ & 10:13:46.842 & -00:54:51.27 & $4.1^{4.7}_{3.5}$ & 0.660 & 0.0424 & A957 & $631^{+43}_{-40}$ & 0.0451 & (2,4,6,12,18)\\ 
JO$206$ & 21:13:47.410 & +02:28:35.50 & $9.1^{10.0}_{8.2}$ & 0.703 & 0.0513 & IIZW108 & $575^{+33}_{-31}$ & 0.0486 & (1,4,6,10,13,18,19)\\ 
JW$39$ & 13:04:07.719 & +19:12:38.41 & $17^{20}_{14}$ & 0.650 & 0.0650 & A1668 & 654 & 0.0634 & (14,18,19,21)\\
\multirow{2}{*}{JW$100$} & \multirow{2}{*}{23:36:25.054} & \multirow{2}{*}{+21:09:02.64} & \multirow{2}{*}{$29^{36}_{22}$} & \multirow{2}{*}{0.530} & \multirow{2}{*}{0.0602} & \multirow{2}{*}{A2626} & \multirow{2}{*}{$650^{+53}_{-49}$} & \multirow{2}{*}{0.0548} & (4,6,7,10,11,\\
&&&&&&&&& 15,17,18,19,20)\\
\enddata
\tablecomments{Columns are: GASP ID of the galaxy as in \citealt{Poggianti2016} (1), RA and Dec of the galaxy (2 and 3), galaxy stellar mass (4), median value for $\mathrm{H\alpha/(H\alpha+[NII])}$ from the MUSE clumps listed in \cite{Poggianti2019a} (5), galaxy redshift (6), ID of the host cluster (7), cluster velocity dispersion (8), cluster redshift (9), references (10). References are: 1) \cite{Poggianti2017b}, 2) \cite{Gullieuszik2017}, 3) \cite{Bellhouse2017}, 4) \cite{Poggianti2017a}, 5) \cite{George2018}, 6) \cite{Moretti2018b}, 7) \cite{Poggianti2019b}, 8) \cite{Bellhouse2019}, 9) \cite{George2019}, 10) \cite{Radovich2019}, 11) \cite{Moretti2020}, 12) \cite{Deb2020}, 13) \cite{Ramatsoku2020}, 14) \cite{Bellhouse2021}, 15) \cite{Tomicic2021a}, 16) \cite{Campitiello2021}, 17) \cite{Ignesti2022a}, 18) \cite{Tomicic2021b}, 19) \cite{Ignesti2022b}, 20) \cite{Sun2021}, 21) \cite{Peluso2022}. Masses are taken from \cite{Vulcani2018}. Cluster redshifts and velocity dispersions are taken from \cite{Biviano2017} and \cite{Cava2009}.}
\label{gal:table}
\end{deluxetable*}

The galaxies were observed using the WFC3/UVIS on board of the Hubble Space Telescope (\textit{HST}), using 4 broad-band filters (F275W, F336W, F606W, F814W), which cover a spectral range from UV- to I-band restframe. In addition, galaxies were also observed with a narrow-band filter, F680N, in order to collect the H$\alpha$ emission at the redshift of these galaxies. Details about observations, data reduction, calibration and analysis, estimate of the standard deviation of the background in each band ($\sigma$, hereafter) and H$\alpha+\mathrm{[NII]}$ extraction from the F680N band are described in \cite{Gullieuszik2023}, but here we summarise the most important properties.

All images collected have pixel angular size of $0.04\arcsec$. The UVIS PSFs in all the 5 filters do not change significantly and have a FWHM of $0.07\arcsec$\footnote{https://hst-docs.stsci.edu/wfc3ihb/chapter-6-uvis-imageing-with-wfc3/6-6-uvis-optical-performance}, corresponding to $\sim 70\,\mathrm{pc}$ at the redshifts of the clusters hosting these galaxies ($0.0424-0.0650$, see Table \ref{gal:table}).
Images were reduced and calibrated using \textsc{AstroDrizzle}\footnote{https://drizzlepac.readthedocs.io/en/latest/CHANGELOG.html}. 
To obtain the H$\alpha+\mathrm{[NII]}$ maps, the continuum emission is modelled by linearly interpolating the emission coming from the broad-band filters F606W and F814W and then subtracted to the F680N images.
Contamination from emission lines within the two broad-band filters is expected to have only a small ($<10\%$) impact on our H$\alpha$ intensity estimates (see \citealt{Gullieuszik2023}).

For our purposes, in some cases we are going to work also with denoised versions of these \textit{HST} images.
Denoising is performed using a Python software package called \textsc{PySAP}\footnote{\url{https://cea-cosmic.github.io/pysap/index.html}} \citep{Farrens2020}. This algorithm expands the image in Fourier series and we set its parameters in order to remove the high-frequency components, which are typically due to noise. We removed the component with the highest frequency, equal to 2 pixels.
This procedure allows us to detect also fainter regions without being dominated by noise, but does not yield reliable sizes. This is the reason for not working with denoised images alone.

Throughout this paper, we work on a smaller squared field-of-view (see Table \ref{gal:fov}) with respect to the entire \textit{HST}-WFC3/UVIS images ($2.67\arcmin\times2.67\arcmin$), still sufficient to cover the entire extension of the galaxies and their tails.

\begin{table}
\centering
\caption{Properties of the \textit{HST} images sub-FOV used to detect clumps. For each galaxy (ID$_{P16}$) the center coordinates ($\mathrm{RA_{center}}$ and $\mathrm{Dec_{center}}$) and the width of the sub-FOV (width) are listed. Sub-FOV are always squared.}
\begin{tabular}{cccc}
\toprule\toprule
ID$_{P16}$ & $\mathrm{RA_{center}}$ & $\mathrm{Dec_{center}}$ & width\\\hline
JO175 & 20:51:17.6169 & -52:49:47.189 & 2.00'\\
JO201 & 0:41:31.7471 & -9:16:02.613 & 2.67'\\
JO204 & 10:13:47.2256 & -0:54:46.617 & 1.60'\\
JO206 & 21:13:44.1417 & +2:28:06.535 & 3.27'\\
JW39 & 13:04:08.8132 & +19:12:17.151 & 2.20'\\
JW100 & 23:36:23.0832 & +21:04:47.508 & 1.93'\\
\bottomrule
\end{tabular}
\label{gal:fov}
\end{table}

\subsection{MUSE data}\label{sec:musedata}
Throughout this work, we also exploit the information obtained from the GASP survey to remove regions powered by AGN or shocks using BPT maps \citep{Baldwin1981}, confirm the redshift of star-forming clump candidates and correct the F680N filter for the line emission of NII. All galaxies were observed in service mode with the Multi Unit Spectroscopic Explorer (MUSE, \citealt{Bacon2010}). MUSE is an integral-field-unit spectrograph with a $1\arcmin \times 1\arcmin$ field of view, sampled with $0.2\arcsec \times 0.2\arcsec$ pixels. The typical seeing of the MUSE observations is $1\arcsec$ ($0.7-1.3\,\mathrm{kpc}$ at the redshifts of these galaxies, Table \ref{gal:table}). Furthermore, MUSE spectra cover a spectral range going from 4500 to 9300 $\mathrm{\AA}$, sampled at $\sim 1.25\,\mathrm{\AA/pixel}$ and with a spectral resolution of $2.6\,\mathrm{\AA}$. The data were reduced using the most recent version of the MUSE pipeline available at the time of each observation (\citealt{Bacon2010}, from version $1.2$ to $1.6$), as described in details in \cite{Poggianti2017b}. The datacubes were then corrected for Galactic extinction using the extinction law and the reddending map by \cite{Cardelli1989} and \cite{Schlegel1998} (considering the recalibration introduced by \citealt{Schlafly2011}), respectively. Fluxes, velocities and velocity dispersions of the gas emission lines were obtained using KUBEVIZ \citep{Fossati2016}, after subtracting the stellar-only component derived with SINOPSIS \citep{Fritz2017}.

\section{Definition of disk, extraplanar and tail regions}\label{tail:flag}

Throughout this work, we are interested in studying the effects of local environment on star formation, and thus we aim at distinguishing star-forming regions originating from stripped gas embedded in the cluster environment from those still in the galaxy disk.

\begin{figure*}
\centering
\gridline{\resizebox{\textwidth}{!}{\includegraphics[height=1cm]{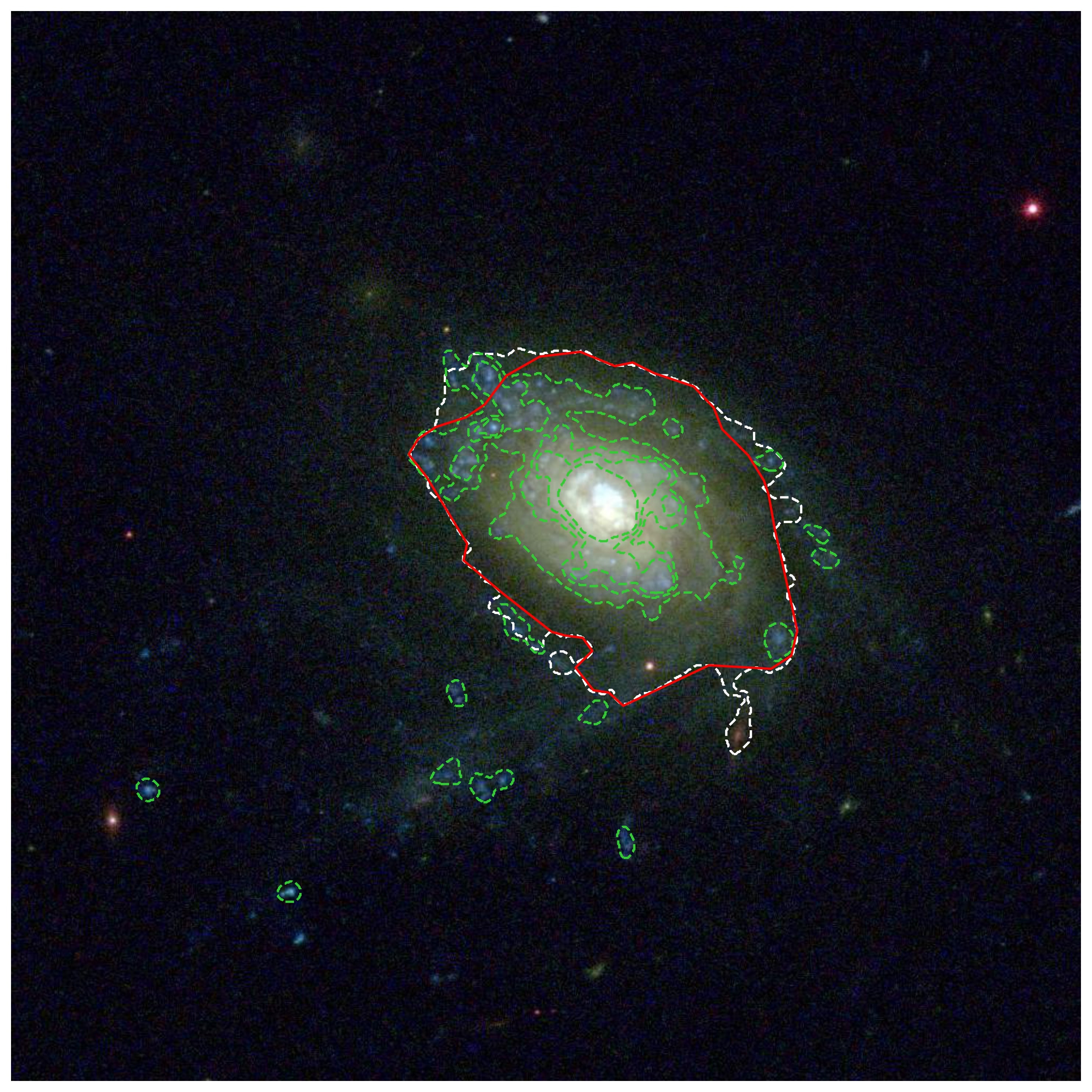}
          \includegraphics[height=1cm]{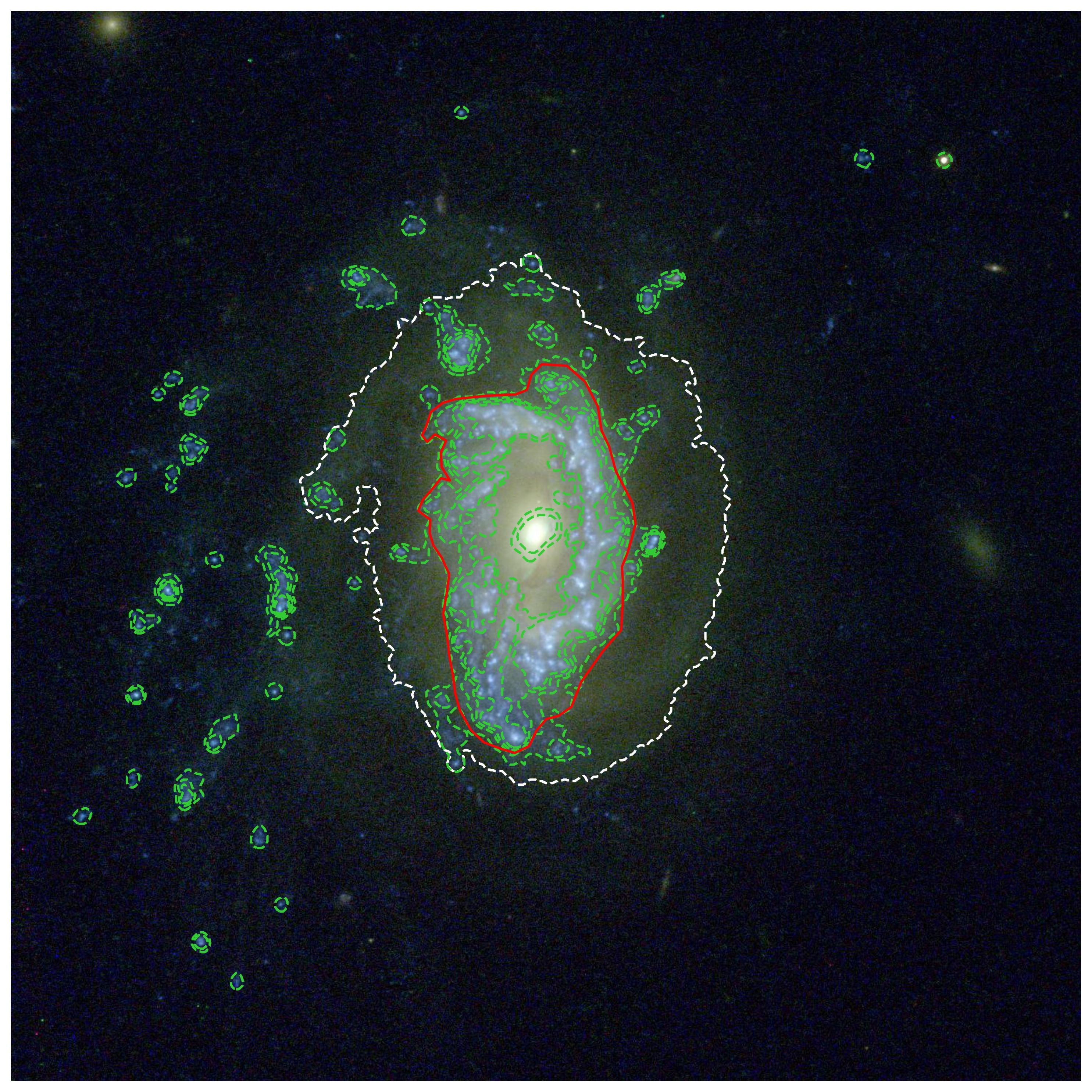}
          \includegraphics[height=1cm]{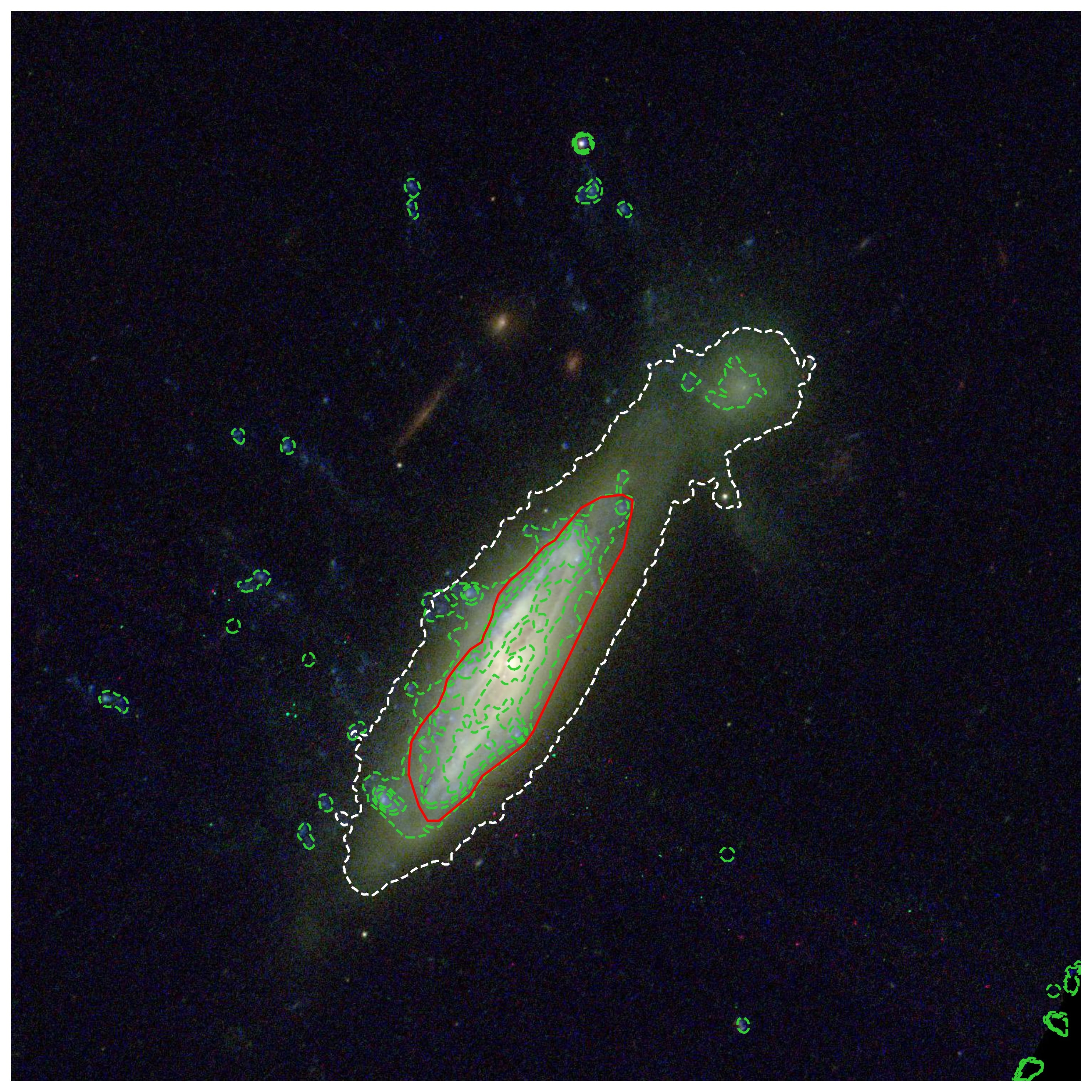}
          }
          }
\gridline{\resizebox{\textwidth}{!}{\includegraphics[height=1cm]{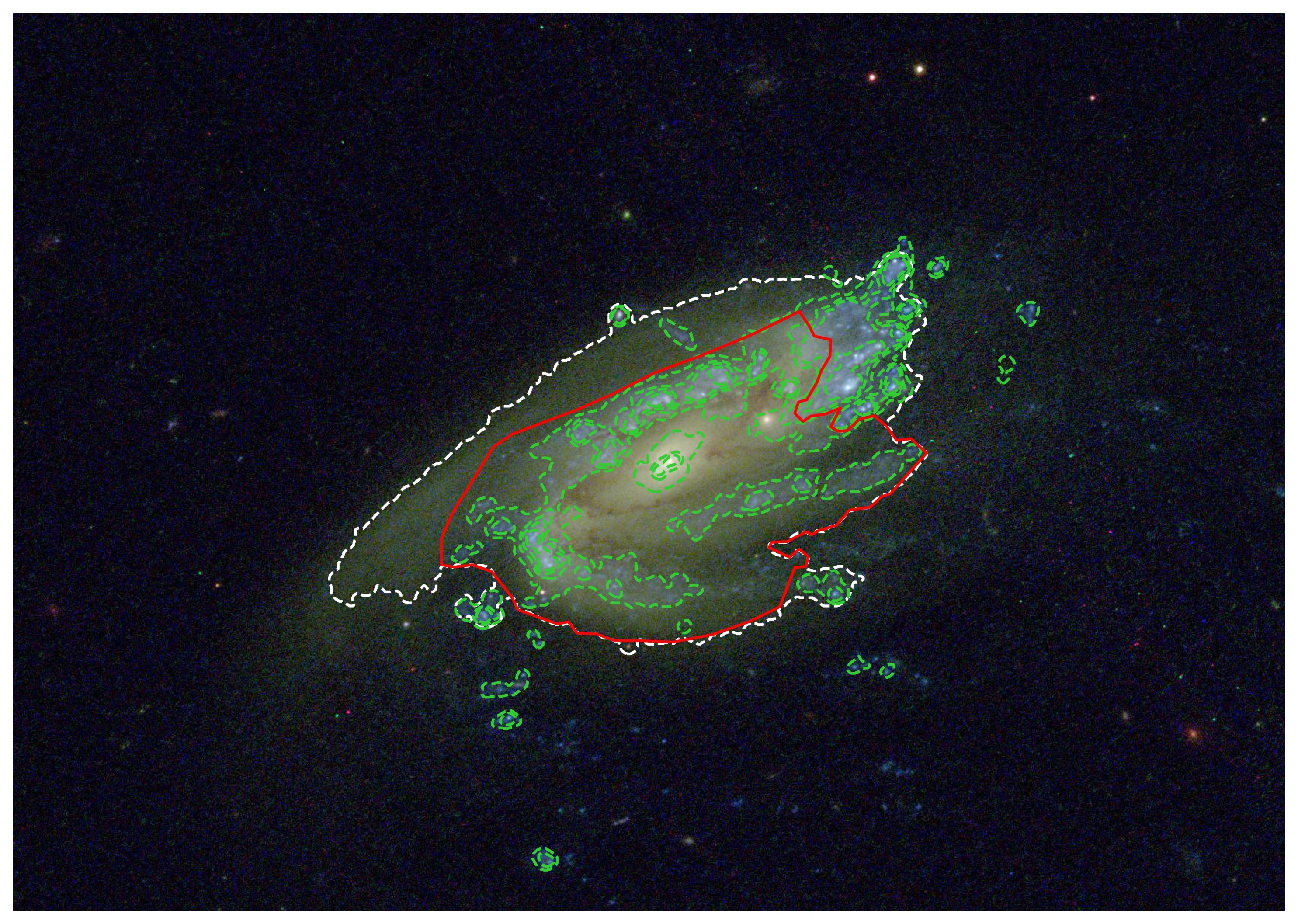}
          \includegraphics[height=1cm]{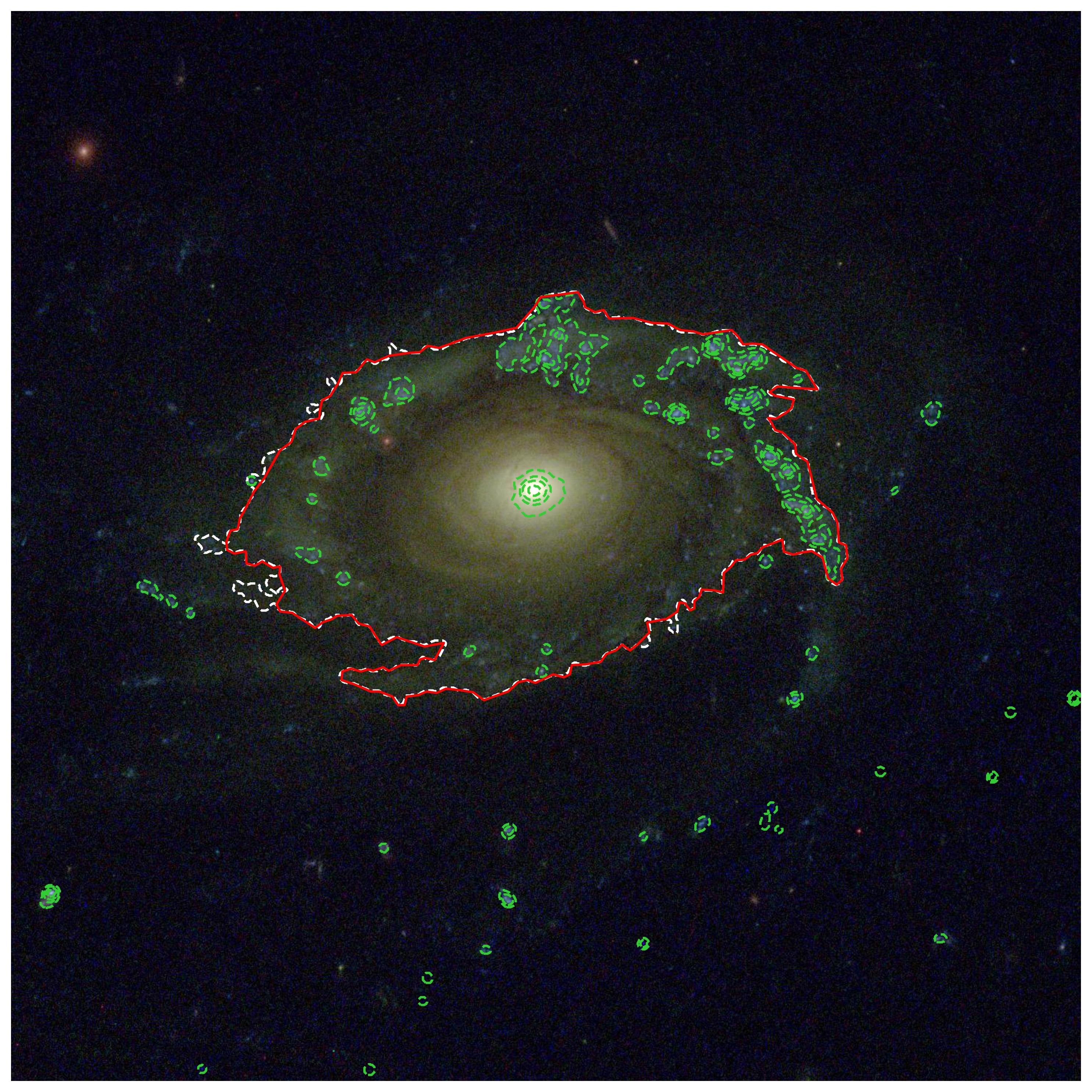}
          \includegraphics[height=1cm]{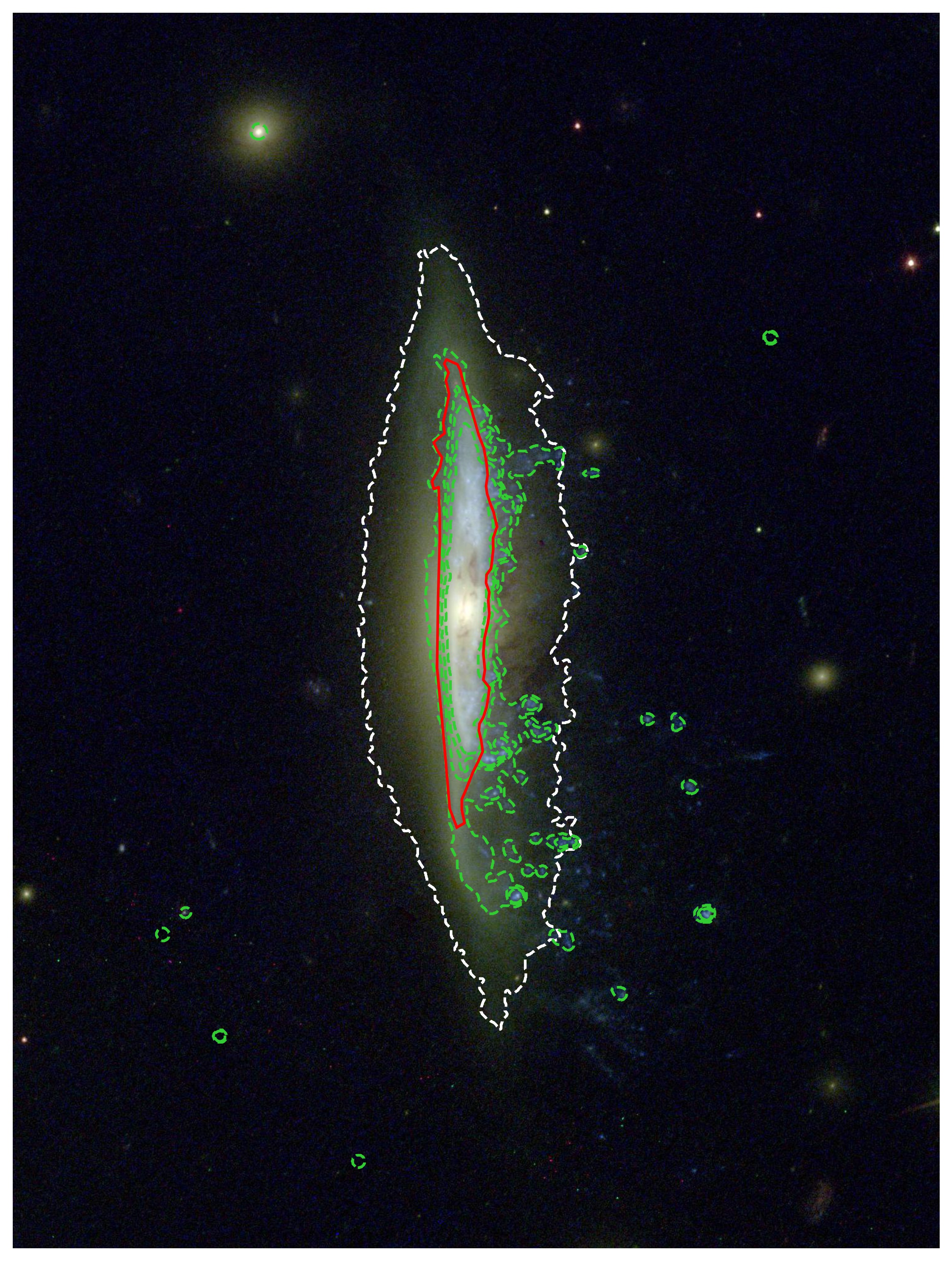}
          }}
\caption{Zoomed-in RGB images of JO$175$, JO$201$, JO$204$, JO$206$, JW$39$, JW$100$ (from top left to bottom right). The three colours of the RGB images are: F$814$W (red), F$606$W (green), a combination of F$275$W, F$606$W and F$814$W (blue). Details are given in \citealt{Gullieuszik2023}. The white dashed contours are the optical disks, defined as the $2\sigma$ contour in F$814$W. The green dashed contours are the $1\sigma$, $2\sigma$, $3\sigma$ and $5\sigma$ UV (F275W+F336W) contours (smoothed for clarity). The red line is the inner disk contour, traced as described in Sect. \ref{tail:flag}.
}
\label{RGB_disk_contour}
\end{figure*}

In analogy with what was done for the MUSE observations \citep{Gullieuszik2020}, the starting point to define the stripped tails is the definition of the galaxy stellar disk. As already noted in \cite{Gullieuszik2023} (see also Fig. \ref{RGB_disk_contour}), the high spatial resolution of \textit{HST} allows us to characterize the galaxy substructures and the stellar disk in more detail than what is possible with MUSE.
We used the $2\sigma$ contour of the reddest photometric band available (F814W) to draw the most external boundary of the stellar optical disk. $2\sigma$ values range from $2.14$ to $2.67\times 10^{-21}\,\mathrm{erg/s/cm^2/\AA}$ per pixel.
We will refer to this contour as the galaxy optical contour (white dashed lines in Fig. \ref{RGB_disk_contour}) and we define as \textit{tail} the region beyond it.

In the disks of these galaxies there are some regions particularly bright in UV (band F275W), faint in optical (band F814W), elongated and aligned in the same direction of the tails. Therefore, they are likely to be young stellar populations formed in gas already stripped by ram pressure, but still inside the galaxy optical contour because of projection effects or because RPS is at an early stage: we call these regions \textit{extraplanar}. In order to separate the extraplanar regions from those still in the disk, we visually inspected the RGB images and traced an inner disk contour (red solid lines in Fig. \ref{RGB_disk_contour}), using the UV contours of the images as a guide (green contours in Fig. \ref{RGB_disk_contour}) to separate clumps with an elongated appearance and aligned along the likely stripping direction from those who do not. We define as \textit{disk} the region within the inner disk contour and as \textit{extraplanar} the region within the galaxy optical contour but outside the inner disk contour.
We point out that this process cannot completely separate undisturbed and stripped gas, since it is done via visual inspection and projection effects may prevent a perfect separation of these two categories of gas.

\section{Clumps and complexes detection}\label{clu:sam}

This section presents the procedure we developed to detect star-forming clumps and to measure their properties. This procedure was applied independently to both the F275W and H$\alpha$ images (Sec. \ref{sf:clumps}), in order to trace star formation on different timescales ($\sim200$ Myr and $\sim10$ Myr, respectively, \citealt{Kennicutt1998a,Kennicutt2012,Haydon2020}). In addition, a different version of it is applied to the F606W images to fully recover the stellar content in the galaxy tails (Sec. \ref{sf:compl}).

\subsection{Preliminary steps and \textsc{Astrodendro} performance}\label{clu:det}
As a first step, foreground and background sources are masked out. This is done using, when available, the spectroscopic information from MUSE and by visually inspecting the RGB images constructed as described in \citep{Gullieuszik2023}, looking for red elliptical or blue spiral-armed sources, likely to be early-type and spiral galaxies, respectively.\par
\begin{figure}
\centering
\includegraphics[width=0.5\textwidth]{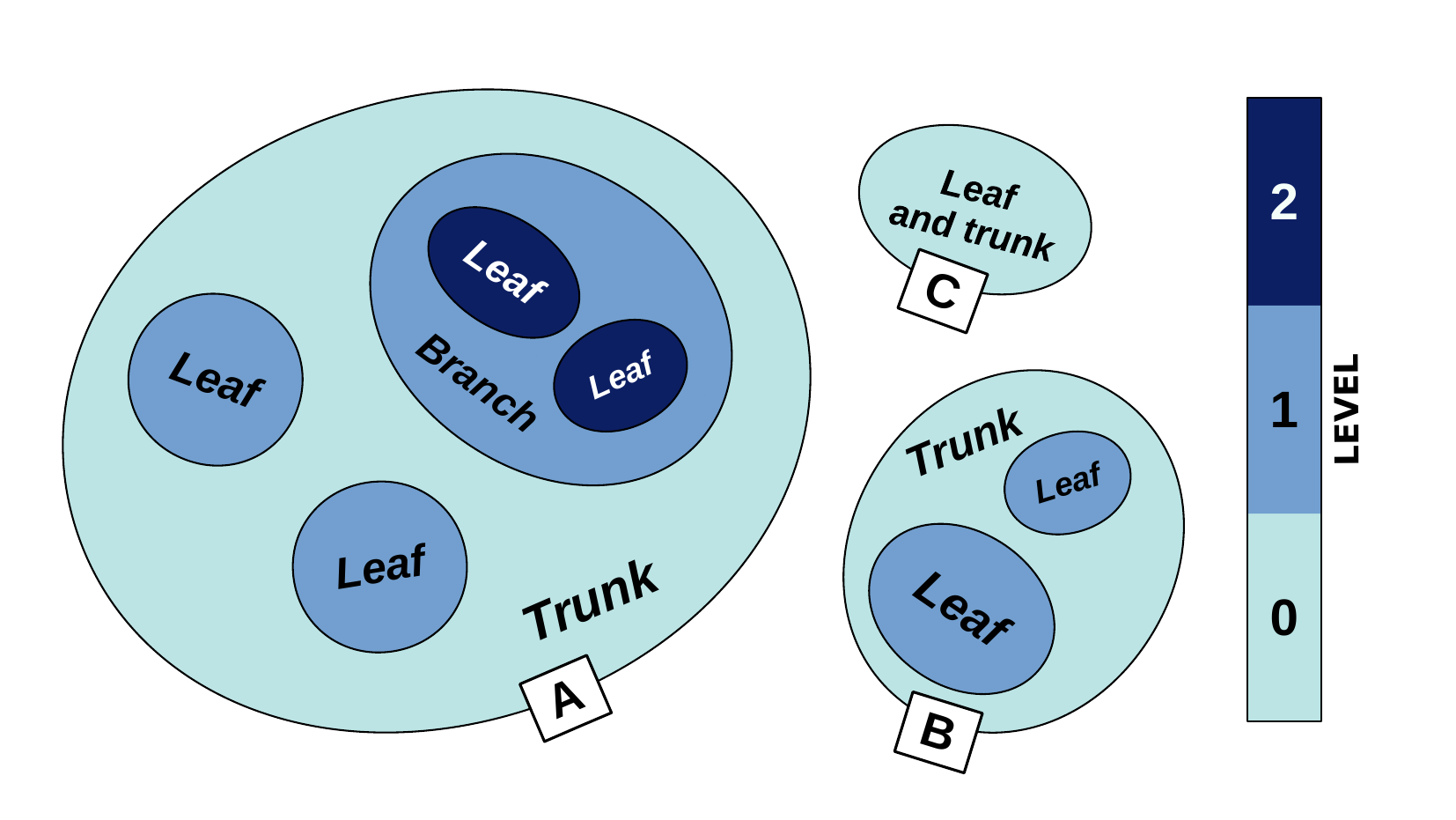}
\caption{Illustration of the dendrogram structures built by \textsc{Astrodendro}. Each clump is labelled with its position (trunk, branch and leaf) and colored according to its level in the tree hierarchy (from 0 to 2).}
\label{fig:dendro_illustration}
\end{figure}

The clump detection is performed using \textsc{Astrodendro}\footnote{http://www.dendrograms.org/}, a software package created to compute dendrograms of observed or simulated Astronomical data, classifying them in a hierarchical tree structure. With this software we are able to detect not only bright clumps, but also sub-clumps inside them.
Fig. \ref{fig:dendro_illustration} shows an illustration of 3 possible structures that \textsc{Astrodendro} can generate.

Clumps are defined as local maxima on the image; then the image is analyzed at fainter and fainter flux levels and the clumps grow by including fainter pixels. Eventually, at some point, adjacent clumps might blend together. In this case, those clumps stop growing and are defined as children of a common parent clump; for the following steps, when fainter flux levels are considered, only the parent clump keeps growing. When the flux threshold reaches a given value (see \textsc{min\_value} in Appendix \ref{appendix}), the algorithm stops and the tree structure is built: starting from the clumps at the base of the tree (i.e. the most extended ones), to which a level equal to 0 is assigned. \textsc{Astrodendro} retraces the tree and assigns to the sub-clumps a level equal to the level of their parent clump $+1$. It also generates a mask to define all pixels corresponding to each clump.\par
The naming convention used to define the position of the clump in the tree hierarchy is as follows:

\begin{itemize}
    \item{\textit{trunk}: clump with $\mathrm{level}=0$, regardless of whether it contains sub-clumps or not;}
    \item{\textit{branch}: clump with $\mathrm{level}>0$ and parent of other clumps;}
    \item{\textit{leaf}: clump with no children sub-clumps. Notice that, according to this definition, a trunk can be also a leaf.}
\end{itemize}

\subsubsection{Observed properties}\label{sec:lumsizedef}
For all the detected clumps, the following quantities are computed by \textsc{Astrodendro}:

\begin{itemize}
    \item{the intensity-weighted mean position of the clump in the plane of the sky, hereafter adopted as clump center;}
    \item{semi-major and semi-minor axes computed as standard deviations of the flux distribution of the clump 
    in the direction of greatest elongation in the plane of the sky;}
    \item{the radius $\rcore$ computed as the geometric mean of the major and minor axes;}
    \item{the exact area of the clump on the plane of the sky $A$.}
\end{itemize}

\begin{figure}
\gridline{\includegraphics[width=0.48\textwidth]{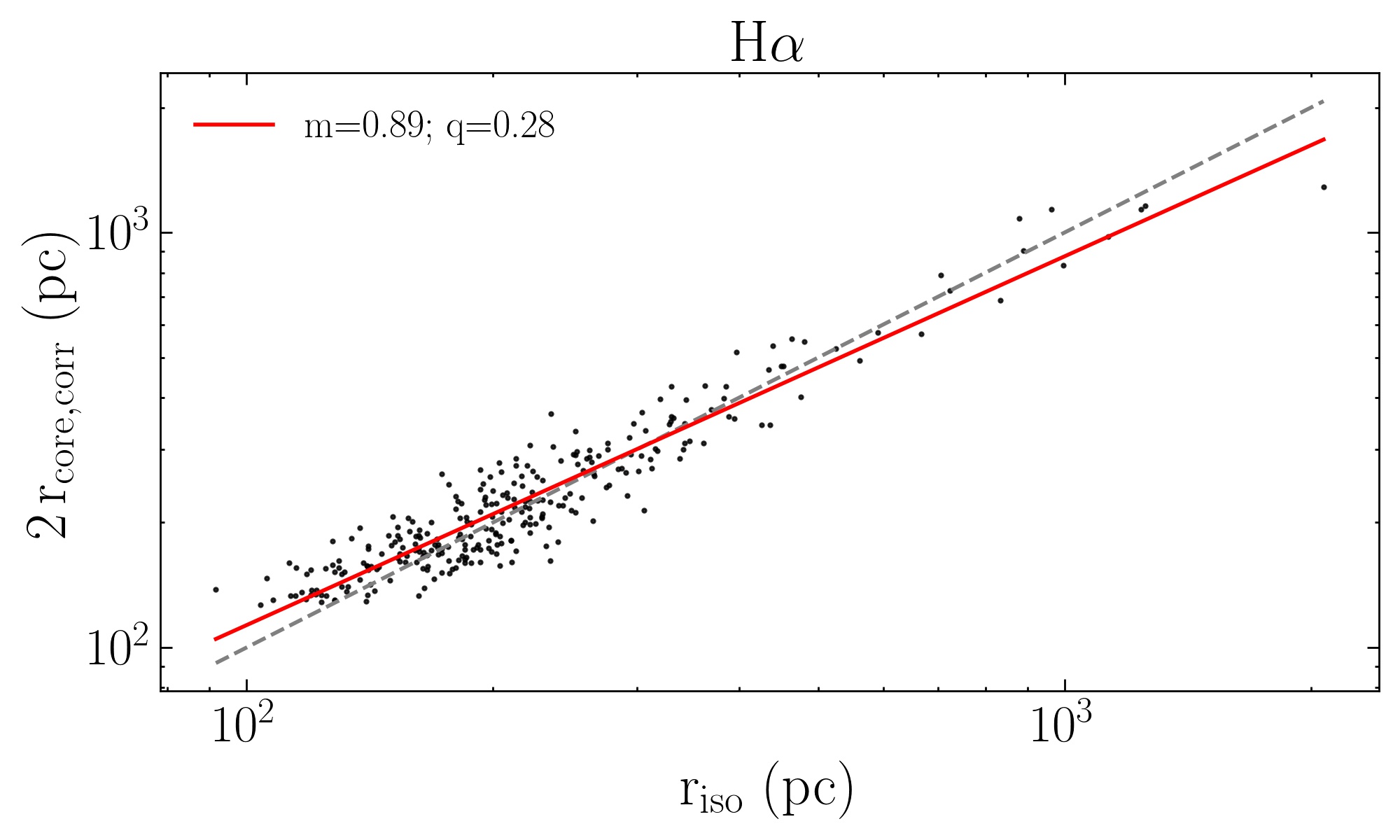}}
\gridline{\includegraphics[width=0.48\textwidth]{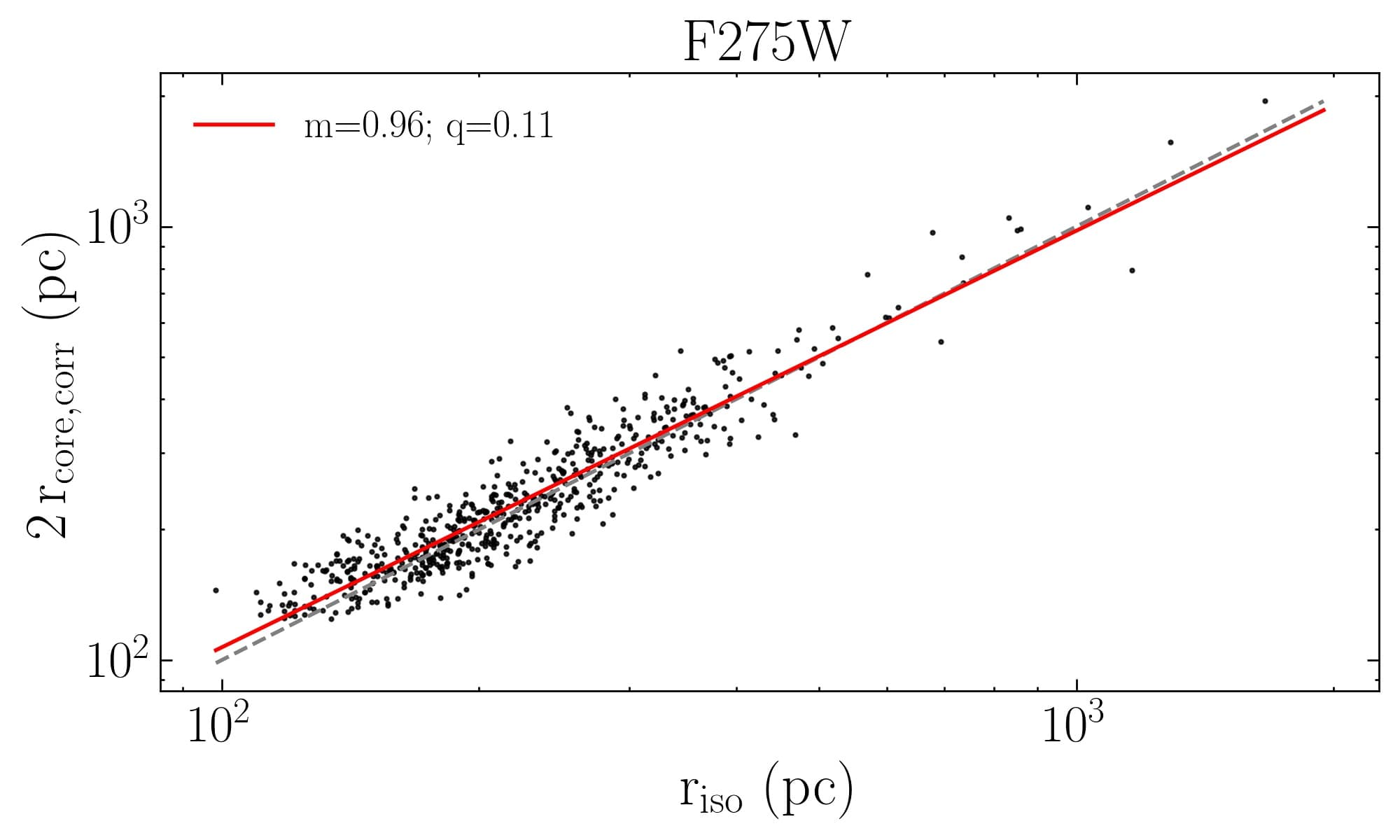}}
\caption{Comparison between the isophotal radius ($r_{\mathrm{iso}}$) and twice the PSF-corrected core radius ($2\rcorecorr$), defined in Sec. \ref{sec:lumsizedef}, both for H$\alpha$-resolved clumps (left panel) and UV-resolved clumps (right panel). The grey dashed line is the $1:1$ relation, while the red solid line is the best-fitting line. The best-fitting line is in good agreement with the $1:1$ relation, both for H$\alpha$- and UV-resolved clumps.}
\label{coreVSiso}
\end{figure}

In addition, we computed the following quantities:

\begin{itemize}
    \item{the flux densities for all the photometric bands, integrated over the clump area $A$. The flux uncertainties are computed summing two contributes in quadrature: the background noise, computed as a function of the clump area as described in \cite{Gullieuszik2023}, and the Poissonian uncertainty on the source counts converted then into flux considering the conversion factor PHOTFLAM, the exposure time and the Milky-Way dust attenuation;}
    \item{Luminosity: calculated from the flux densities using the redshift of the cluster hosting the galaxy (column 8 in Table \ref{gal:table}). In order to get H$\alpha$ luminosities, we compute $\mathrm{H\alpha/(H\alpha+[NII])}$ for the H$\alpha$ clumps detected with MUSE in the galaxies of our sample \citep{Poggianti2019a}. The median values obtained for each galaxies are listed in Tab. \ref{gal:table} and used to correct the \linefilter flux for the NII emission lines;}
    \item{$\rcorecorr$: PSF-corrected core radius. It is computed by subtracting in quadrature the $\sigma$ of the PSF ($\mathrm{FWHM}/2.35\simeq0.03\arcsec$, see Sec. \ref{sec:hstdata}) to $\rcore$ and it is converted in physical scale according to the redshift of the hosting cluster of each galaxy;}
    \item{the isophotal radius, defined as
    
    \begin{equation}
        r_{\mathrm{iso}}=\sqrt{\frac{A}{\pi}}.
    \end{equation}
    }
    \item{size: defined as $2\rcorecorr$. This choice is supported by the fact that $\rcorecorr$ is defined by the flux distribution of the clump, therefore it is less sensitive to the flux threshold above which clumps are detected \citep{Wisnioski2012}. Similarly to \cite{Wisnioski2012}, 
    in Fig. \ref{coreVSiso} we plot twice the PSF-corrected core radius against the isophotal radius, to show that these two quantities almost follow a 1:1 relation.\footnote{In our size range, also \cite{Wisnioski2012} found that their isophotal radii are larger than core radii by a factor $\sim 2$.}}
\end{itemize}

\subsection{Star-forming clumps}\label{sf:clumps}

Star-forming clumps are identified in the F275W (UV-selected clumps) and $\rm H\alpha$ (H$\alpha$-selected clumps) images running \textsc{Astrodendro} with a flux threshold of $2.5\sigma$ on the original images and $2\sigma$ on the denoised images\footnote{The $2\sigma$ threshold varies from galaxy to galaxy, ranging from $1.30$ to $1.62\times 10^{-20}\,\mathrm{erg/s/cm^2/\AA}$ for the F275W, and from $1.78$ to $2.20\times10^{-18}\,\mathrm{erg/s/cm^2}$ for the H$\alpha$. More details on how these values are computed can be found in \cite{Gullieuszik2023}}.
The two samples are computed independently, meaning that, in principle, some UV- and H$\alpha$-selected clumps may overlap if the same region is bright enough in both the filters.
Details about the parameters set for \textsc{Astrodendro} and the methods can be found in Appendix \ref{appendix}.
Throughout the paper, we use only leaf and trunk clumps (\textit{LT sample}), unless otherwise stated, to avoid considering the same region too many times.

\begin{figure*}
\centering
\includegraphics[width=0.8\textwidth]{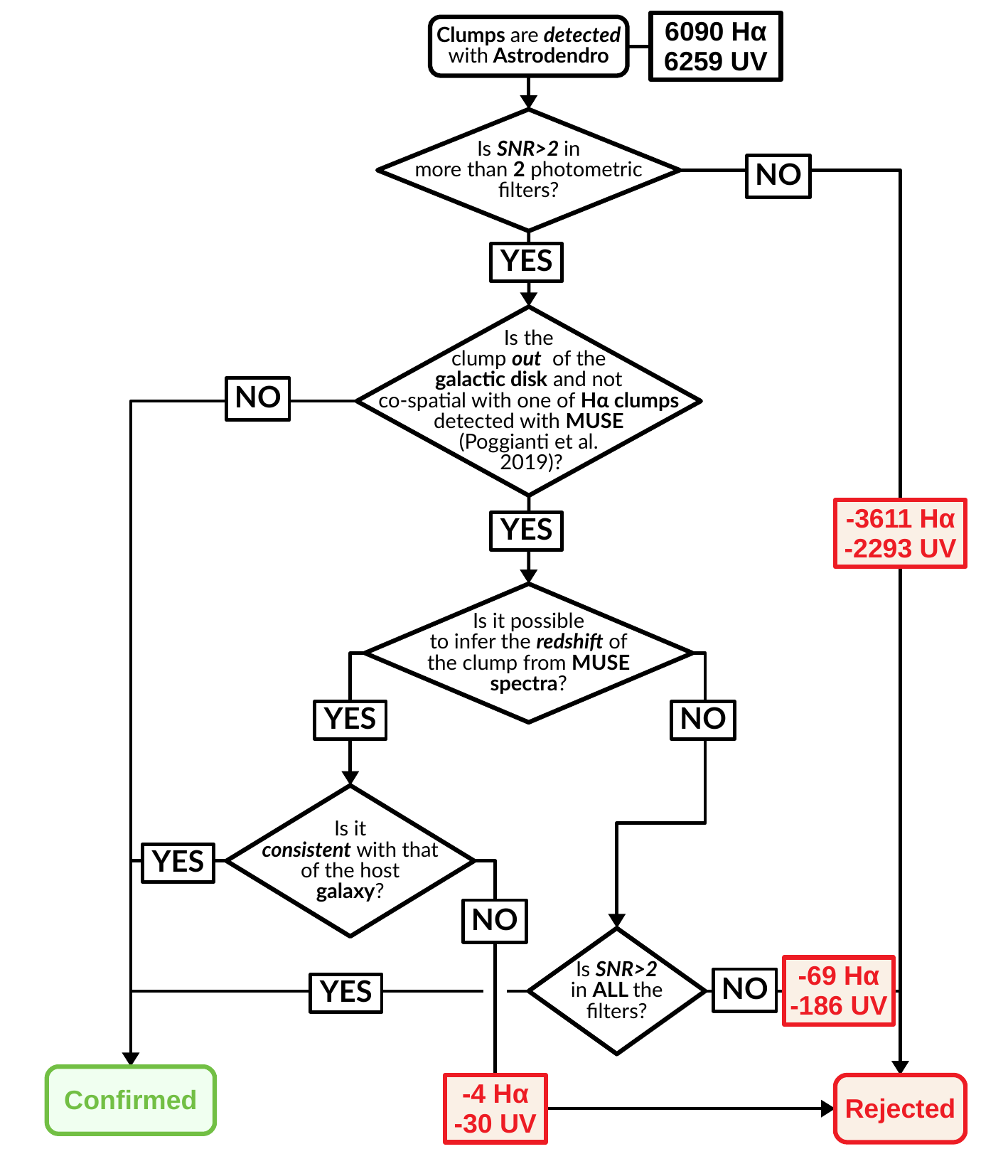}
\caption{Flow chart summarizing the selection procedure adopted in this paper to confirm (green checkmark) or reject (red cross) clump candidates detected by \textsc{Astrodendro}.}
\label{fig:clump_selection_flow_chart}
\end{figure*}

\textsc{Astrodendro} detected an initial total of 6090 H$\alpha$ and 6259 UV candidates.
To minimize the number of spurious detections we adopted the following procedure, which is schematised in the flow chart shown in Fig. \ref{fig:clump_selection_flow_chart}.

\begin{figure*}
\centering
\gridline{\includegraphics[width=0.5\textwidth]{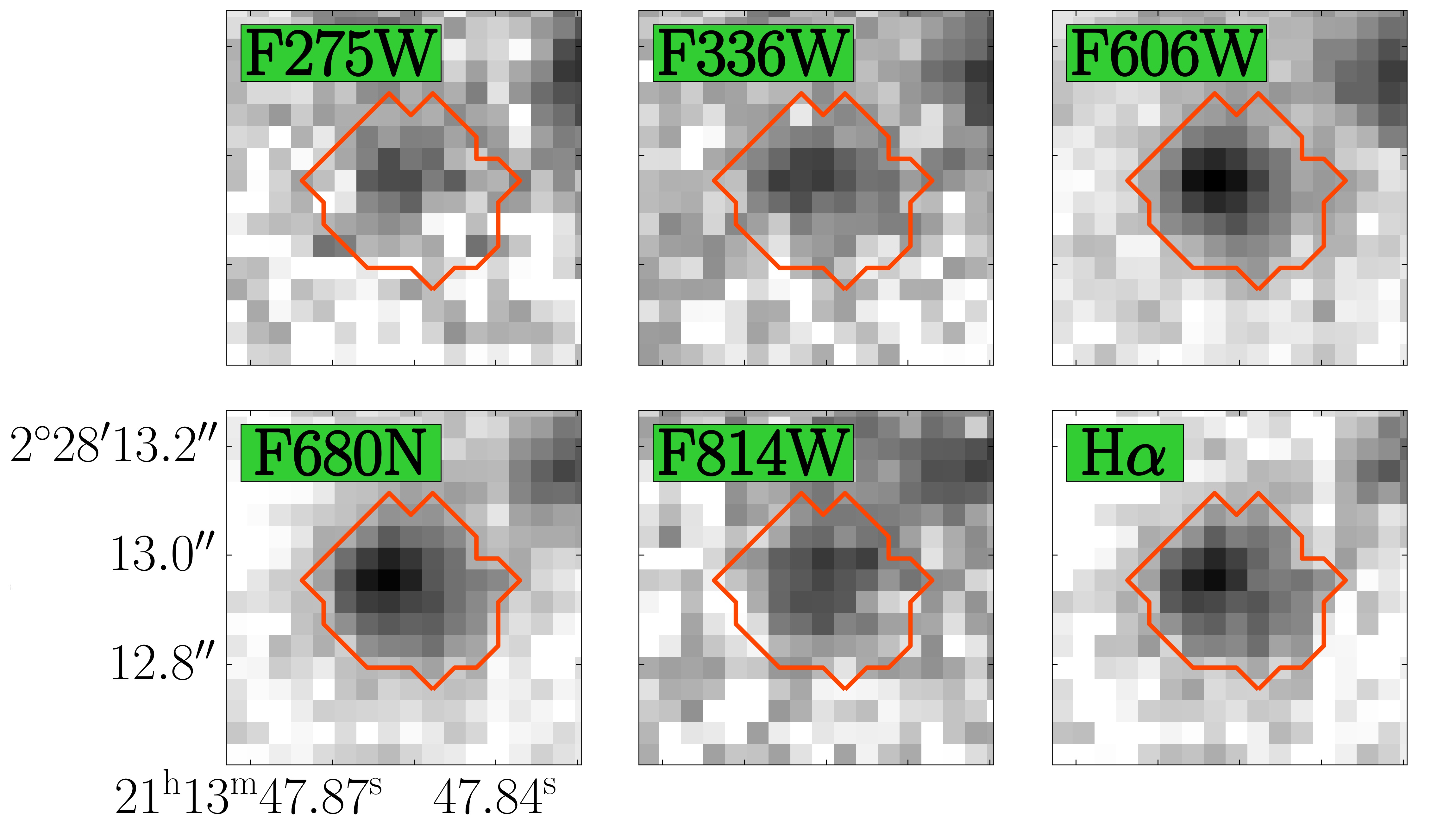}
    \includegraphics[width=0.5\textwidth]{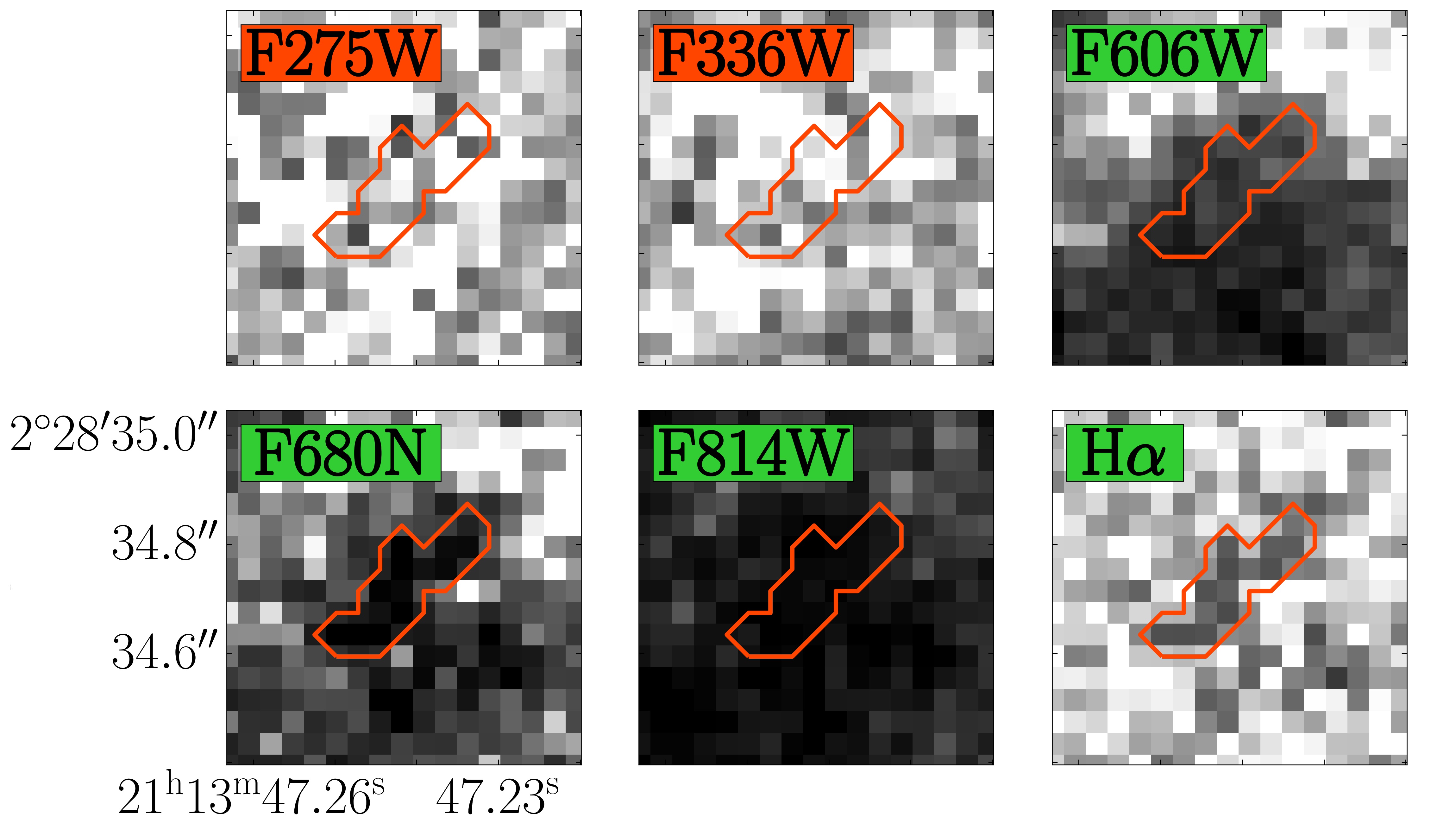}}
\gridline{\includegraphics[width=0.5\textwidth]{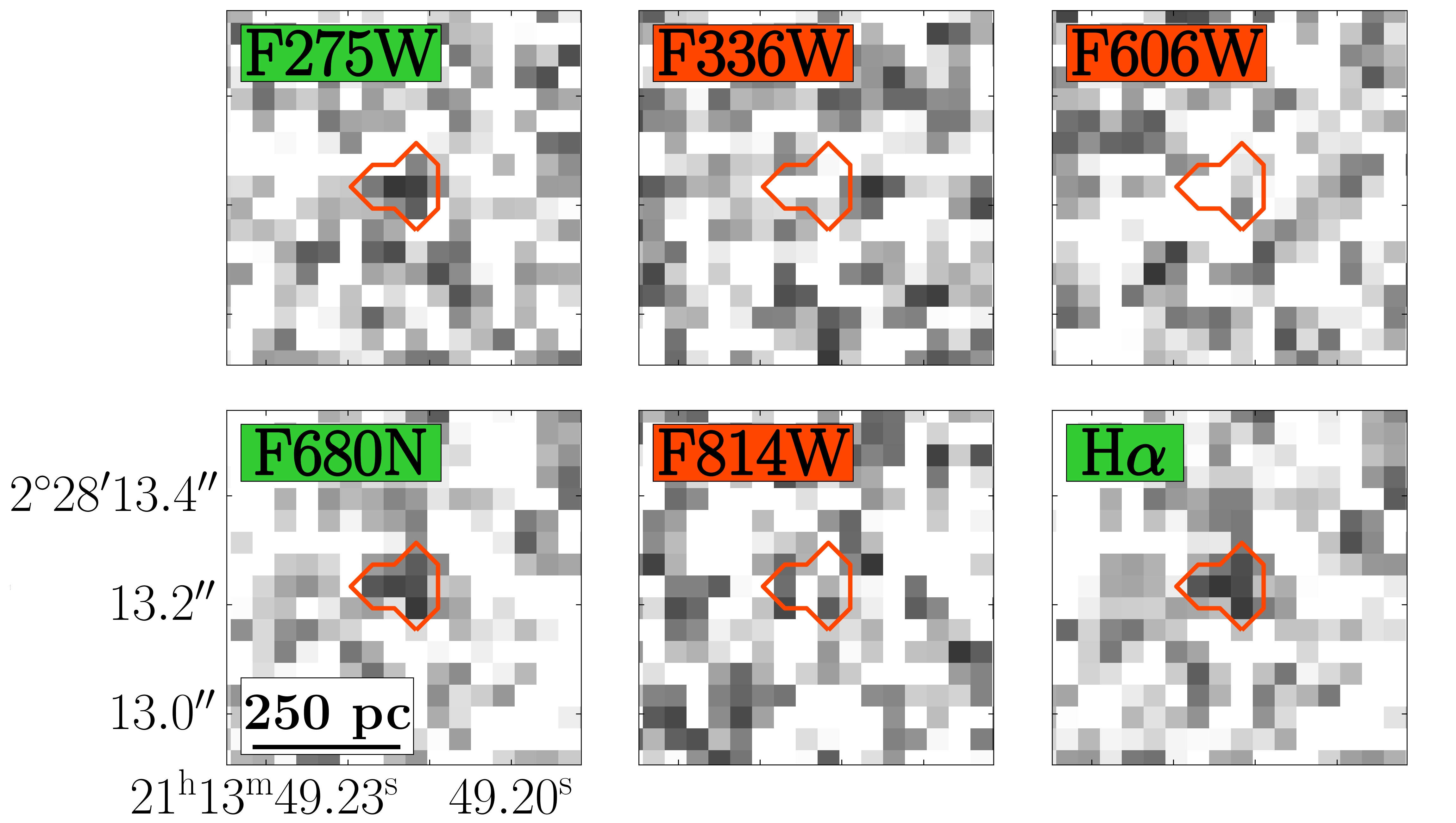}
    \includegraphics[width=0.5\textwidth]{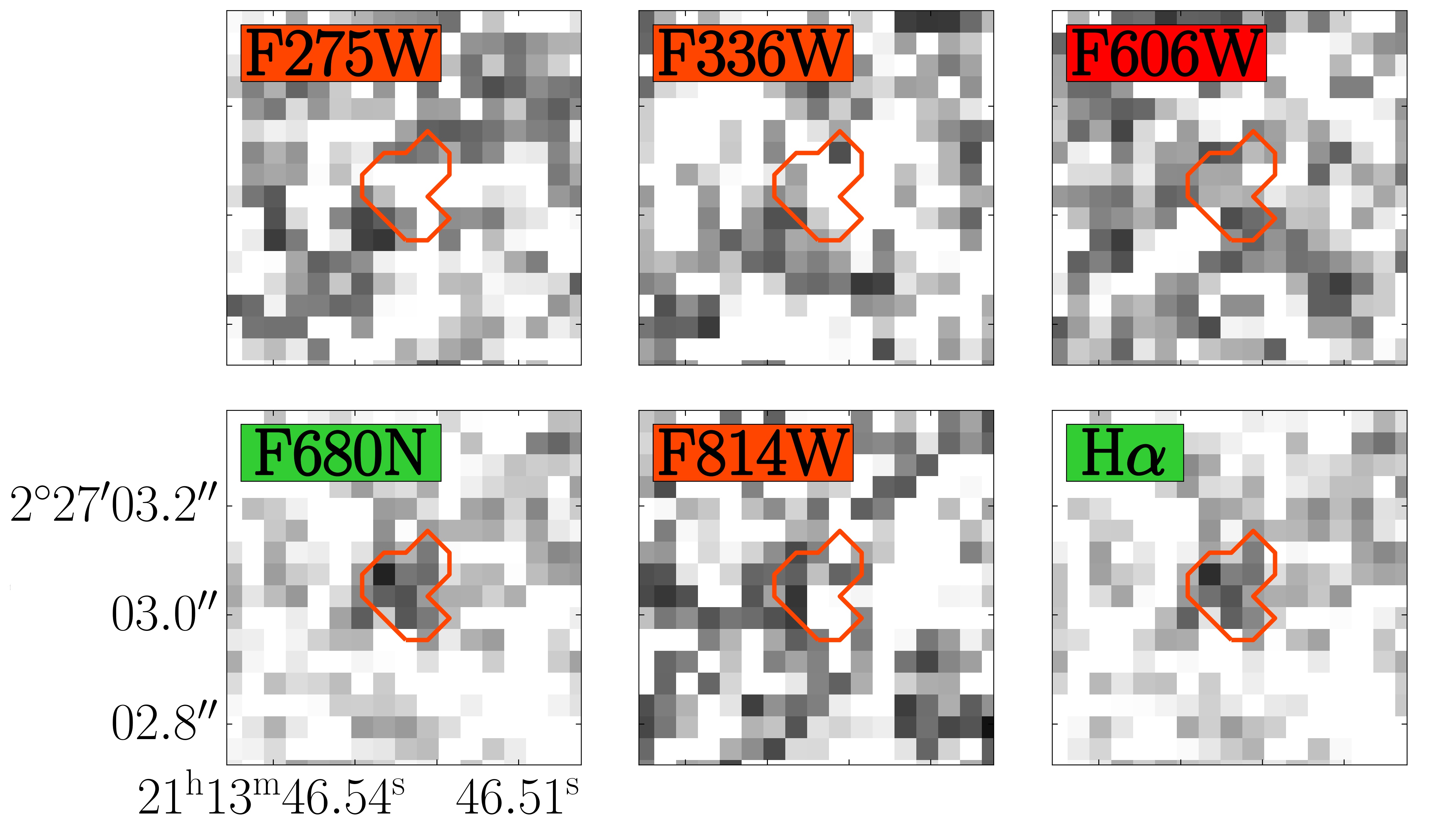}}
\caption{Images of 4 H$\alpha$ clumps of JO206 in all the filters. Each clump is shown in 6 different filters, which are (from upper left to lower right): F275W, F336W, F606W, F680N, F814W, H$\alpha$. Each filter is labelled in green or red whether we have a detection or not, according to our definition (Sec. \ref{sf:clumps}). The FOV is constant for all the clumps (a lengthscale equal to 250 pc is plotted in lower left panel of the third clump).
}
\label{clumps_images}
\end{figure*}

Firstly, for each of the 5 photometric bands, we flagged a clump as detected if its flux has a signal-to-noise ratio  SNR\footnote{Defined as the ratio between the total flux of the clump and the noise of the image in an area as large as that of the clump.} higher than 2.
We then exclude all clumps that were not detected in at least 3 photometric bands or in both F275W and F680N\footnote{The reason for this is that a star-forming clump might be in principle bright in UV and H$\alpha$ only.}. These criteria yield a reliable detection of clumps, as confirmed by subsequent visual inspection. A total 3611 H$\alpha$ and 2293 UV spurious detections were removed.
As an example, in Fig \ref{clumps_images} we show the images in the 5 filters and in H$\alpha$ of four H$\alpha$-selected clump candidates of JO201: the first one (upper left panel) is clearly detected in all images; the second one (upper right panel) does not show UV emission but is detected in three optical filters and in H$\alpha$; the third one (lower left panel) is detected only in F680N, F275W and H$\alpha$; all these three are therefore confirmed star-forming clumps. The last one (lower right panel) shows emission only in F680N and H$\alpha$ and was therefore rejected.

Outside the stellar disk, \textsc{Astrodendro} is more prone to detect residual cosmic rays and noise peaks as clump candidates in the tails. Cosmic rays and noise peaks are typically compact and bright, like the star-forming clumps we aim to study.
For these reasons, we perform an additional check for clumps in the tails. Both the H$\alpha$- and UV-selected tail clumps are matched with the corresponding catalogs of H$\alpha$ clumps detected with MUSE observations and described in \cite{Poggianti2019a}. If a match is found, the \textit{HST} clump is validated; if not, the clump is validated only if either the redshift from the MUSE spectrum at the corresponding position of the clump is consistent with that of the galaxy, or no redshift can be inferred from the MUSE spectrum, but the clump is detected in all \textit{HST} filters. To infer the redshift, emission lines such as the $\mathrm{[NII]}\,6548,6583-\mathrm{H}\alpha$ and $\mathrm{H}\beta-\mathrm{[OIII]}\,4958,5006$ triplets and the $\mathrm{[SII]}\,6716,6730$ doublet are fitted to the MUSE spectra, obtained within a circular aperture as close as possible to that of the clump. 73 H$\alpha$ and 216 UV candidates were rejected after this selection.\par
Finally, 5 UV-selected trunk clumps in the disks of the JO201, JO204, JO206 and JW100 are removed as their sizes and morphologies are such that they cannot be considered clumps, rather than more likely entire parts of the stellar disks.

For studying sizes, we define a sub-sample (\textit{resolved sample}) of resolved clumps\footnote{When possible, we substitute unresolved leaf clumps with its smallest, resolved, parent branch clump, if it does not contain another resolved leaf clump.} by selecting those objects whose PSF-corrected core radius, $\rcorecorr$, exceeds the PSF FWHM ($0.07\arcsec$), which corresponds to $\sim140$ pc at the typical redshifts of our targets.

Furthermore, when specified in the following, we removed regions whose emission is powered by an AGN.
In order to do that, we used the BPT maps \citep{Baldwin1981} of the MUSE images of the corresponding galaxies \citep{Poggianti2017a}. Adopting the boundary lines by \cite{Kauffmann2003}, \cite{Kewley2001} and \cite{Sharp2010}, the MUSE spaxels were flagged as star forming, composite, AGN or LINER regions according to the line ratios $\log(\mathrm{[NII]}/\mathrm{H}\alpha)$ and $\log(\mathrm{[OIII]}/\mathrm{H}\beta)$ (for the spaxels with S/N$>3$ for each line). The \textit{HST} clumps are flagged as the MUSE spaxels they fall into and in the following we remove those flagged as AGN or LINER when studying the luminosities of the H$\alpha$-selected clumps.\par

\subsection{Star-forming complexes}\label{sf:compl}

UV- and $\rm H\alpha$-selected clumps probe the emission coming from or due to stars younger than $\sim 10^8$ yr and $\sim 10^7$ yr, respectively.
The contribution from stellar components older than such timescales can be detected in other optical bands used in this analysis, in order to trace the whole stellar populations formed from the stripped gas in the tails.

Therefore we decided to run \textsc{Astrodendro} also on the F606W filter images, which are deeper than the UV images (see \citealt{Gullieuszik2023}) and are sensitive to older stellar populations with respect to F275W and H$\alpha$. 
Details of the \textsc{Astrodendro} run on the F606W images are given in Appendix \ref{appendix}.
Only tail trunk clumps are considered, and we retain only F606W clumps overlapping with at least one pixel of any star-forming clump in either the H$\alpha$-selected or the UV-selected samples.
In the following, we call \textit{star-forming complex} the union of a F606W clump and each star-forming clump matched to it.

\section{Number of clumps: disk, extraplanar and tail clumps}\label{sec:results}

\begin{figure*}
\centering
\includegraphics[width=\textwidth]{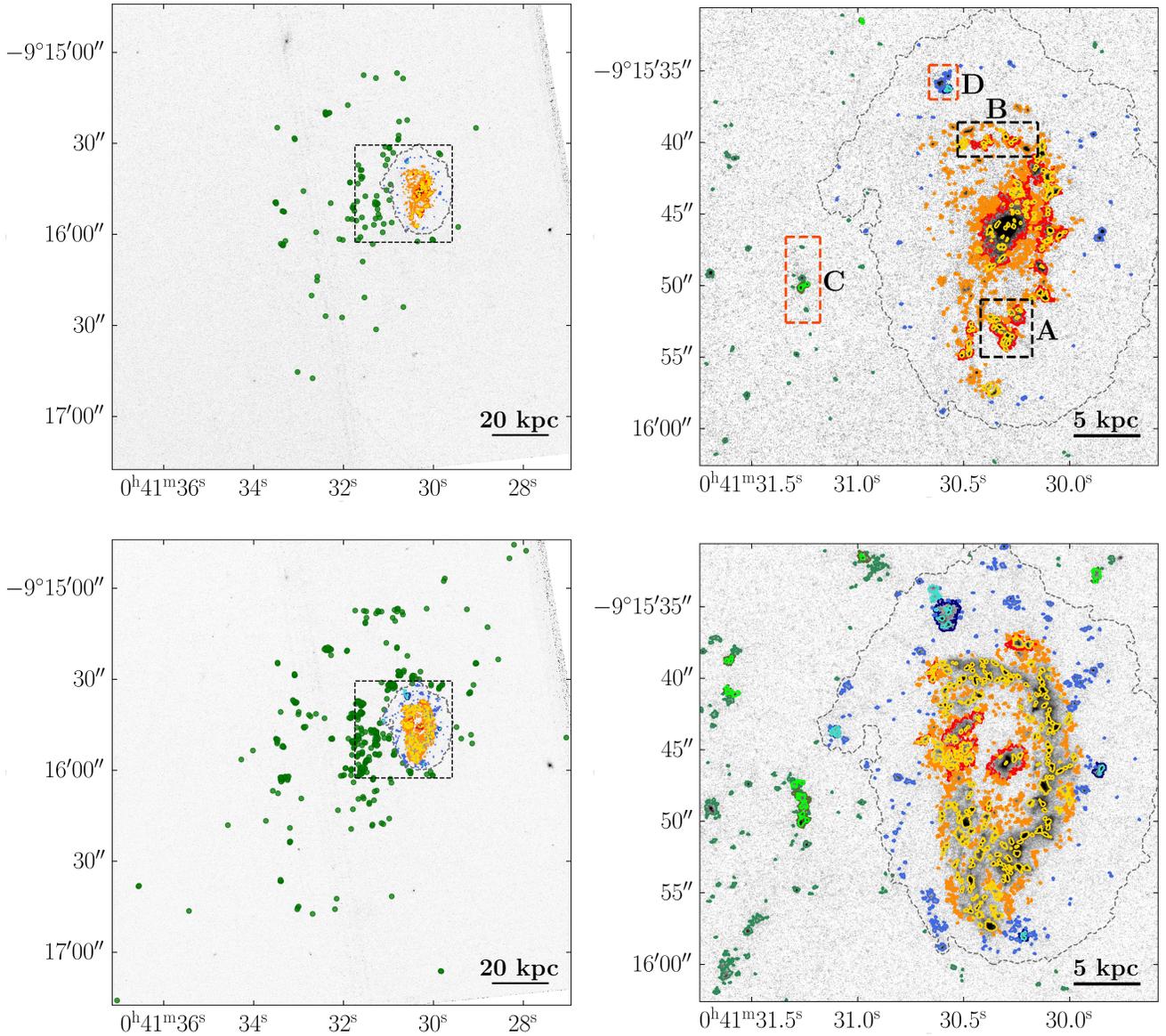}
\caption{Map of the clumps detected in JO201, superimposed onto the image in the filter used for the detection. Upper panels: H$\alpha$-selected clumps. Lower panels: UV-selected clumps. Left panels: field of view including all the clumps. Right panel: zoomed-in version on the vicinity of the disk (highlighted in the left panel with the black dashed rectangle). Colors in the right panels represent the spatial category and the tree structure (Sec. \ref{tail:flag} and \ref{clu:det}): disk clumps are plotted in red (trunks which are not leaves), orange (trunks which are leaves) and yellow (leaves which are not trunks). Similarly, extraplanar clumps are plotted in dark blue, blue and light blue, and tail clumps are plotted in dark green, green and light green in right panels. The grey dashed contour is the galaxy disk contour (see Sect. \ref{tail:flag}. In the left panels, for the sake of clarity, the tail clumps are plotted as green dots of fixed size. The regions highlighted and labelled as A, B, C and D are shown in figure \ref{JO201_clumps_zoom}.\\
\\
The complete figure set (6 images) including also the other galaxies is available in the online journal.}
\label{fig:detected_clumps}
\end{figure*}

\begin{figure}
\centering
\includegraphics[width=0.5\textwidth]{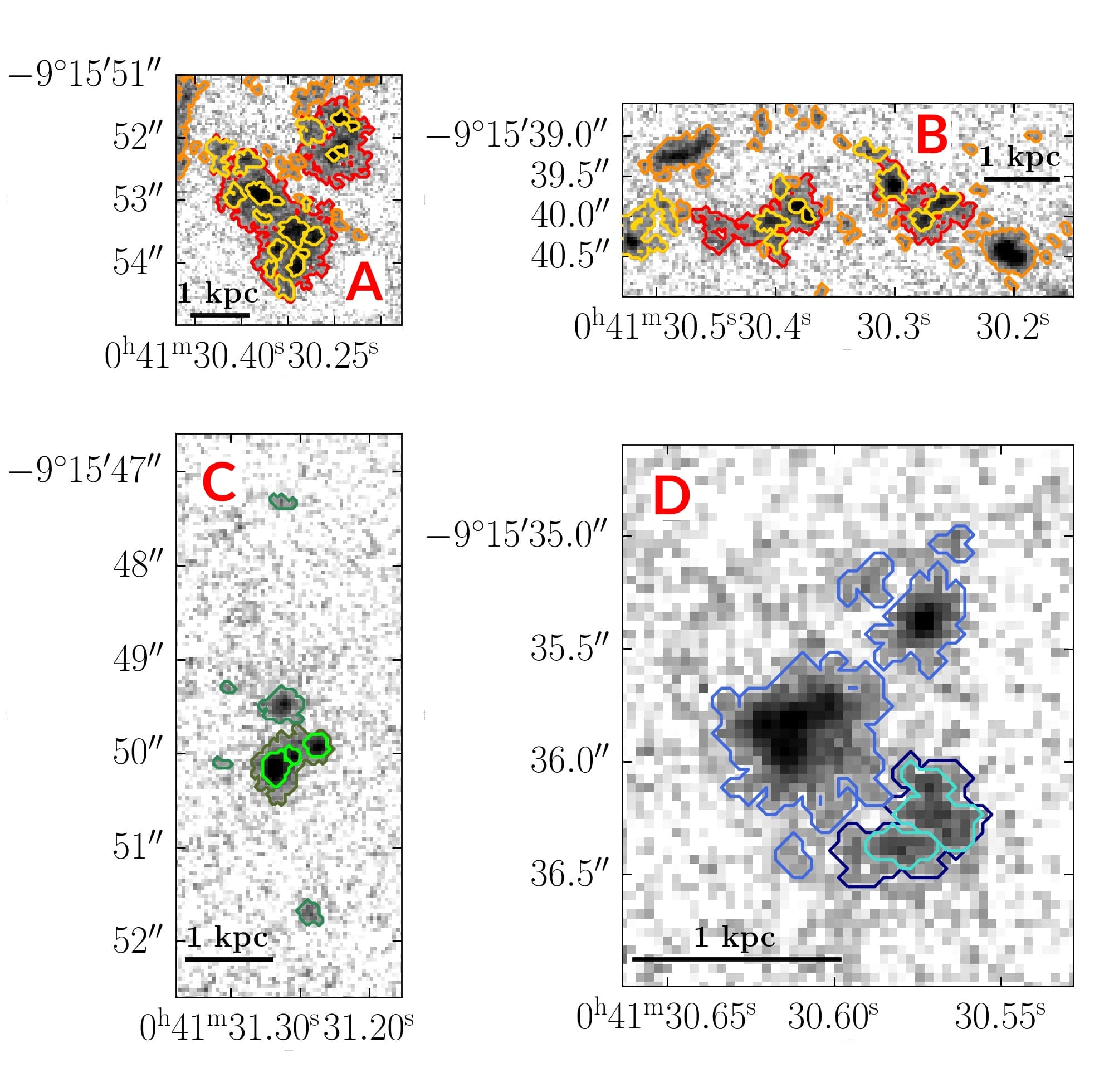}
\caption{H$\alpha$ images of the JO201 regions highlighted in the upper right panel of figure \ref{fig:detected_clumps}. The colors are the same of figure \ref{fig:detected_clumps}.}
\label{JO201_clumps_zoom}
\end{figure}

All the clumps are shown in Fig. \ref{fig:detected_clumps} (the complete figure set is available in the online journal). Here the disk, extraplanar and tail clumps can be seen in different colors, and their hierarchical, tree structure can be appreciated from the color shading. 
In Fig. \ref{JO201_clumps_zoom}, we show zoomed-in examples of H$\alpha$-selected clumps in JO201, to illustrate the hierarchical structure and the irregular morphologies of these clumps.

\figsetstart
\figsetnum{6}
\figsettitle{H$\alpha$- and UV-selected clumps detected in our sample of galaxies.}

\figsetgrpstart
\figsetgrpnum{6.1}
\figsetgrptitle{JO201}
\figsetplot{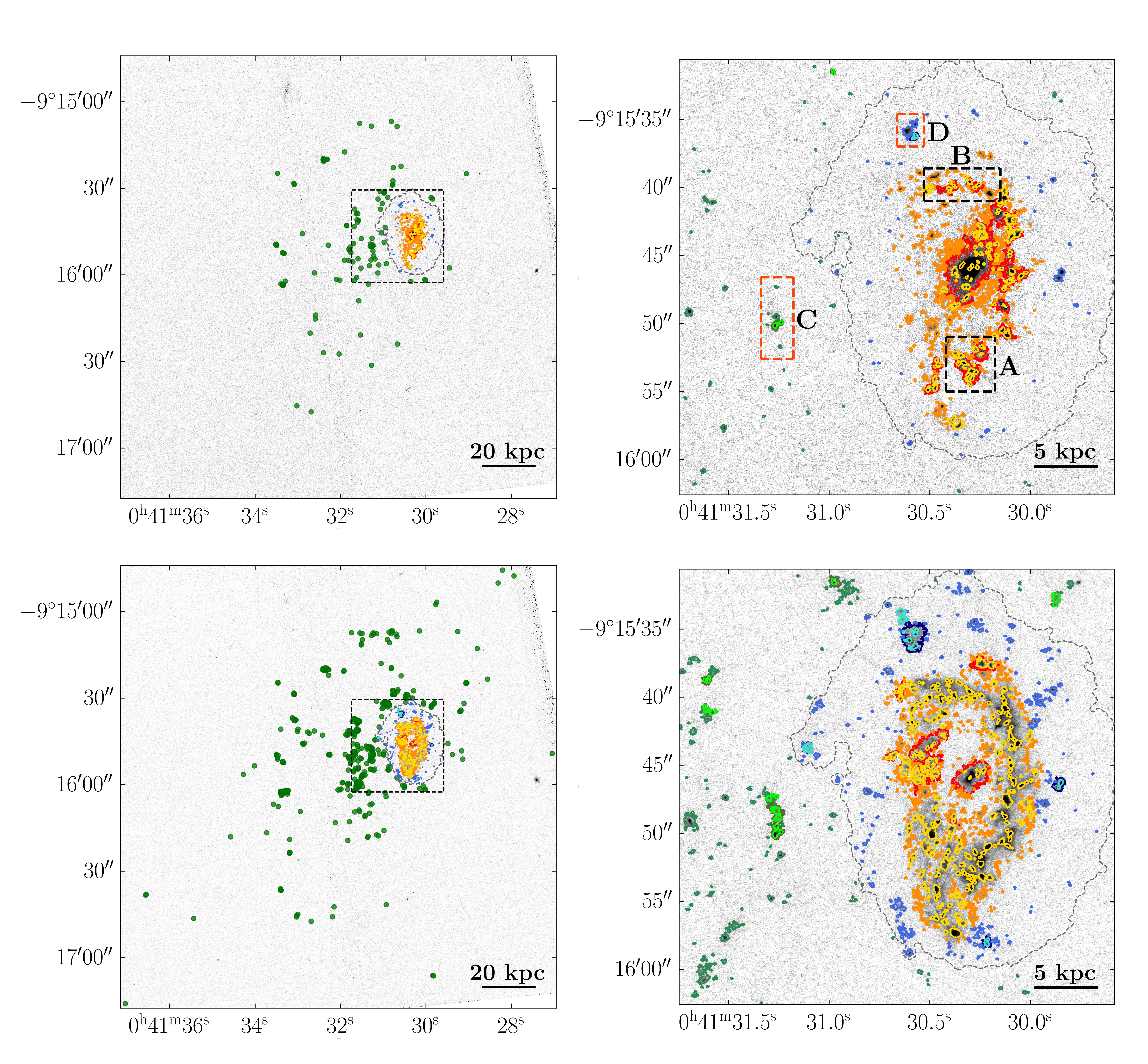}
\figsetgrpnote{Star-forming clumps detected in JO201.}
\figsetgrpend

\figsetgrpstart
\figsetgrpnum{6.2}
\figsetgrptitle{JO175}
\figsetplot{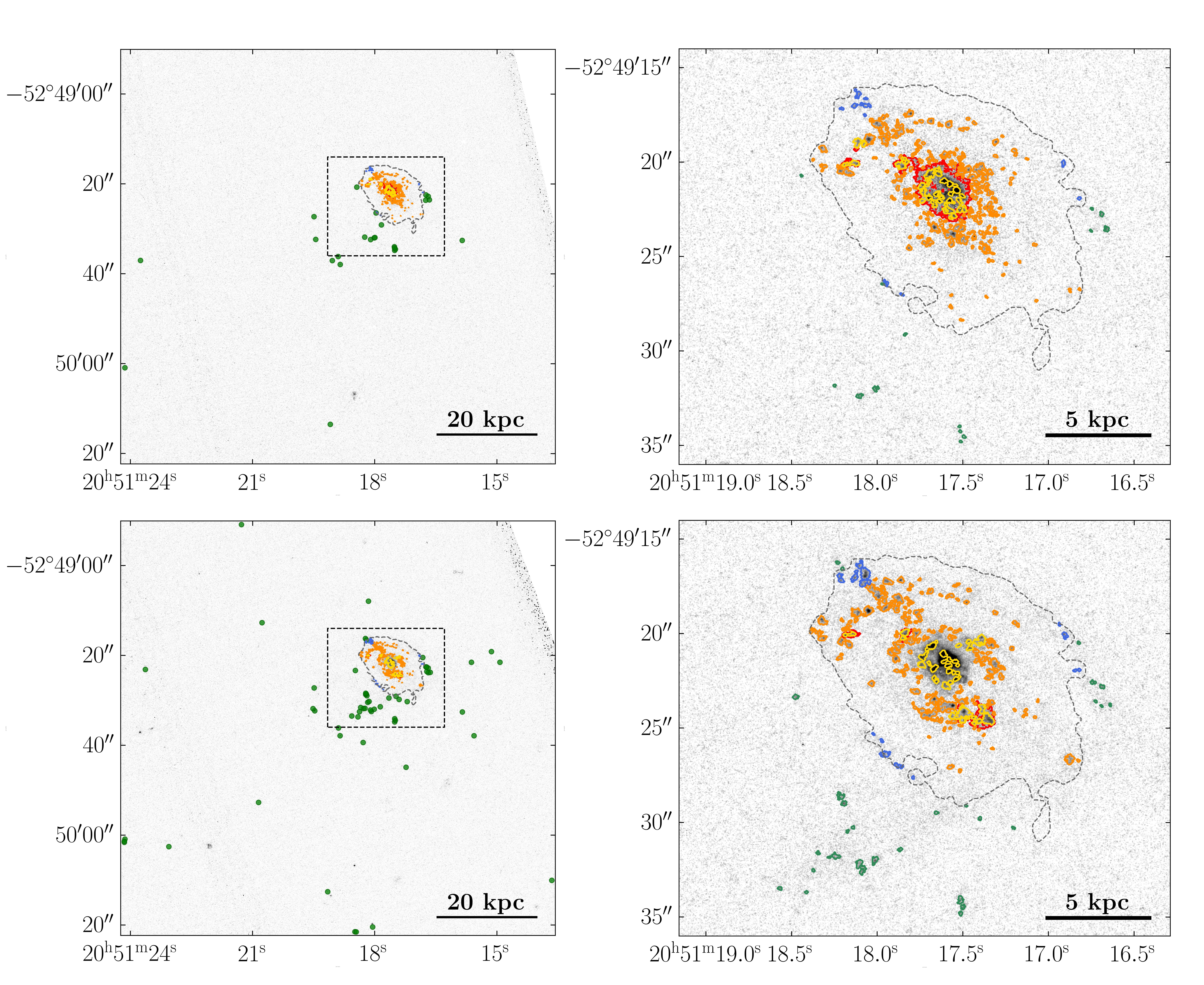}
\figsetgrpnote{Star-forming clumps detected in JO175.}
\figsetgrpend

\figsetgrpstart
\figsetgrpnum{6.3}
\figsetgrptitle{JO204}
\figsetplot{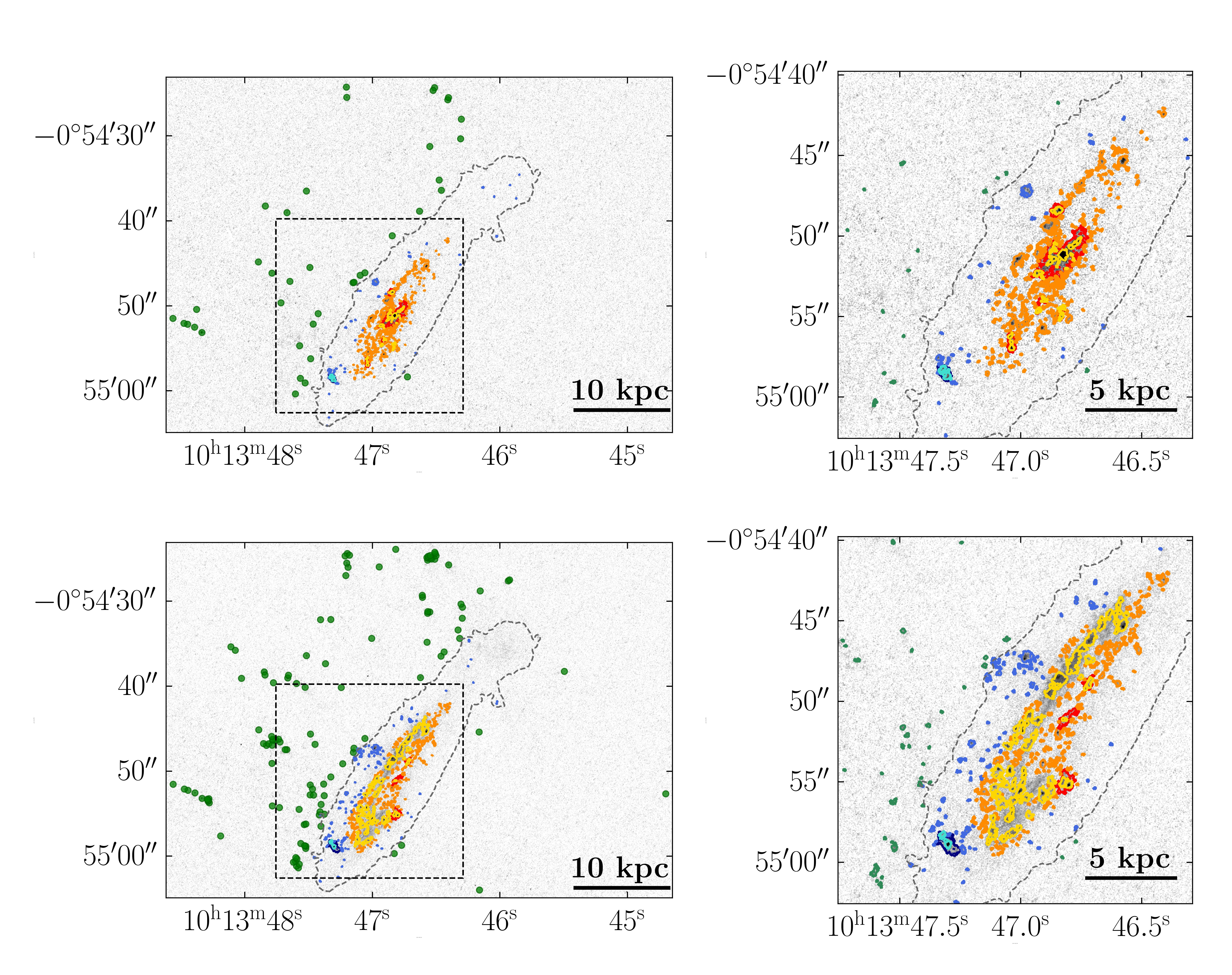}
\figsetgrpnote{Star-forming clumps detected in JO204.}
\figsetgrpend

\figsetgrpstart
\figsetgrpnum{6.4}
\figsetgrptitle{JO206}
\figsetplot{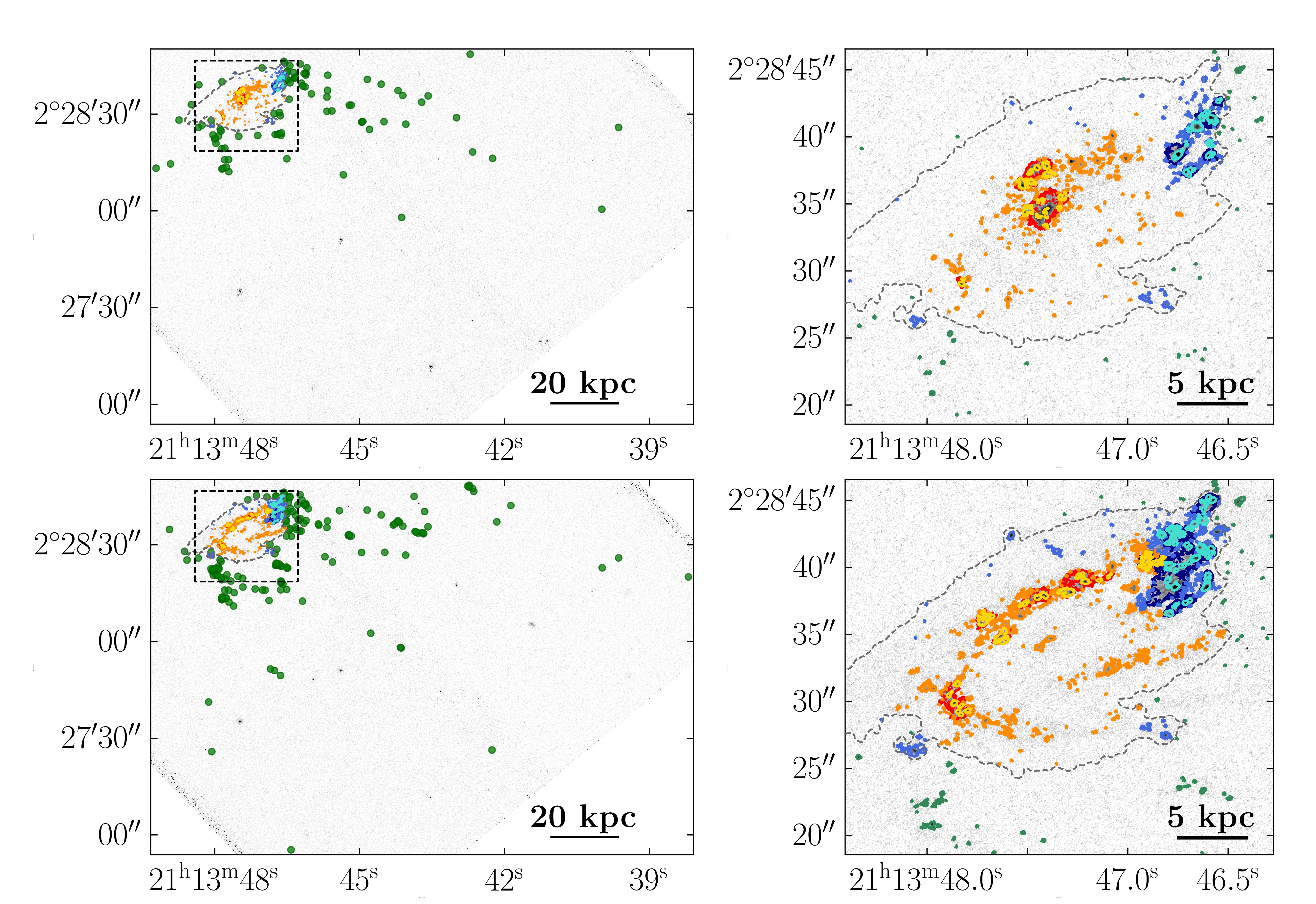}
\figsetgrpnote{Star-forming clumps detected in JO206.}
\figsetgrpend

\figsetgrpstart
\figsetgrpnum{6.5}
\figsetgrptitle{JW39}
\figsetplot{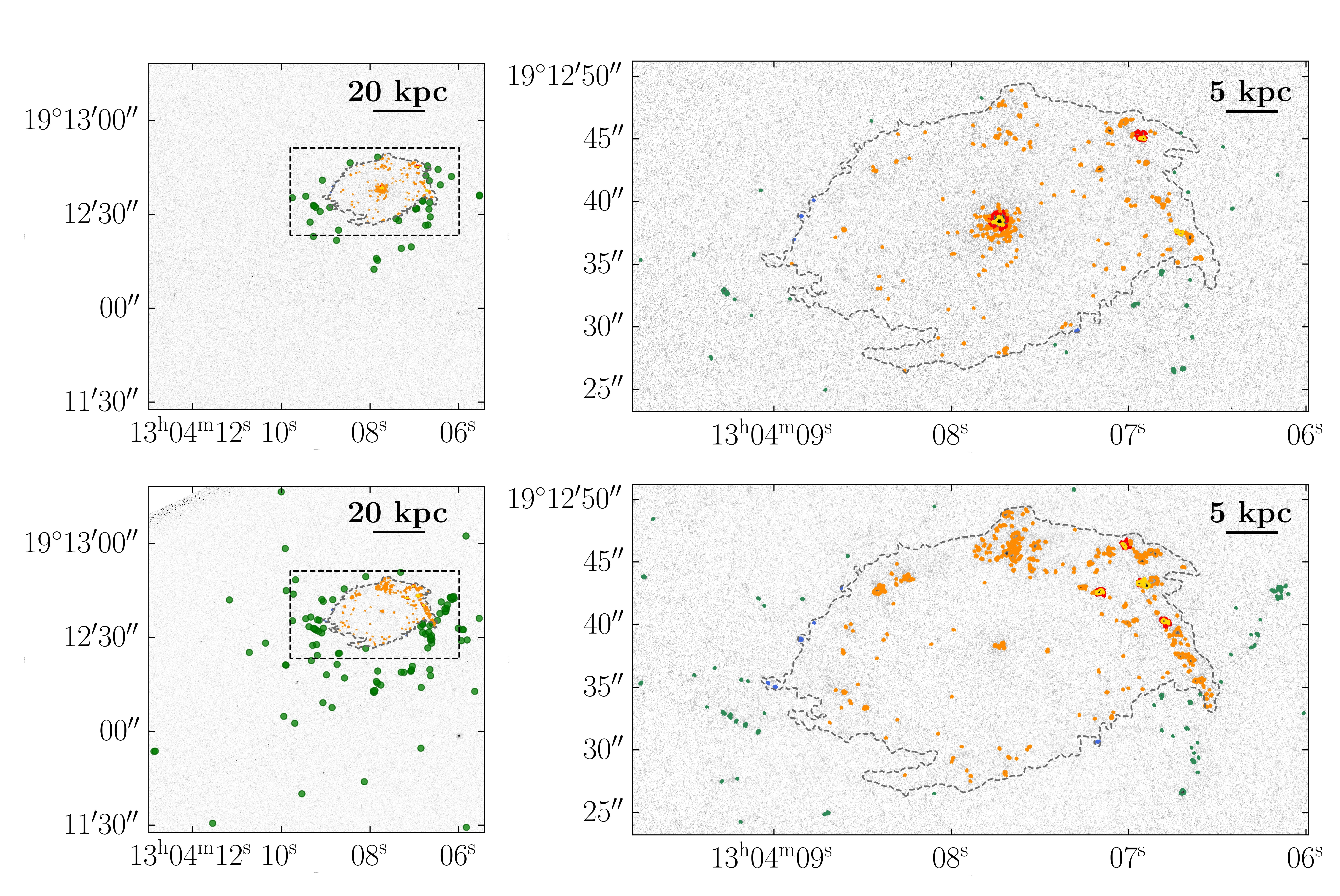}
\figsetgrpnote{Star-forming clumps detected in JW39.}
\figsetgrpend

\figsetgrpstart
\figsetgrpnum{6.6}
\figsetgrptitle{JW100}
\figsetplot{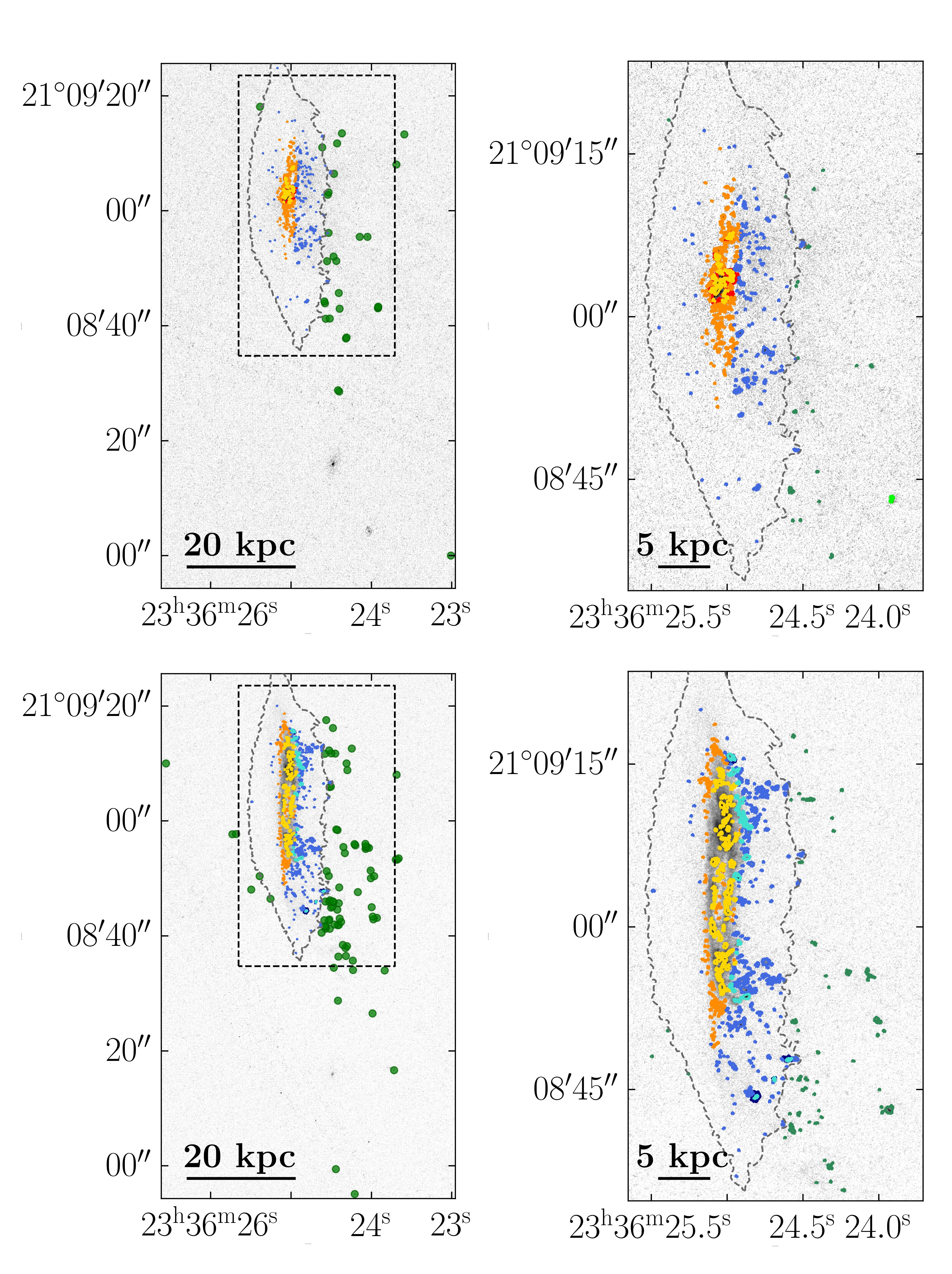}
\figsetgrpnote{Star-forming clumps detected in JW100.}
\figsetgrpend

\figsetend

The largest clumps are found in the disks of JO175, JO201 and JO206 and in the extraplanar region of JO206. As shown by Fig. \ref{JO201_clumps_zoom}, large disk clumps (red contours) typically contain several sub-clumps (yellow contours), while extraplanar and especially tail clumps often have only one level. One can also appreciate the effects of RPS on extraplanar clumps, like the filamentary structures in JO206 and JW100, which are particularly bright in UV (lower right panels in Figs. 6.4 and 6.6).

In the tails, clumps are often aligned in extended linear or arched structures, suggesting the presence of many sub-tails in each galaxy (as already noticed in \citealt{Bellhouse2021} and \citealt{Franchetto2020}, who found sub-tails in these galaxies from MUSE images). Whether clump and complex properties correlate with distance from the galaxy or along its sub-tails and how they are influenced by the properties of the hosting galaxy are beyond the scope of this paper and will be investigated in future works.

\begin{figure}
\centering
\includegraphics[width=0.45\textwidth]{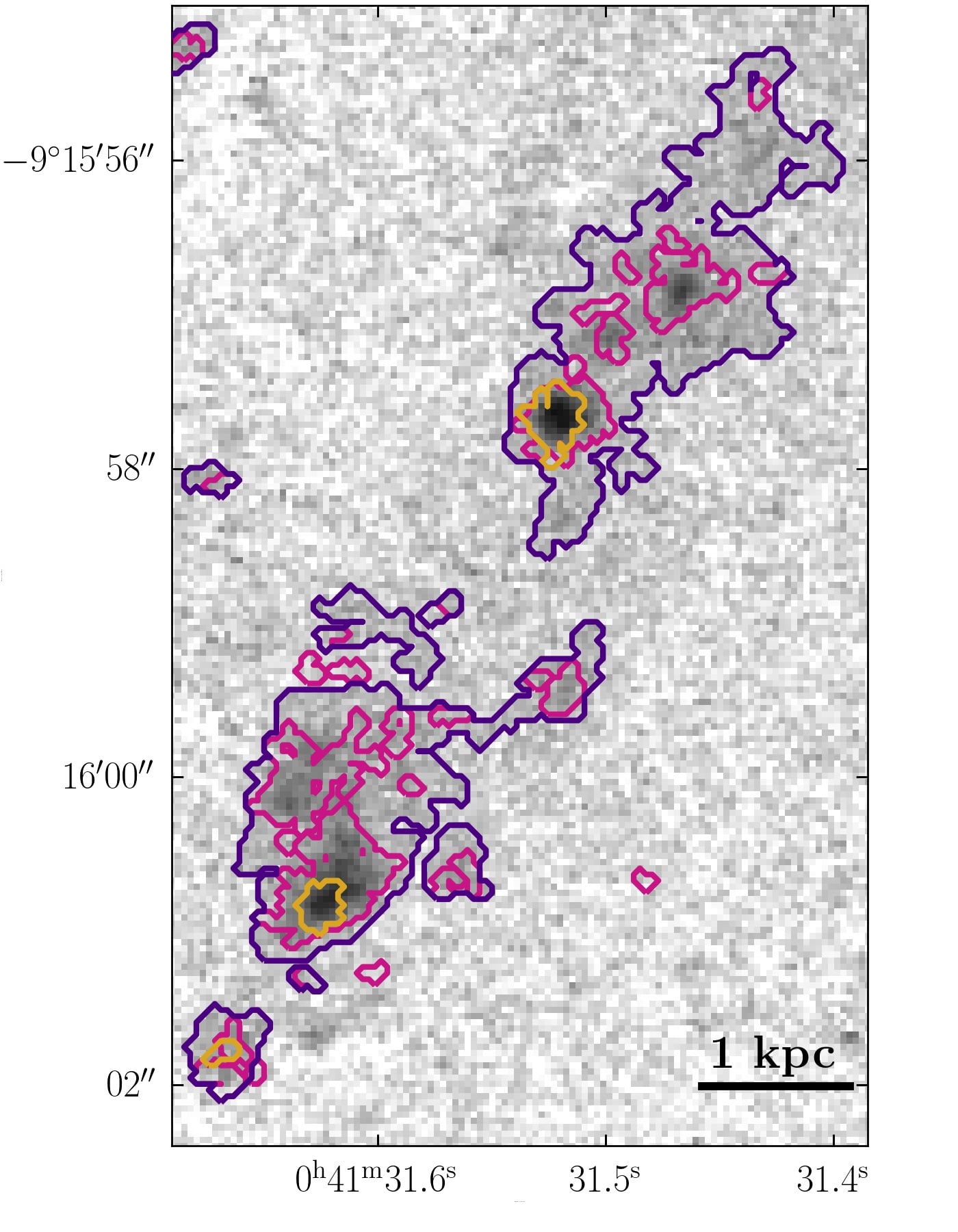}
\caption{Zoomed-in F606W image of some star-forming complexes and clumps in JO201. H$\alpha$-selected clumps are plotted in dark orange, UV-selected clumps in violet and complexes in dark violet.}
\label{JO201_complexes}
\end{figure}

Fig. \ref{JO201_complexes} shows a zoomed-in F606W image of some structures in the tails of JO201, to better appreciate the different spatial distributions of star-forming complexes (dark violet contours), H$\alpha$-selected clumps (dark orange) and UV-selected clumps (violet).
Typically, an H$\alpha$-selected clump has a corresponding UV-selected clump, while the viceversa is not true. Indeed, the number of UV-selected clumps is higher than the number of H$\alpha$-selected ones (see below). Furthermore, the corresponding UV-selected clump is generally bigger and almost completely encompasses the H$\alpha$-selected clump. Similarly, star-forming complexes contain many UV- and H$\alpha$-selected clumps, embedded in fainter, optical regions.

The number of star-forming clumps and complexes per galaxy is given in Table \ref{clu:table} and Table \ref{complexes:table}, respectively. In total, including all galaxies, our LT sample comprises 2406 H$\alpha$-selected clumps (1708 disk clumps, 375 extraplanar clumps and 323 tail clumps), 3745 UV-selected clumps (2021 disk clumps, 825 extraplanar clumps and 899 tail clumps) and 424 star-forming complexes. Typically, 98-99\% of the selected clumps are leaves (including also simple trunks with no substructures inside), while the trunks containing leaves represent only 1-2\% of the whole sample (the fraction increases to 7-14\% when restricting the analysis only to resolved clumps).\par

Avoiding AGN areas and including both resolved and unresolved clumps, $\sim 21$\% of the $\rm H\alpha$-selected and $\sim 7$\% of the UV-selected clumps get excluded. The percentage is smaller in the latter, indicating that UV-selected clumps are more preferentially located out of AGN regions than H$\alpha$-selected clumps. Most of these are disk clumps, as expected, but a few of them can be found in the ionization cone of the AGN, whose extension can reach into the extraplanar region. The exact numbers are listed in brackets in Table \ref{clu:table}.

\begin{deluxetable*}{ll||c|ccc||c|ccc}
\tablecaption{Number of clumps in each sub-sample.}
\tablewidth{0pt}
\tablehead{
\colhead{} & \colhead{} & \multicolumn{4}{c}{LT sample} & \multicolumn{4}{c}{resolved sample} \\\cmidrule(lr){3-6}\cmidrule(lr){7-10}
\colhead{Filter} & \colhead{gal} & \colhead{$\mathrm{N_{LT}}$} & \colhead{$\mathrm{N_{d}}$} & \colhead{$\mathrm{N_{e}}$} & \colhead{$\mathrm{N_{t}}$} & \colhead{$\mathrm{n_{res}}$} & \colhead{$\mathrm{n_{d}}$} & \colhead{$\mathrm{n_{e}}$} & \colhead{$\mathrm{n_{t}}$}
}
\decimalcolnumbers
\startdata
    & JO$175$ & 290(290) & 252(252) & 14(14) & 24(24) & 38(38) & 37(37) & 0(0) & 1(1)\\
    & JO$201$ & 663(476) & 507(321) & 51(49) & 105(105) & 115(88) & 96(69) & 4(4) & 15(15)\\
    & JO$204$ & 373(296) & 290(219) & 44(38) & 39(39) & 40(30) & 32(23) & 5(4) & 3(3)\\
H$\alpha$ & JO$206$ & 438(377) & 234(173) & 117(117) & 87(87) & 44(37) & 24(17) & 19(19) & 1(1)\\
    & JW$39$ & 235(168) & 192(125) & 4(4) & 39(39) & 14(12) & 11(9) & 0(0) & 3(3)\\
    & JW$100$ & 407(284) & 233(139) & 145(116) & 29(29)& 35(18) & 24(9) & 9(7) & 2(2)\\\cmidrule(lr){2-10}
    & tot & 2406(1891) & 1708(1229) & 375(339) & 323(323) & 286(223) & 224(164) & 37(34) & 25(25)\\\cmidrule(lr){1-10}
    & JO$175$ & 287(287) & 211(211) & 17(17) & 59(59) & 47(47) & 38(38) & 2(2) & 7(7)\\
    & JO$201$ & 1244(1100) & 659(518) & 213(209) & 372(372) & 233(211) & 143(122) & 29(28) & 61(61)\\
    & JO$204$ & 531(475) & 302(258) & 110(102) & 119(115) & 82(78) & 63(59) & 9(9) & 10(10)\\
UV & JO$206$ & 741(733) & 392(384) & 186(186) & 163(163) & 106(104) & 51(49) & 41(41) & 14(14)\\
    & JW$39$ & 355(349) & 243(237) & 6(6) & 106(106) & 31(31) & 26(26) & 0(0) & 5(5)\\
    & JW$100$ & 587(533) & 214(178) & 293(274) & 80(80) & 92(78) & 48(36) & 37(35) & 7(7)\\\cmidrule(lr){2-10}
    & tot & 3745(3476) & 2021(1787) & 825(794) & 899(895) & 591(549) & 369(330) & 118(115) & 104(104)\\
\enddata
\tablecomments{Number of clumps detected in each galaxy and depending on the spatial category. From left to right: photometric band in which the clumps were detected (column 1), name of the galaxy (2), number of LT clumps (3), number of disk-LT clumps (4), number of extraplanar-LT clumps (5), number of tail-LT clumps (6), number of resolved clumps (7), number of resolved-disk clumps (8), number of resolved-extraplanar clumps (9), number of resolved-tail clumps (10). In brackets, the number of clumps in the each sample, but selected in order to avoid regions powered by AGN emission (see Sect. \ref{sf:clumps}).}
\label{clu:table}
\end{deluxetable*}

\begin{table}
\centering
\caption{Number of star-forming complexes detected in the tails of each galaxy.}
\begin{tabular}{cc}
\toprule\toprule
gal & N\\\hline
JO175 & 31\\
JO201 & 129\\
JO204 & 53\\
JO206 & 92\\
JW39 & 73\\
JW100 & 46\\\hline
tot & 424\\
\bottomrule
\end{tabular}
\label{complexes:table}
\end{table}

Only 12\% of H$\alpha$-selected and 16\% of UV-selected clumps are spatially resolved, which means that the majority of the clumps have diameters smaller than $\sim 140$ pc. Most of the resolved clumps are star-forming according to the BPT, except in the disk where about 25\% of the H$\alpha$-resolved clumps are flagged as AGN or LINER. 

\begin{figure*}
\centering
\includegraphics[width=0.9\textwidth]{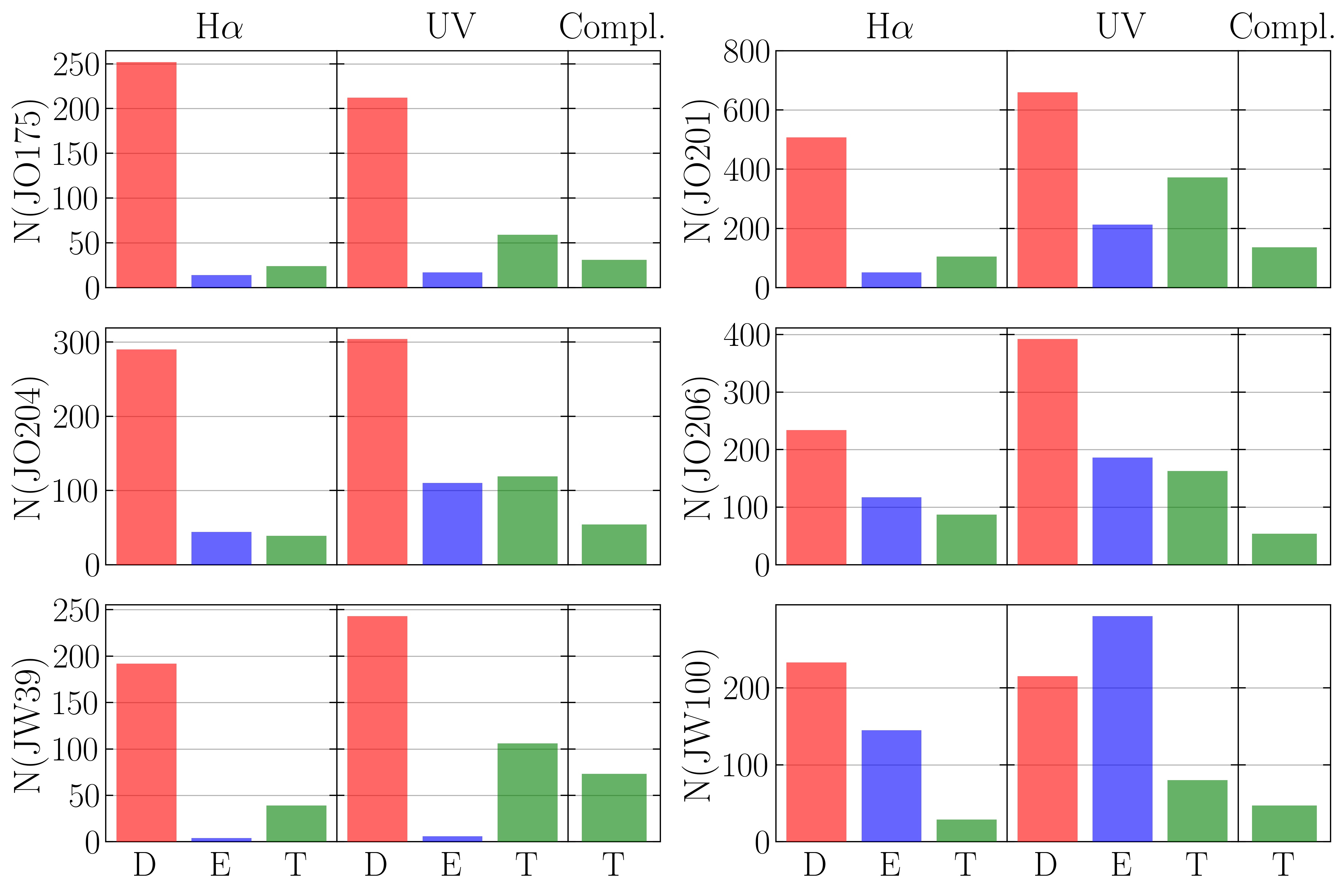}
\caption{Histograms of the number of star-forming clumps and complexes in each galaxy, divided according to the selection filter and the spatial category. For each galaxy three panels are shown, with the number of H$\alpha$-selected clumps (left panel), UV-selected clumps (middle panel) and star-forming complexes (right panel), divided according to their spatial category: disk (red), extraplanar (blue) and tail (green).}
\label{hist_gals}
\end{figure*}

In Fig. \ref{hist_gals} we plot the histograms of the number of clumps per galaxy, divided according to the selection band (H$\alpha$ or UV) and the spatial category (disk, extraplanar and tail), together with the number of complexes.

In most cases disk clumps are much more numerous than extraplanar and tail clumps, regardless of the selection filter, with the only exception being the UV-selected clumps in JW100, which is seen edge-on and is stripped mostly on the plane of the sky \citep{Poggianti2019b}, thus in the most favorable conditions to appreciate the extraplanar clumps. For what concerns the number of extraplanar and tail clumps, the prevalence of one over the other depends on the galaxy: in JW100 the number of extraplanar clumps is much larger than that of tail clumps in both the selection filters, in JO204 and JO206 they are almost of the same number, while in the other galaxies tail clumps are more numerous than extraplanar. The number of clumps in each category clearly depends on both the disk inclination and the stripping direction with respect to the line of sight.

Furthermore, with the only exception being disk clumps in JO175, for the same spatial category there are more UV-selected clumps than H$\alpha$-selected ones. This indicates that there are a number of stellar-only clumps, with little or no ionized gas left.

The number of star-forming complexes in the tails of the galaxies is generally smaller than the number of tail UV-selected clumps, but larger than that of tail H$\alpha$-selected ones, with the only exception being JO206, suggesting that many complexes are matched only to UV-selected clumps, without any H$\alpha$ counterpart.

\section{Distribution functions}\label{sec:df}
The luminosity (size) distribution function (LDF and SDF hereafter) is defined as the number of sources per luminosity (size) bin normalized by the width of the luminosity (size) bin itself and by the total number of sources in the sample and is a useful tool to study the statistical properties of the star-forming clumps. As described is Sec. \ref{sec:intro}, they are typically well described by a power law, and seem to be good proxies of the environmental effects on the star-formation process and on the clustering properties of the clumps.

\subsection{Luminosity distribution functions}\label{lum:dis}
\begin{figure*}
\centering
\gridline{\resizebox{\textwidth}{!}{\includegraphics[height=1cm]{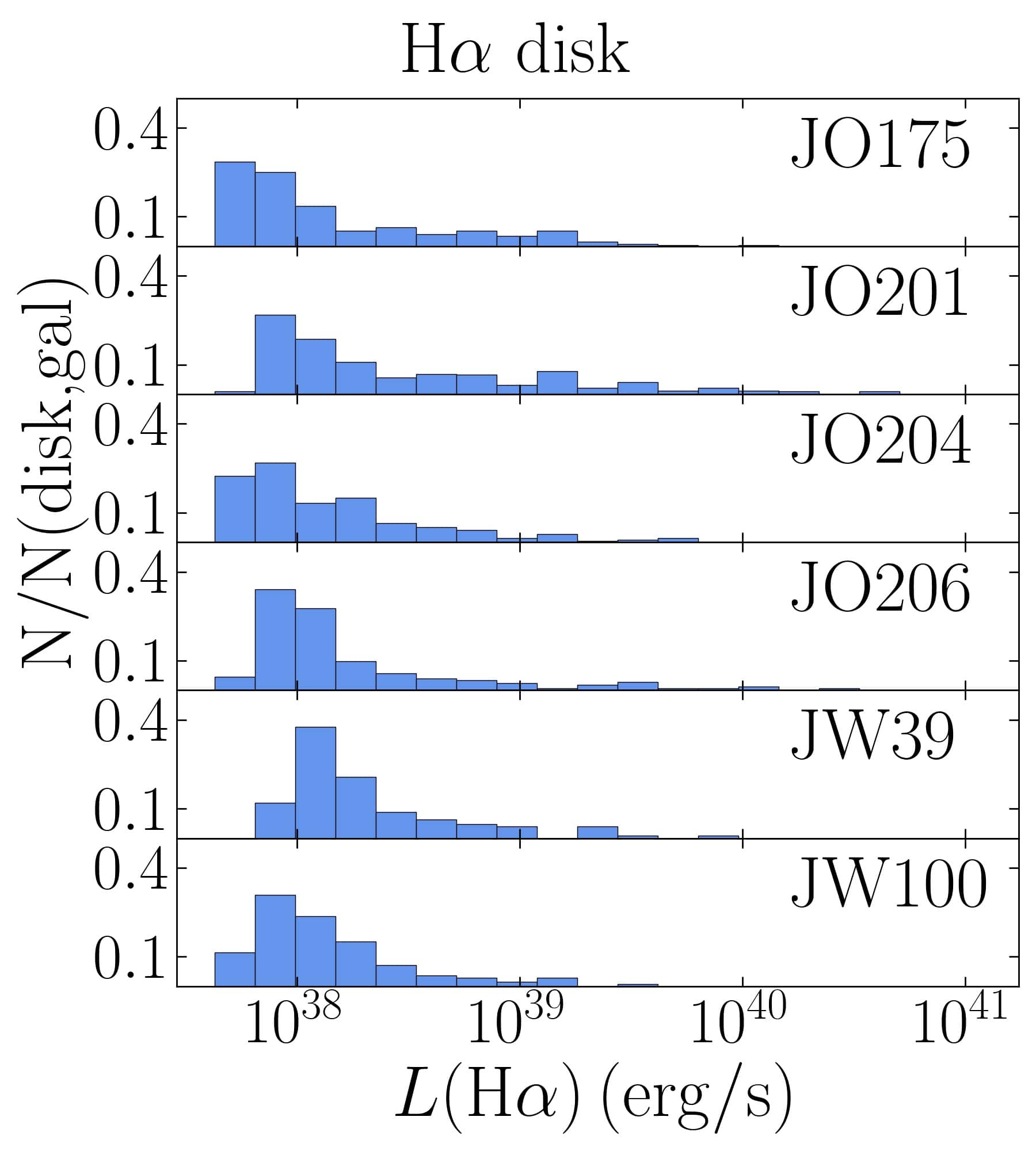}
          \includegraphics[height=1cm]{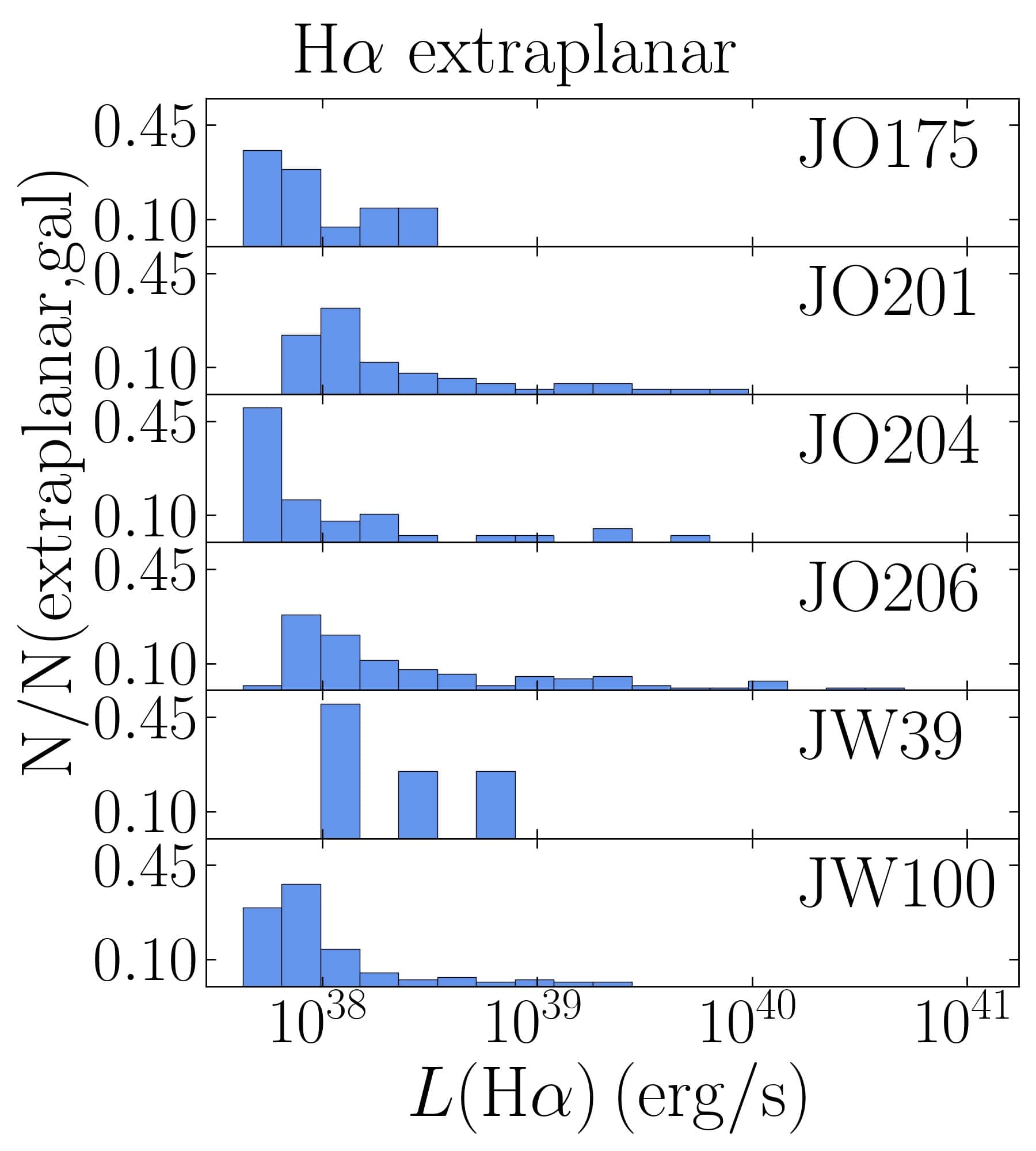}
          \includegraphics[height=1cm]{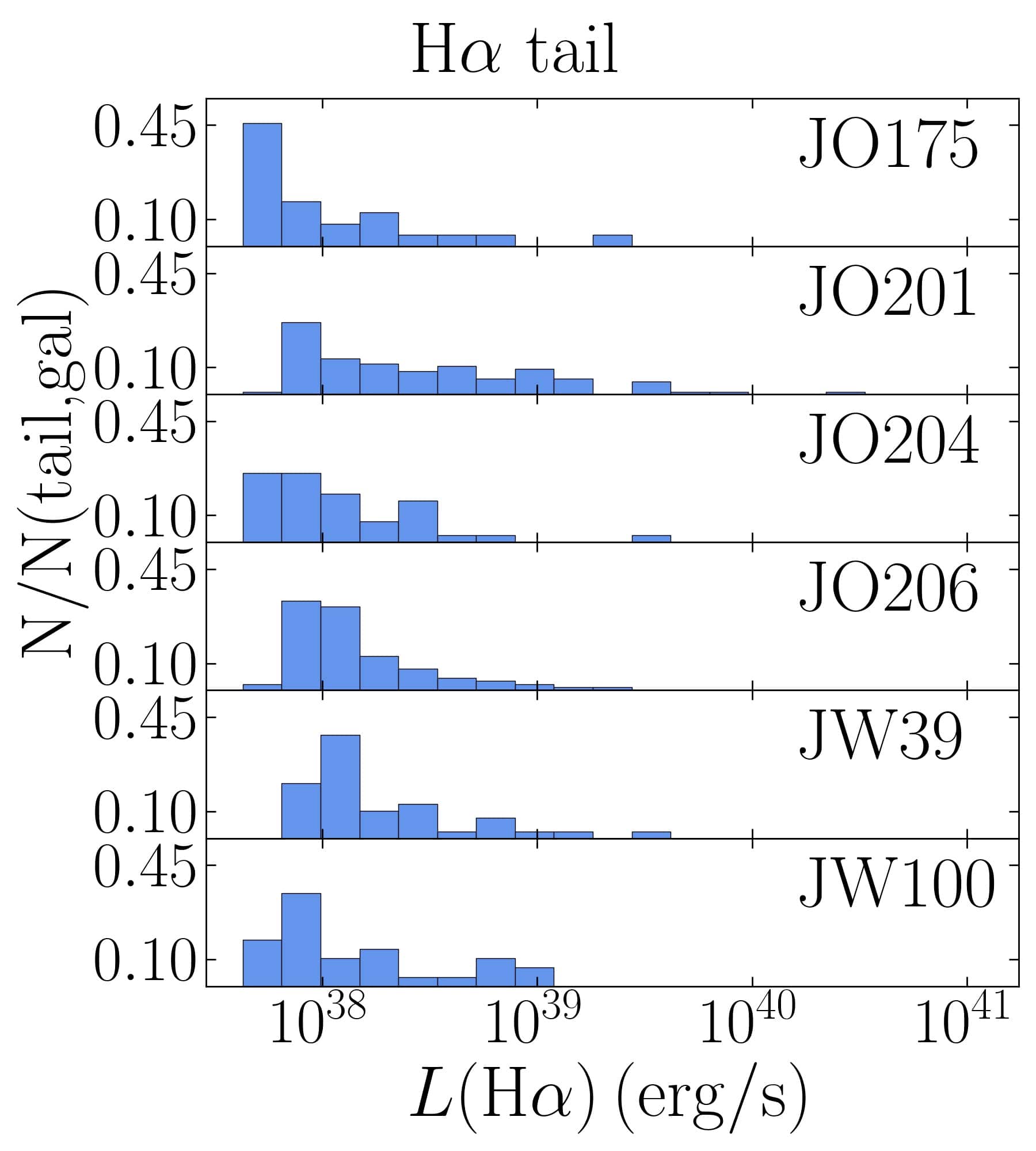}
          }}
\gridline{\resizebox{\textwidth}{!}{\includegraphics[height=1cm]{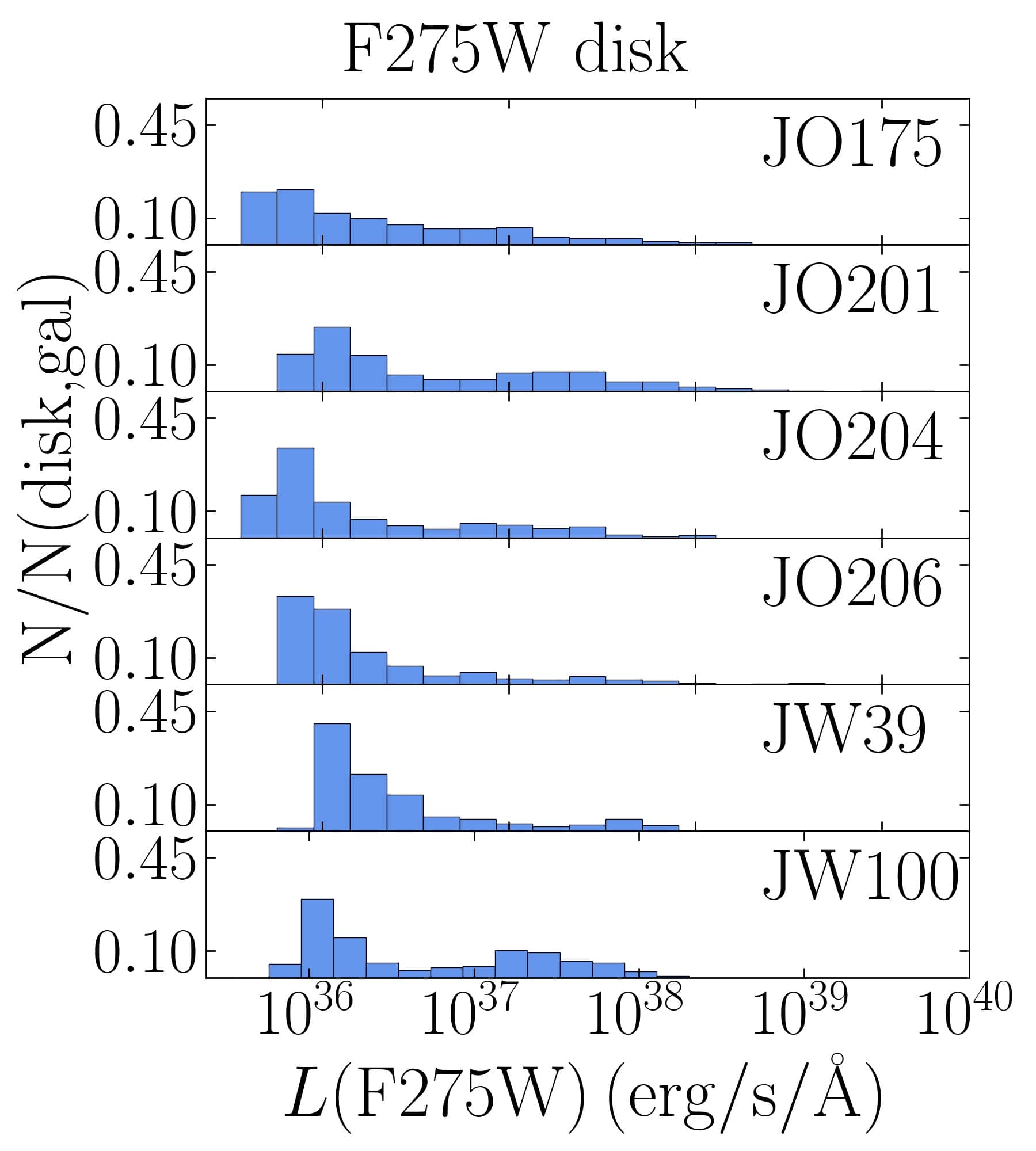}
          \includegraphics[height=1cm]{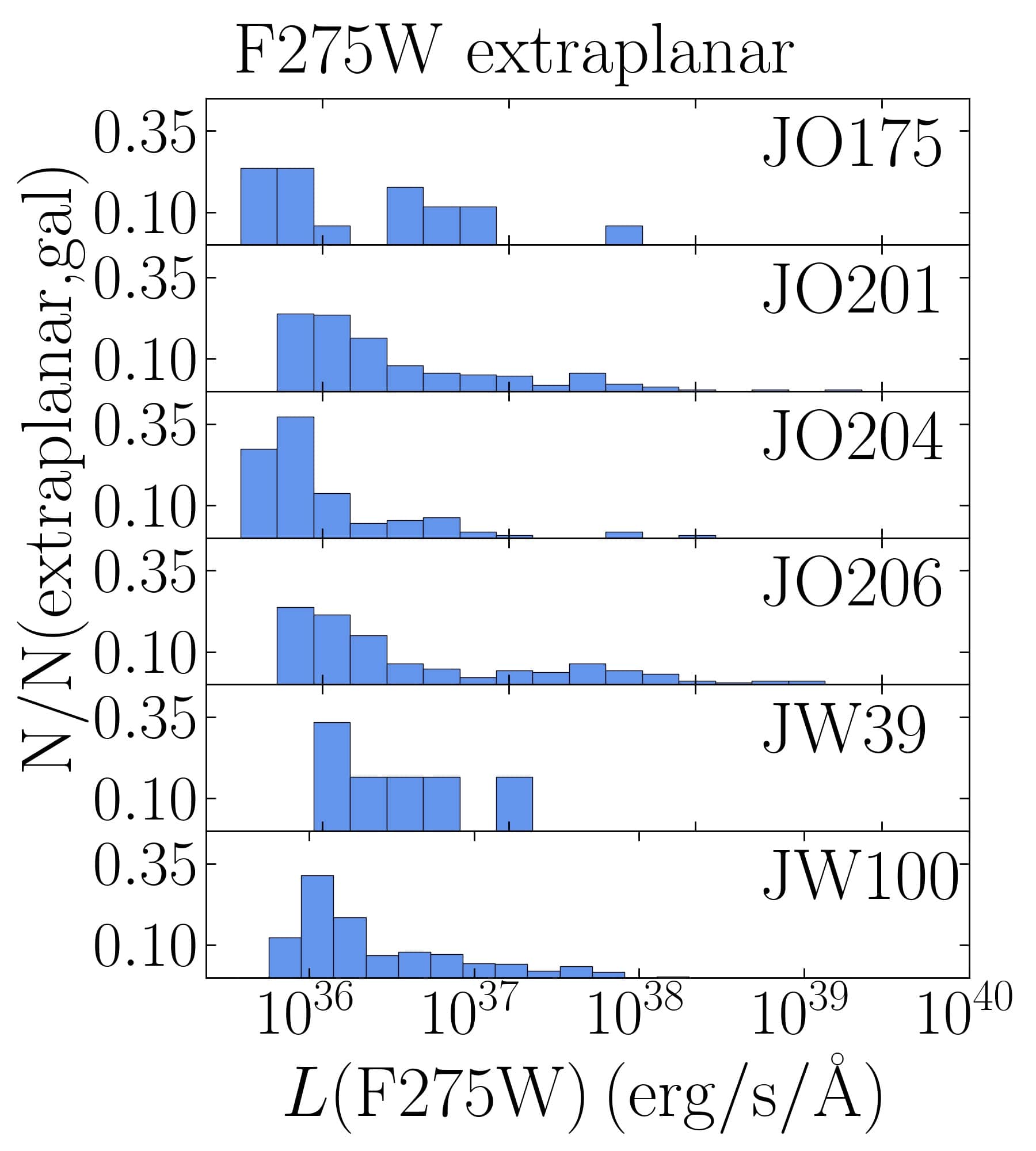}
          \includegraphics[height=1cm]{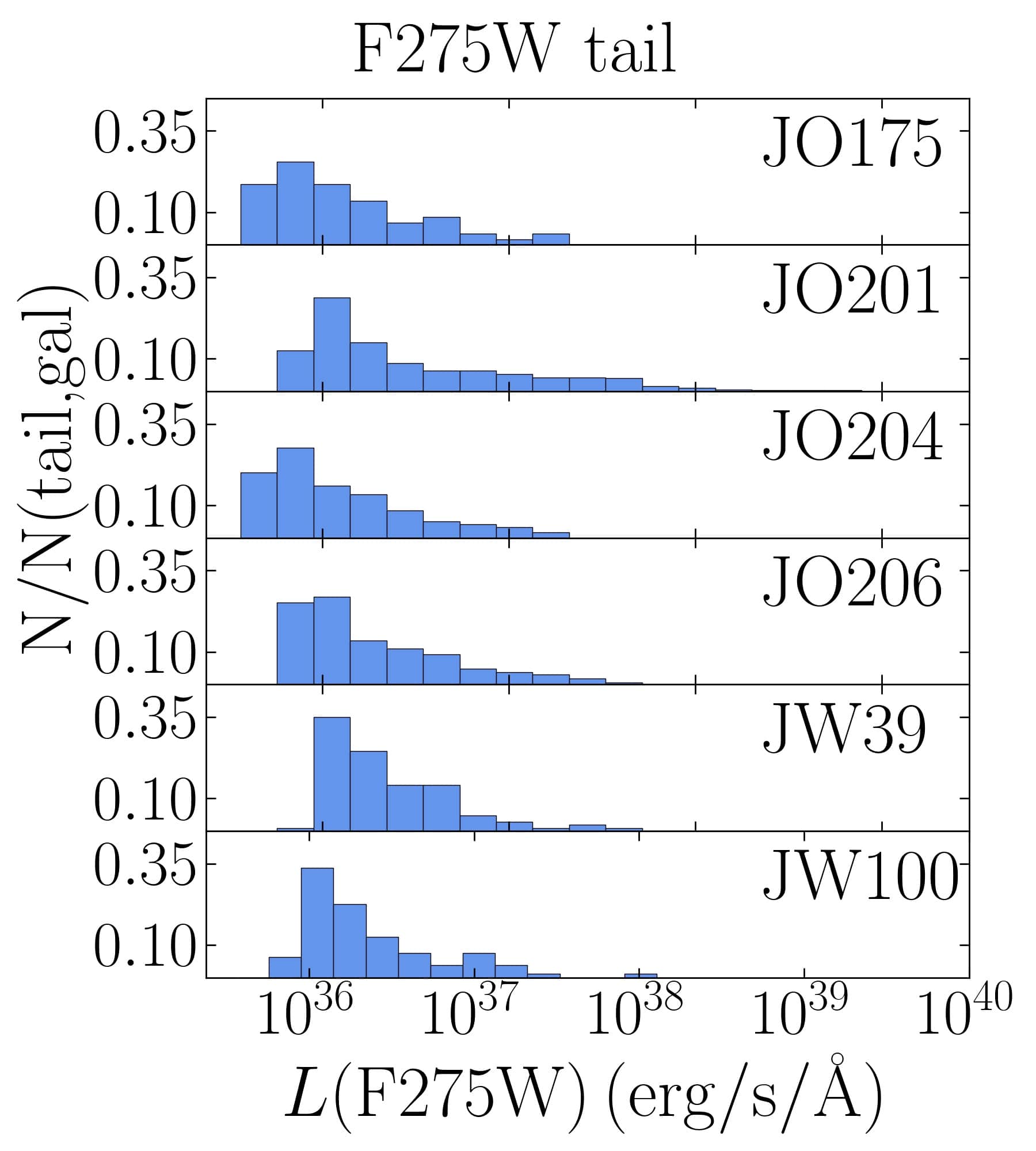}
          }}
\caption{Fraction of H$\alpha$-selected (top row) and UV-selected (bottom row) clumps per spatial category and per galaxy. In each row, from left to right: disk, extraplanar and tail. Y-axis is normalized for the number of clumps in the galaxy and in the spatial category. Notice that the H$\alpha$ luminosity of the H$\alpha$-selected clumps is the integrated emission of the H$\alpha$ line, therefore in erg/s, while the UV luminosity of the UV-selected clumps is in erg/s/$\mathrm{\AA}$.}
\label{lum_hist}
\end{figure*}

\begin{figure}
\centering
\includegraphics[width=0.4\textwidth]{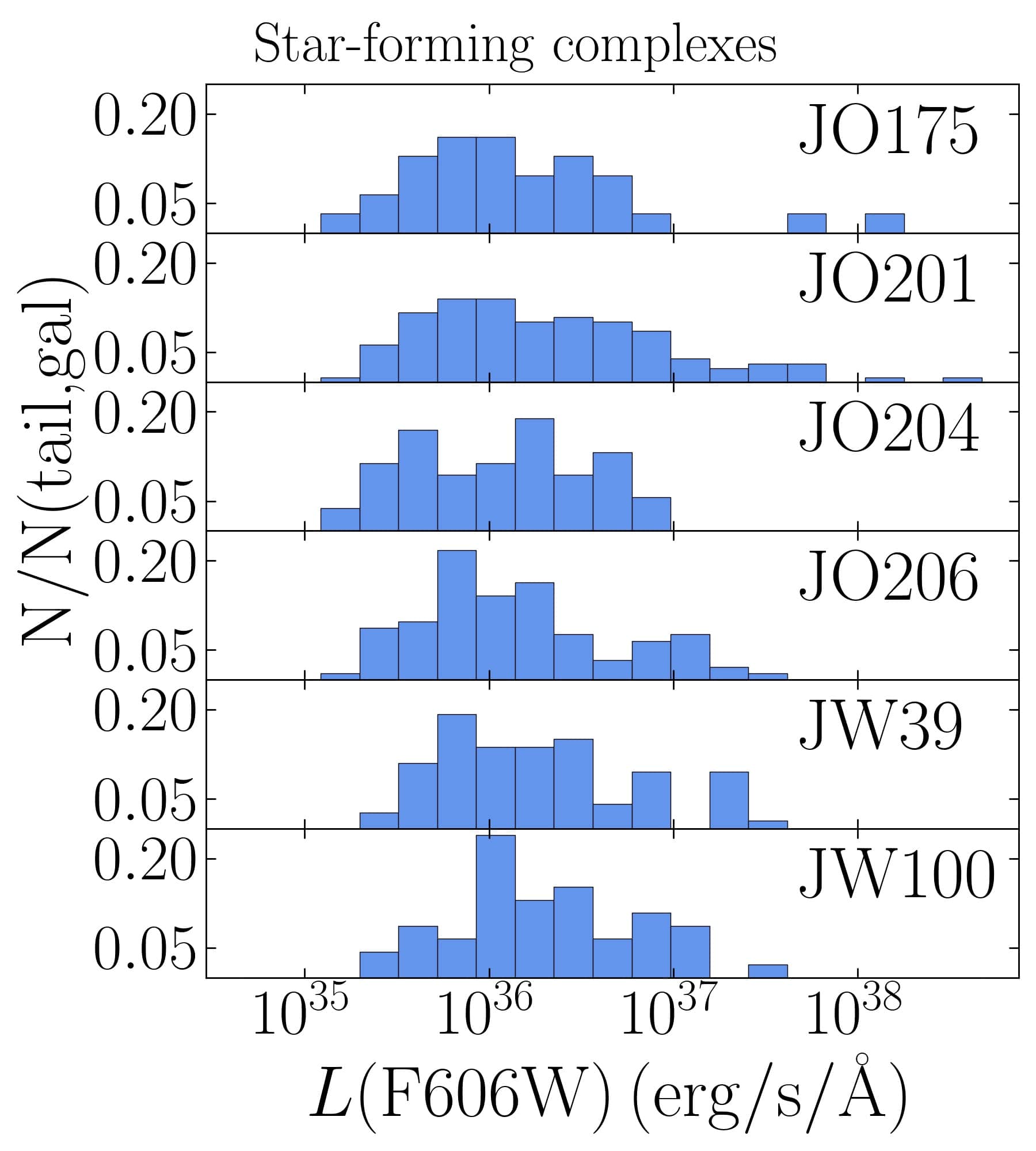}
\caption{Same as Fig. \ref{lum_hist}, but for star-forming complexes.}
\label{lum_hist_compl}
\end{figure}

Figs. \ref{lum_hist} and \ref{lum_hist_compl} show the histograms of the clumps in each spatial category and in each galaxy, binned in luminosity. The $y$-axis of plots are normalized by the total number of clumps in the spatial category and in the galaxy. Most H$\alpha$-selected clump distributions are peaked at values fainter than $\sim 10^{38}$ erg/s, at the faint-end of the luminosity dynamical range. The luminosities of the H$\alpha$-selected clumps are consistent with those of "giant" HII regions (like the Carina Nebula), whose H$\alpha$ luminosities $L(\mathrm{H}\alpha)$ are typically $10^{37-39}$ erg/s, and "super giant" HII regions (like 30 Doradus in the Large Magellanic Cloud), with $L(\mathrm{H}\alpha)>10^{39}$ erg/s \citep{Lee2011}.
As expected, the faintest clumps are observed mostly in the closest galaxies of our sample (Table \ref{gal:table}).
JO201 stands out for its population of bright
H$\alpha$-selected clumps, both in the disk and in the tail, while in the extraplanar regions the brightest clumps are those of JO206 (located in the crest to the top right of the disk, see Fig. 6.4). Similar trends are found for UV-selected clumps. Also, we point out the hint for a bimodality in the distributions of the disk UV-selected clumps of JO201 and JW100 and of the extraplanar clumps of JO206. Finally, the star-forming complexes distributions are very different from the others, since they do not peak in the faint-end of the distribution.

As done in \cite{Cook2016}, throughout this work the datapoints of the luminosity distribution functions $\mathrm{d\widetilde{N}/d}L$ (LDF) are computed fixing the number of objects while varying the bin size, in order to obtain a robust representation of the distribution function. For our LDFs we choose 20 sources per bin. The luminosity of each bin is the central luminosity of the bin. Datapoints brighter than a given peak luminosity $L_{\mathrm{peak}}$ are fitted\footnote{Varying the number of clumps per bin between 5 and 50, the slope of the fit to the LDF does not change significantly. Using a different fitting method which does not depend on the binning (\textsc{PowerLaw}, \citealt{Alstott2014,Clauset2009,Klaus2011}), results do not change significantly either.} by a power law

\begin{equation}\label{power:law}
    p(L)=K\,L^{-\alpha}\quad\mathrm{with}\,\,L\geq L_{\mathrm{peak}},
\end{equation}

\noindent where $K$ is the normalization and $\alpha$ is the slope of the power law. $L_{\mathrm{peak}}$ is chosen for each sub-sample starting from the peak value of the LDF and, if necessary, varying it in order to avoid noisy regions of the LDF.

H$\alpha$-selected clumps, UV-selected clumps and star-forming complexes are fitted independently in each spatial category, in order to study variations in the properties of the LDFs as a consequence of RPS.
We used the whole UV-selected and star-forming complexes samples, but only the BPT-selected H$\alpha$-selected clumps, in order to avoid AGN- and LINER-powered regions (see Sec. \ref{sf:clumps}). Fits were performed using the \textsc{curve\_fit} method implemented in the \textsc{SciPy}\footnote{https://docs.scipy.org/doc/scipy/index.html} Python package, with uncertainties on the LDF computed as the Poisson noise of the number of objects in the bin.

\begin{figure*}
\centering
\gridline{\includegraphics[width=\textwidth]{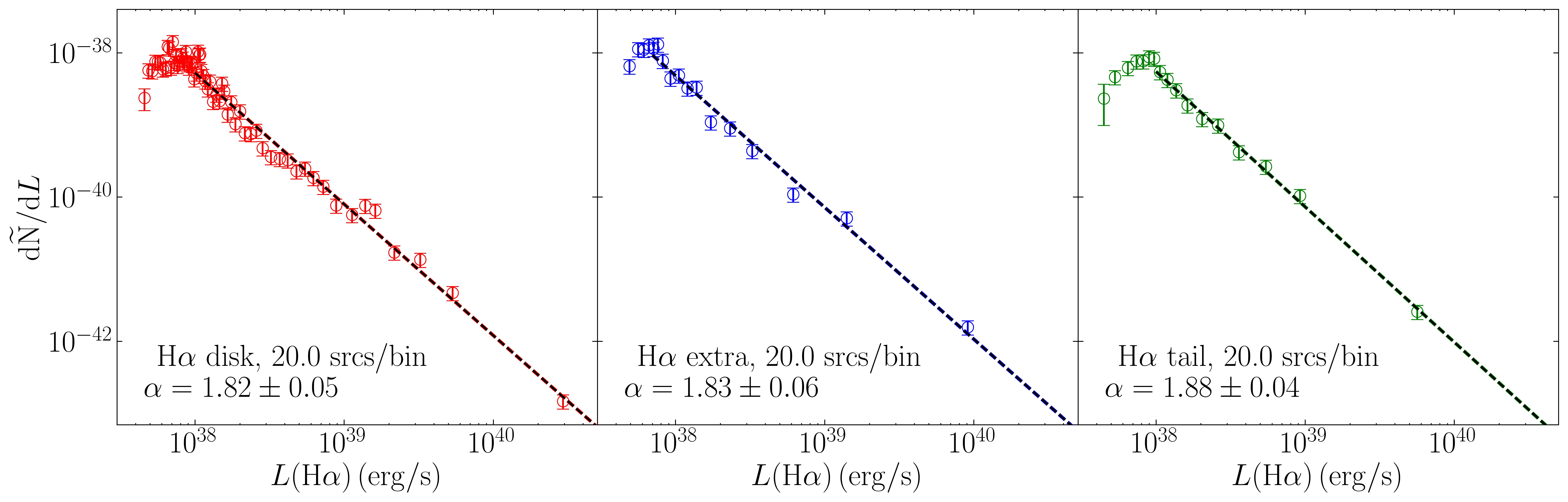}}
\gridline{\includegraphics[width=\textwidth]{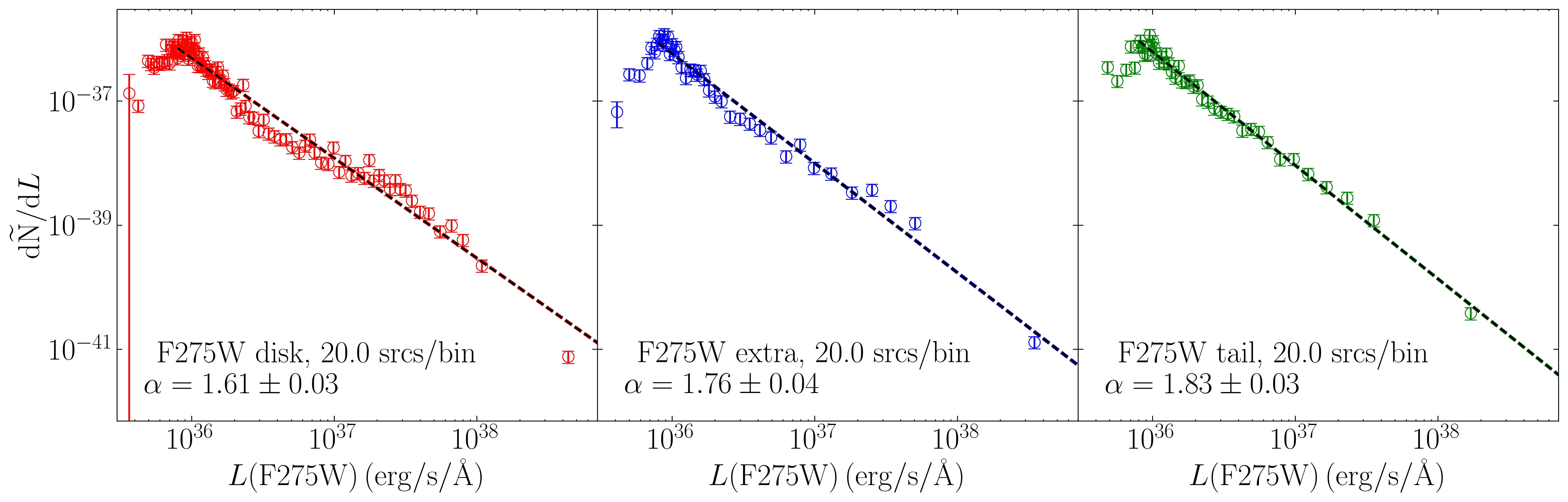}}
\caption{Luminosity distribution functions $\mathrm{d\widetilde{N}/d}L$ of H$\alpha$-selected (upper panels) and UV-selected (lower panels) clumps. Clumps are divided according to their spatial category: disk (left panels, in red), extraplanar (middle panels, in blue) and tail (right panels, in green). For each plot we show: the empirical LDF of the corresponding sample (open circles with errorbars), generated with equal-number bins (i.e. each bin contains the same number of objects, see \citealt{Cook2016}), and the best-fitting line (dashed line). Notice that the H$\alpha$ luminosity of the H$\alpha$-selected clumps is the integrated emission of the H$\alpha$ line, therefore in erg/s, while the UV luminosity of the UV-selected clumps is in erg/s/$\mathrm{\AA}$.}
\label{lum_distr}
\end{figure*}

\begin{figure}
\centering
\includegraphics[width=0.4\textwidth]{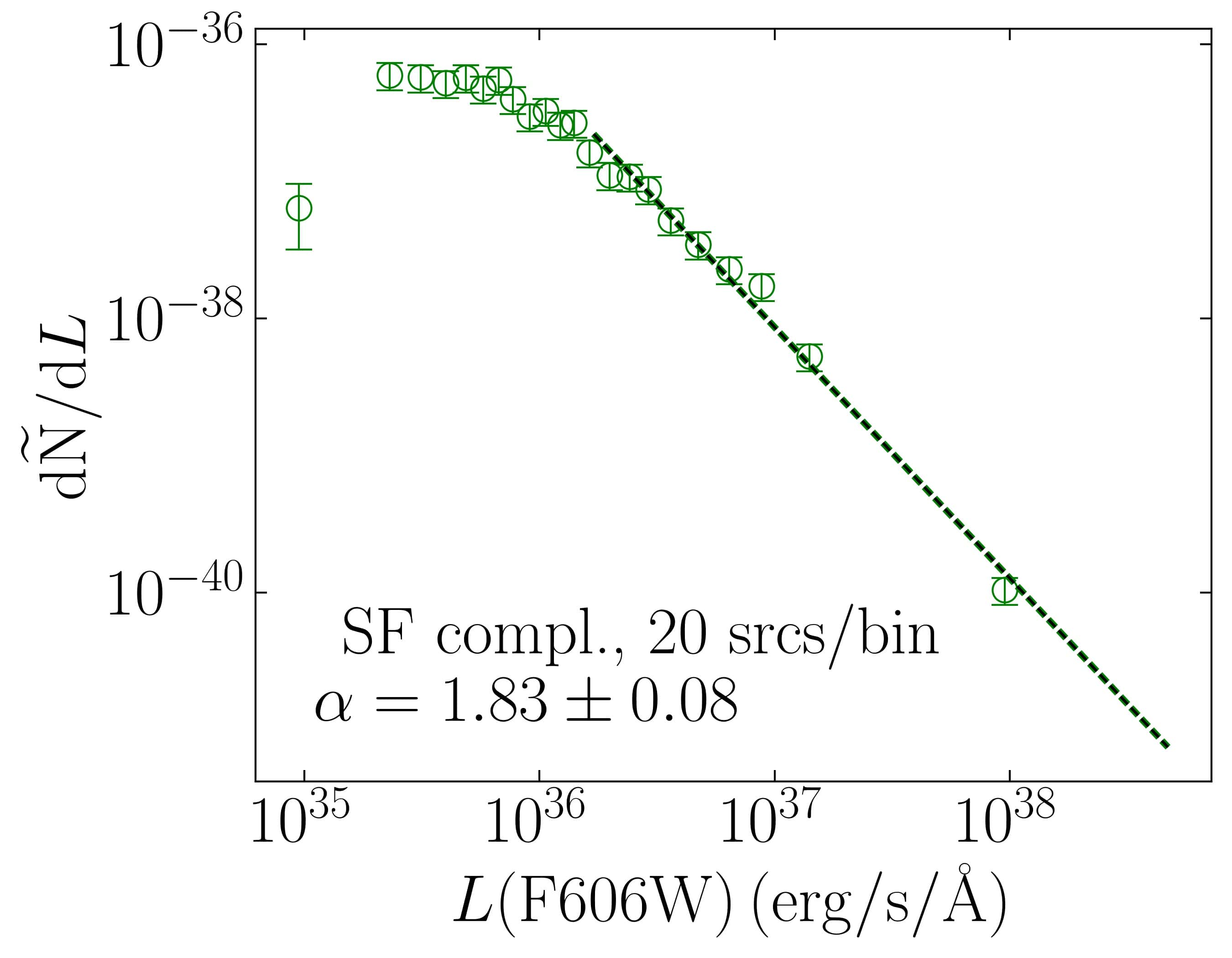}
\caption{Same as Fig. \ref{lum_distr}, but for star-forming complexes.}
\label{lum_distr_compl}
\end{figure}

In Figs. \ref{lum_distr} and \ref{lum_distr_compl} we plot the observed LDFs together with the corresponding best-fitting power laws. Tail LDFs seem to be well described by a single power law, both for H$\alpha$- and UV-selected clumps.

In Table \ref{powerlaw:res} we list the best-fitting values of the slopes $\alpha$ and the normalizations $K$, together with the peak luminosities $L_{\mathrm{peak}}$. 
Considering all the cases, the value of the slope $\alpha$ is in the range from $1.61$ to $1.88$ (thus always smaller than 2), with a mean value of $1.79\pm 0.09$ ($1.84\pm 0.03$ for H$\alpha$-selected clumps and $1.73\pm 0.09$ for UV-selected clumps). In order to rule out the possibility that the inclusion in the sample of trunk clumps with sub-clumps can bias the results, we performed the same fits to the LDFs excluding them. Since this kind of trunks is $\sim 2\%$ of the whole sample, excluding them does not affect significantly the results and the leaf-only slopes are always consistent within 1$\sigma$ with those obtained including both trunks and leaves.

\begin{deluxetable*}{lc|ccc|ccc}
\tablecaption{Luminosity and size distribution functions best-fitting parameters.}
\tablewidth{0pt}
\tablehead{
\colhead{} & \colhead{} & \colhead{$\alpha$} & \colhead{$K_L$} & \colhead{$L_{\mathrm{peak}}$} & \colhead{$\alpha_{\mathrm{s}}$} & \colhead{$K_{\mathrm{s}}$} & \colhead{size$_{\mathrm{peak}}$}\\
\colhead{} & \colhead{} & \colhead{} & \colhead{} & \colhead{$\mathrm{[erg/s(/\AA)]}$} & \colhead{} & \colhead{} & \colhead{pc}
}
\decimalcolnumbers
\startdata
& D & $1.82\pm 0.05$ & $30.9\pm 1.8$ & $1\times 10^{38}$ & $2.8\pm 0.2$ & $4.2\pm 0.4$ & 150\\
H$\alpha$ & E & $1.83\pm 0.06$ & $31\pm 2$ & $7\times 10^{37}$ & $3.6\pm 0.6$ & $6.4\pm 1.4$ & 200\\
& T & $1.88\pm 0.04$ & $33.0\pm 1.6$ & $1\times 10^{38}$ & $4.4\pm 0.8$ & $8\pm 2$ & 160\\\hline
& D & $1.61\pm 0.03$ & $21.8\pm 1.2$ & $8\times 10^{35}$ & $2.95\pm 0.16$ & $4.6\pm 0.4$ & 190\\
UV & E & $1.76\pm 0.04$ & $27.3\pm 1.6$ & $8\times 10^{35}$ & $2.9\pm 0.3$ & $4.4\pm 0.7$ & 200\\
& T & $1.83\pm 0.03$ & $26.9\pm 1.1$ & $8\times 10^{35}$ & $3.5\pm 0.4$ & $5.9\pm 0.9$ & 175\\\hline
\multicolumn{2}{c|}{Complexes} & $1.83\pm 0.08$ & $30\pm 3$ & $1.7\times 10^{36}$ & - & - & -\\
\enddata
\tablecomments{List of the best-fitting values of the LDFs and SDFs when fitted to the different samples of star-forming clumps and complexes. Column (1) refers to the clump selection photometric band. Column (2), from top to bottom: disk (D), extraplanar (E) and tail (T) sub-samples of H$\alpha$-selected and UV-selected clumps; the last row refers to star-forming complexes, which are only in the tails by construction. Columns (3), (4) and (5) contain the values of the best-fitting slopes $\alpha$, the best-fitting normalization $K_L$ and the peak luminosity $L_{\mathrm{peak}}$ arbitrarily chosen, over which datapoints are fitted (Eq. \ref{power:law}). Notice that $L_{\mathrm{peak}}$ is in erg/s for H$\alpha$-selected clumps, whereas it is in erg/s/$\mathrm{\AA}$ for UV-selected clumps and star-forming complexes. Columns (6), (7) and (8) list the same quantities (best-fitting slope $\alpha_s$, best-fitting normalization $K_s$ and peak size size$_{\mathrm{peak}}$), but for the SDFs.
}
\label{powerlaw:res}
\end{deluxetable*}

Our LDF slopes are consistent with previous results for HII regions (\citealt{Kennicutt1989}: $2.0\pm 0.5$) and are smaller than 2. Similar results have been found by \cite{Santoro2022} for HII regions and \cite{Cook2016} for UV-selected young stellar clusters.

We also performed a \textit{KS}-test on the luminosity distributions for pairs of spatial categories, for H$\alpha$- and UV-selected clumps separately, to infer whether the LDF significantly changes from one region to another.
We compare the distributions above the maximum $L_{\mathrm{peak}}$ value above which we can assume all three sub-samples to be complete.
The resulting P values are listed in the Appendix (Table \ref{ks:test}) and are consistent with what one would expect when comparing the slopes of the LDFs. For H$\alpha$-selected clumps, where the slopes are consistent with each other within the errors, the KS-test cannot exclude that each pair of distributions are identical. For UV-selected clumps, the KS-test confirms significant differences for the pairs disk-extraplanar and disk-tail.

\begin{figure}
\centering
\gridline{\includegraphics[width=0.45\textwidth]{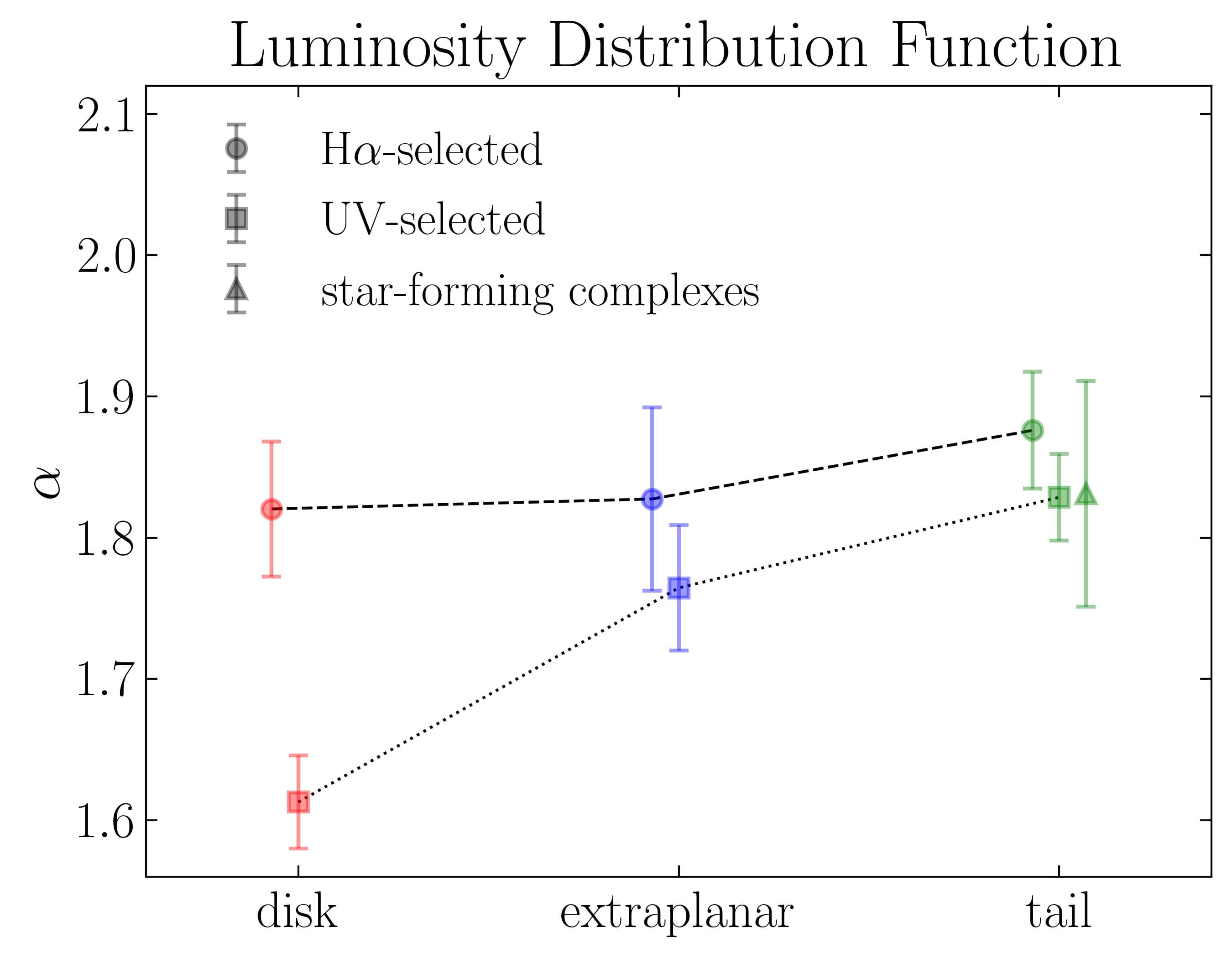}}
\gridline{\includegraphics[width=0.45\textwidth]{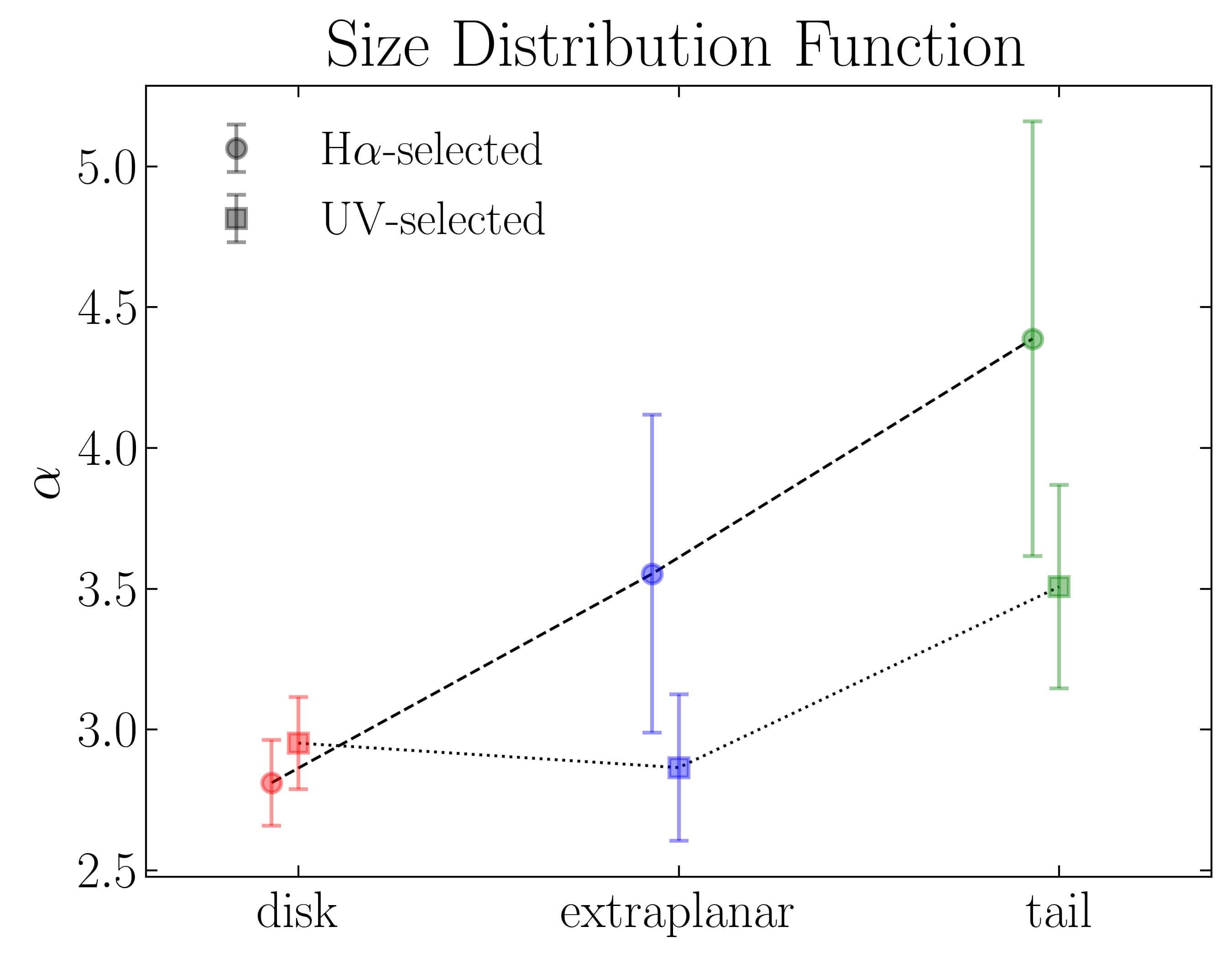}}
\caption{Comparison of the slopes of the LDFs (left panel) and SDFs (right panel) of star-forming clumps and star-forming complexes, both as a function of the selection band and the spatial category (see Table \ref{powerlaw:res}). Circles are H$\alpha$-selected clumps, squares are UV-selected clumps and triangles are star-forming complexes. Colors refers to the spatial category: red for disk, blue for extraplanar and green for tail.}
\label{slope_comparison}
\end{figure}

In the top panel of Fig. \ref{slope_comparison} we show the comparison among the best-fitting slopes, as a function of the selection band and the spatial category. Both the UV and H$\alpha$ slopes steepen going from disk, to extraplanar to tail regions, where the closest match with the expected slope $\alpha=2$ is found.

Shallower LDFs are found in galaxies with high sSFR \citep{Santoro2022}, such as all our jellyfish galaxies \citep{Vulcani2018}, which may explain why our slopes are smaller than 2. Furthermore, as described in Sec. \ref{sec:intro}, past works \citep{Cook2016,Messa2018b,Santoro2022} find flatter LDFs in environments with a high SFR surface density $\Sigma_\mathrm{SFR}$. Whether tails and disks are characterized by different $\Sigma_\mathrm{SFR}$ is a matter of future works, where masses and SFR od the clumps will be found by SED fitting.
Projection effects and blending, which are more likely to affect the disk than the tails, have also been demonstrated to flatten the LDF (as demonstrated by \citealt{Mirka2018} in the case of mass distribution function).
Flatter slopes are also found in simulations that include the ageing effect of the most massive clumps \citep{Gieles2009,Fujii2015}, which would be consistent with the fact that the slopes of the H$\alpha$-selected clumps (circles in Fig. \ref{slope_comparison}) are larger than those of the UV-selected clumps (squares) of the corresponding spatial category. It would be also confirmed by the slope of the star-forming complexes (all of which are located in the tails by construction), which is very close to that of tail UV-selected clumps.

Our analysis therefore suggests that the tails contain proportionally fainter clumps than the disks, and the extraplanar regions are intermediate between the two. However, this difference is statistically significant only when comparing UV-selected disk clumps with the other spatial categories, while for H$\alpha$-selected clumps there are only hints of such trend (Fig. \ref{slope_comparison}).
Furthermore, observational biases could explain the shallower LDF observed in disk clumps, since disk clumps are expected to be more affected by blending effects and underlying disk contamination, while the tail clumps are the least contaminated population, being isolated. Hence their observed LDF should be the closest to the intrinsic one. Indeed, it is the closest to the theoretical expected value of 2 \citep{Elmegreen2006}. Thus, we can conclude that the properties of the gas in which clumps are embedded are likely to play a minor role in influencing the LDF. Nonetheless, this analysis cannot fully exclude effects on other properties of the clumps, like the mass, which we will investigate in future works.

\subsubsection{Deviation from single power law}
Carefully inspecting Fig. \ref{lum_distr}, it is evident that disk (and to some extent, also extraplanar) LDFs show some particular features, such as slope changes, plateaus and secondary peaks, hinting to the need of a more complex model rather than a single power law.
To characterize these different regimes, disk LDFs are divided in three intervals: the faint-end interval, the plateau and the bright-end interval, each fitted with a power-law. Furthermore, for the H$\alpha$-selected LDF we fitted a power law also to datapoints brighter than $1.2\times 10^{39}$ erg/s, in correspondence of a secondary peak of the LDF\footnote{This secondary peak is dominated by clumps in JO201 (the galaxy with the largest amount of disk and tail clumps). Nonetheless, we do not have reasons to think there is a bias in luminosity artificially increasing the number of clumps at such luminosity, therefore it is a matter of interest to characterize this interval, too.} ("secondary-peak interval", hereafter). 
We superimposed the best-fitting disk power laws to the extraplanar LDFs, in order to understand if also this spatial category could be characterized by the same regimes (we do not have enough statistics to divide the extraplanar LDFs in intervals and fit a power law in each of them).

\begin{figure*}
\centering
\includegraphics[width=0.8\textwidth]{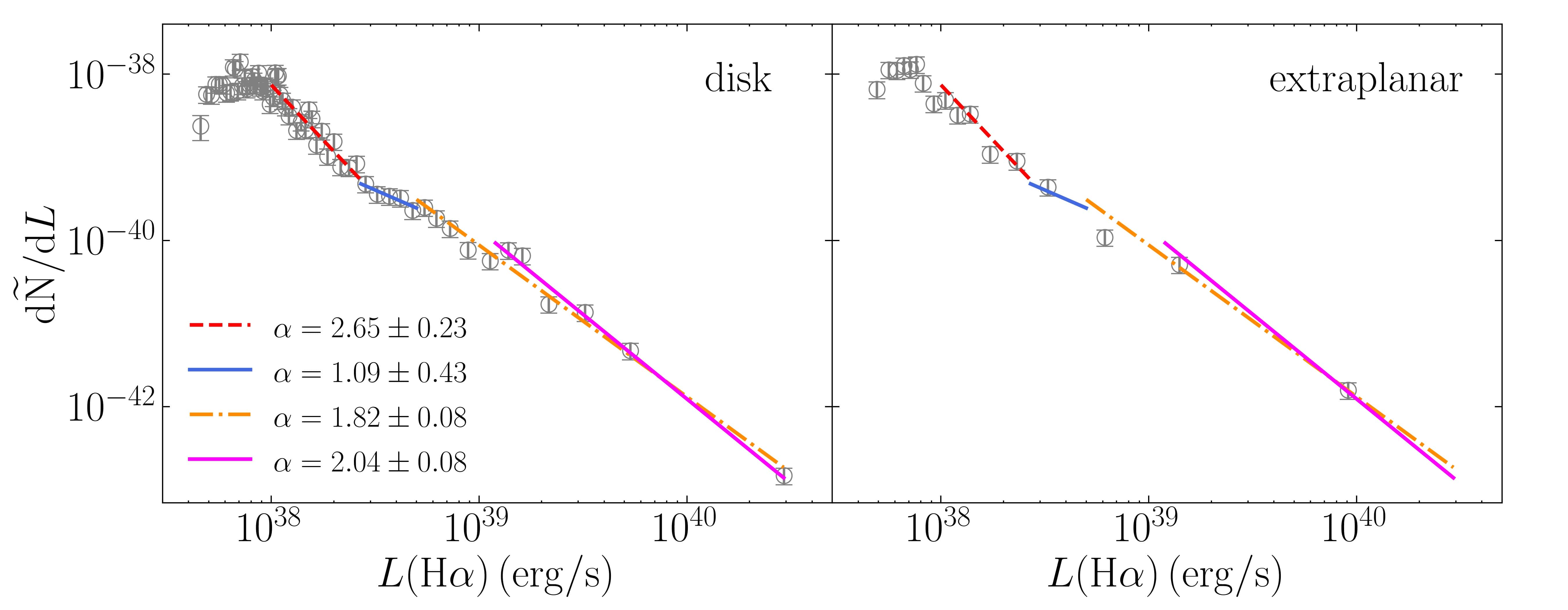}\\
\includegraphics[width=0.8\textwidth]{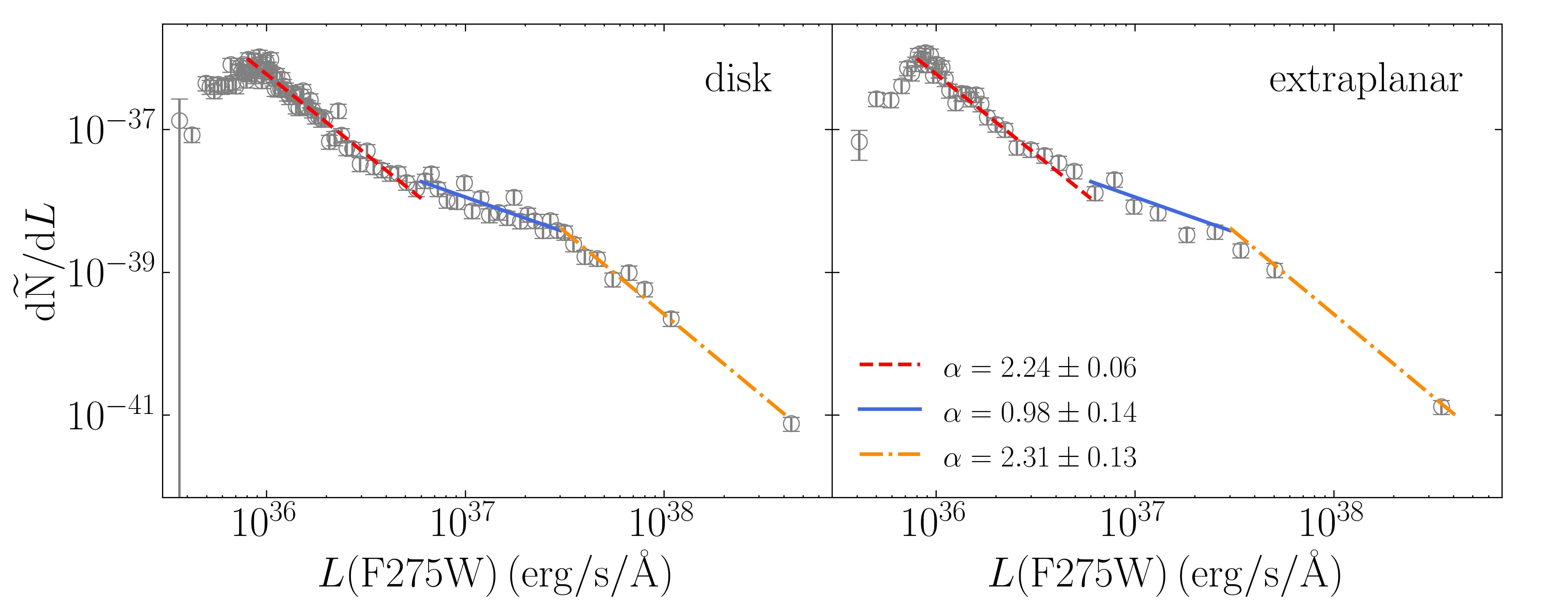}
\caption{Luminosity distribution functions $\mathrm{d\widetilde{N}/d}L$ of H$\alpha$-selected (upper panels) and UV-selected (lower panels) clumps. Clumps are divided according to their spatial category: disk (left panels, in red), extraplanar (middle panels, in blue) and tail (right panels, in green). For each plot we show: the empirical LDF of the corresponding sample (open circles with errorbars), generated with equal-number bins (i.e. each bin contains the same number of objects, see \citealt{Cook2016}), and the best-fitting line (dashed line). Notice that the H$\alpha$ luminosity of the H$\alpha$-selected clumps is the integrated emission of the H$\alpha$ line, therefore in erg/s, while the UV luminosity of the UV-selected clumps is in erg/s/$\mathrm{\AA}$.}
\label{lum_distr_interval}
\end{figure*}

The best-fitting slopes and the luminosity range boundaries of each interval are shown in Table \ref{broken:pl} and in Fig. \ref{lum_distr_interval} we show the best-fitting power laws superimposed onto the disk and extraplanar LDFs.

\begin{deluxetable*}{cc|cccc}
\tablecaption{Best-fitting slopes to the intervals of the disk LDFs.}
\tablewidth{0pt}
\tablehead{
\colhead{Phot. band} & \colhead{Parameters} & \colhead{Faint-end} & \colhead{Plateau} & \colhead{Bright-end} & \colhead{Secondary-peak}\\
}
\startdata
& $\alpha$ & $2.7\pm 0.2$ & $1.1\pm 0.4$ & $1.82\pm 0.08$ & $2.04\pm 0.08$\\
H$\alpha$ & $L_{\mathrm{min}}$ [erg/s] & $1\times 10^{38}$ & $2.7\times 10^{38}$ & $5\times 10^{38}$ & $1\times 10^{39}$\\
& $L_{\mathrm{max}}$ [erg/s] & $2.7\times 10^{38}$ & $5\times 10^{38}$ & - & -\\ \hline
& $\alpha$ & $2.24\pm 0.06$ & $0.98\pm 0.14$ & $2.31\pm 0.13$ & -\\
UV & $L_{\mathrm{min}}$ [erg/s/$\mathrm{\AA}$] & $8\times 10^{35}$ & $6\times 10^{36}$ & $3\times 10^{37}$ & -\\
& $L_{\mathrm{max}}$ [erg/s/$\mathrm{\AA}$] & $6\times 10^{36}$ & $3\times 10^{37}$ & - & -\\
\enddata
\tablecomments{Results obtained when a set of power laws are fitted to the disk LDFs divided in intervals (Sect. \ref{lum:dis}). From left to right: clump selection photometric band (Phot. band); best-fitting slope $\alpha$ and luminosity range boundaries of the interval $L_{\mathrm{min}}$ and $L_{\mathrm{max}}$ (Parameters); names of the intervals (Faint-end, Plateau, Bright-end and Secondary-peak).}
\label{broken:pl}
\end{deluxetable*}

For what concerns H$\alpha$-selected LDFs, the faint-end interval slope is larger than that of the bright-end interval, hinting to a change in the properties of the clumps before and after the plateau. When considering the secondary-peak interval, the distribution gets steeper than for the bright-end interval, but still flatter than at the faint-end. 
When superimposing these results on the extraplanar LDF (right-end panels in Fig. \ref{lum_distr_interval}), we can notice that the faint- and bright-end best-fitting power laws describe quite well the distribution. On the other hand, the extraplanar LDF seems to lack the plateau and the secondary-peak, even though we do not have enough datapoints in these intervals to exclude this hypothesis.

Concerning the disk UV-selected LDF, the slopes in the faint- and bright-end intervals are consistent within the uncertainties. The plateau covers a wider luminosity range compared to the H$\alpha$ plateau.
The presence of a plateau in UV LDFs has never been observed before.
Furthermore, the extraplanar LDF is well described by the results obtained for the disk, especially in the faint-end interval.

Whether these different regimes are an effect of the ageing or not is not clear yet. The position of the plateau in the disk H$\alpha$ LDF is compatible with a change in the bounding regime (from \textit{density bound} to \textit{ionization bound}) of the HII regions \citep{Beckman2000} at a predicted H$\alpha$ luminosity (at $\sim4\times10^{38}$ erg/s). On the other hand the slopes at the low and high luminosity ends are similar, while the \cite{Beckman2000} model predicts a steepening at bright luminosities, where HII regions are ionization bound. Moreover, our LDFs show the same plateau also in the disk UV-selected clumps, which should not be affected by the changing in the ionization regime.

\subsection{Size distribution functions}\label{siz:dis}
In this section we use the clumps of the resolved sample(s).
The analysis of the size distribution functions of the clumps (SDFs hereafter) is performed in the same way described in Sect. \ref{lum:dis}.
The samples are binned using 15 sources per bin for disk clumps and 5 sources per bin for the extraplanar and tail clumps, because of the low number of clumps in these spatial categories. 
SDFs are qualitatively similar to LDFs. Their intrinsic functional form is a power law, but incompleteness effects introduce a cutoff at small sizes. In analogy with what we did for the LDF (Eq. \ref{power:law}), we define the peak value as size$_{\mathrm{peak}}$ and we fit a power law to datapoints above this value.

\begin{figure*}
\centering
\gridline{\includegraphics[width=\textwidth]{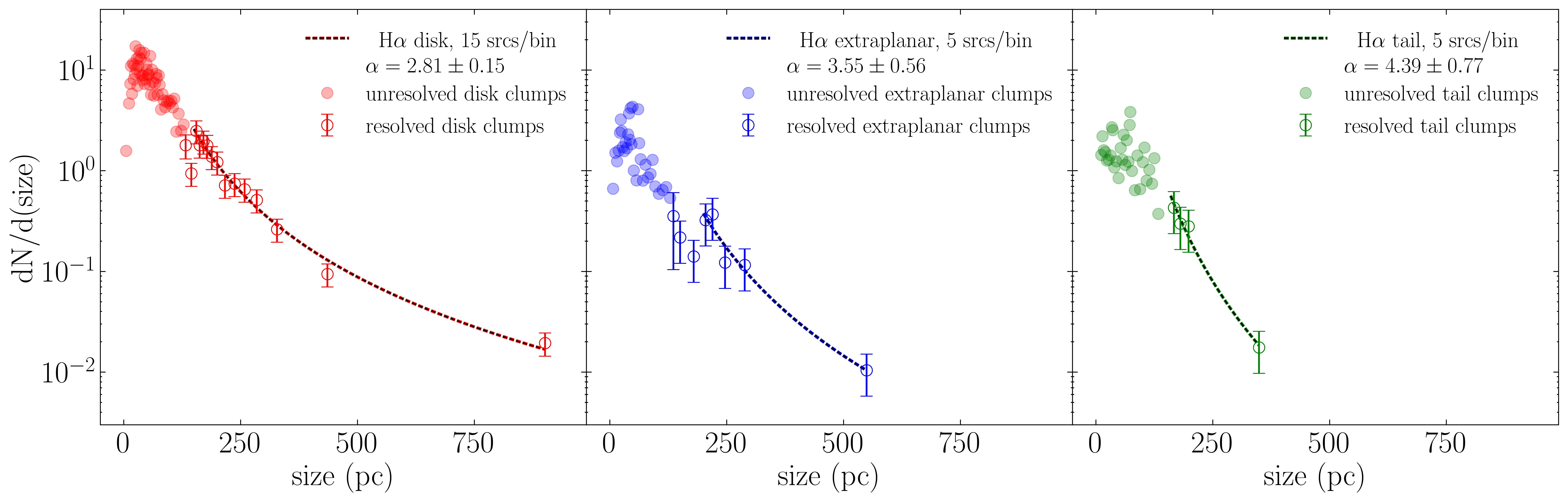}}
\gridline{\includegraphics[width=\textwidth]{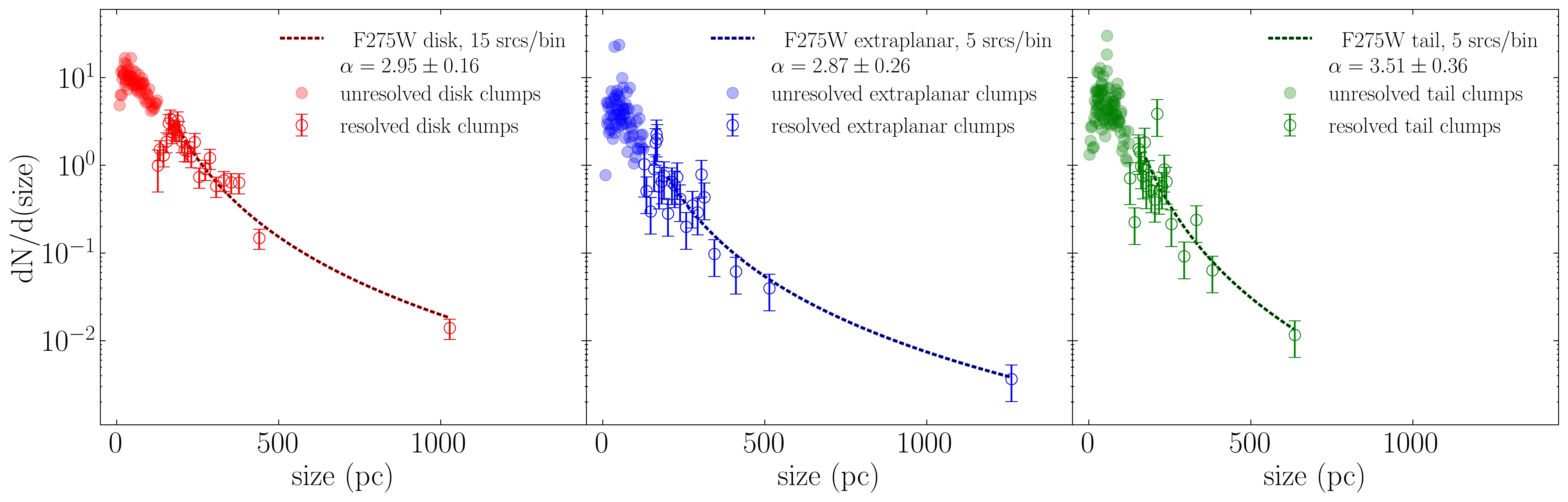}}
\caption{Size distribution functions for disk (red), extraplanar (blue) and tail (green) clumps. Top: $\rm H\alpha$. Bottom: UV. Resolved clumps are shown as empty circles with $1\sigma$ errorbars, while unresolved clumps are plotted as filled circles without errorbars.
In this case, SDFs are not normalized by the total number of clumps and the $x$-axis is in linear scale.
}
\label{size_distr}
\end{figure*}

In Fig. \ref{size_distr} the observed SDFs and the best-fitting model of each sub-sample are shown.
For completeness, we plot also the SDF datapoints of unresolved clumps, for which we have only upper limits for the sizes (filled dots). In order to do that, SDFs are not normalized for the total number of clumps, since the normalization changes when considering unresolved clumps or not. A single power law is likely to be a good representation of the resolved data, especially considering that the sample is about $15\%$ of the one used to constrain the parameters of the LDFs (see Table \ref{clu:table}). The loss of statistics can affect especially the extraplanar and tail sub-samples, for which the regime in which the sample is complete includes just a few datapoints.
The fitted power laws do not seem to well describe the unresolved datapoints, as expected from incompleteness.
These features, together with the fact that unresolved clumps have, by definition, no reliable estimates of their sizes, imply that we cannot draw any conclusion for sizes below $\sim 140$ pc.

The best-fitting slopes and normalizations, and the chosen size$_{\mathrm{peak}}$ of each sub-sample are listed in columns (6), (7) and (8) of Table \ref{powerlaw:res}. The average slope is $3.3\pm 0.6$ ($3.6\pm 0.6$ for H$\alpha$-resolved clumps and $3.1\pm 0.3$ for UV-resolved clumps). Slopes of extraplanar and tail H$\alpha$-resolved clumps are consistent with the one found by \cite{Kennicutt1980} in the disk of a low-z spiral galaxy ($\alpha=4.1$).

As done in Sect. \ref{lum:dis} for LDFs, we computed the P values from the KS statistics comparing the size distributions of pairs of spatial categories, keeping the two selection filters separated.
Results are listed in Appendix (Table \ref{ks:test}). In this case, the KS finds significantly different distributions for all pairs, except for disk vs tail H$\alpha$-resolved clumps. However, both the slopes and the P values have to be taken with caution, due to small numbers, especially in the tail clumps.

We find that these distributions are different from those inferred for the H$\alpha$ clumps of these galaxies detected by \citep{Poggianti2019a} from the MUSE H$\alpha$ luminosities using the luminosity-size relation by \cite{Wisnioski2012}: the expected median size was $440$ pc for clumps in the disks and $320$ pc for clumps in the tails.
Here, the median sizes are $\sim 210$ pc, $\sim 211$ pc, $\sim 180$ pc for disk, extraplanar and tail H$\alpha$-resolved clumps, respectively, and $\sim 215$ pc, $\sim 223$ pc, $\sim 208$ pc for disk, extraplanar and tail UV-resolved clumps. Consistently with what was inferred by \cite{Poggianti2019a}, clumps in the tails are smaller than those in the disk.
Nonetheless, values found in this work are about half the expected size.
The origin of this difference is a direct consequence of the differences between luminosity-size relation by \cite{Wisnioski2012} and the one obtained from our HST observations (see Sect. \ref{lum_size}).

In the bottom panel of Fig. \ref{slope_comparison} the slopes of resolved clumps in each category are plotted. Also in the case of SDFs the slope of UV-resolved clumps are smaller (with the exception of disk clumps), even if consistent within errorbars, than those of H$\alpha$-resolved clumps. Moreover, disk and extraplanar UV-resolved slopes are almost equal, while in H$\alpha$ there are hints of a slope increase from disk to tail regions.

The slope increase can be partially explained based on the work in \cite{Gusev2014}, whose observations of the nearby galaxy NGC 628 demonstrated that the overall slope of SDFs reaches values between 4.5 and 6\footnote{\cite{Gusev2014} studied the slopes of the cumulative distribution functions. Also, their slopes are defined as negative. Therefore the slopes in this work ($\alpha_{\mathrm{s}}$) and the slopes by \cite{Gusev2014} ($\alpha_{\mathrm{G}}$) are connected by $\alpha_{\mathrm{s}}=1-\alpha_{\mathrm{G}}$.} when analysing the smallest structures of the star-forming regions (i.e. what we define as leaves in Sec. \ref{clu:det}) or isolated objects. Instead, the slope decreases up to 2.5 once all the substructures of complex star-forming regions are taken into account.
Our trend is analogous. We find steep slopes ($\sim4.4\pm0.8$, consistent with 4.5) in the H$\alpha$ tails, whose clumps have typically no or few substructures. On the other hand, the slope is smaller in the case of disk clumps, which are more structured than extraplanar and tail clumps. Therefore, including both trunks and leaves in the samples has little effects on the slope of tail clumps, while it may explain the flatter distribution found for disk clumps.
Indeed, we observe steeper disk SDFs when using only the leaves ($3.28\pm0.28$ in H$\alpha$ and $3.22\pm0.23$ in UV).
Alternatively, recent simulations of star forming regions in presence of different ambient pressures \citep{Nath2020} found slopes similar to the one of the disk SDFs, while they suggest the presence of a lower pressure environment in the tail. The pressure producing the measured steepening in the tail SDF would be one order of magnitude lower than the typical ICM pressure of our galaxies \citep{Bartolini2022}. Therefore the variation of the slope of the SDF across different environments seems to be different to that expected from environmental effects.

In conclusion, the largest clumps of our sample are found in the disk and in the extraplanar regions of our galaxies, whether we consider UV- or H$\alpha$-resolved clumps, and (as hinted by the \textit{KS}-test), clumps of different spatial categories are likely to follow different SDFs with different slopes.
The sizes of the clumps seem to be poorly affected by the environment in which they are embedded, ICM in the tails and ISM in the disks, and more linked to their clustering features.

\section{Luminosity-size relations}\label{lum_size}

In this section we study the luminosity-size relation, both for H$\alpha$- and UV-resolved clumps. Here H$\alpha$-resolved clumps are BPT-selected to avoid AGN- and LINER-powered regions.
To calculate the linear regression fits with the inclusion of an intrinsic scatter, we employed the Python software package \textsc{linmix} \citep{Kelly2007}. \textsc{linmix} implements a Markov Chain Monte Carlo (MCMC) algorithm to converge on the posterior and return a sample of sets of parameters drawn from the posterior distribution. The linear relation fitted by \textsc{linmix} is

\begin{equation}\label{lum_size_eq}
\mathrm{Log} L\,=\,m\,\mathrm{Log} (size)\,+\,q\,+\,\mathrm{G}(\varepsilon),
\end{equation}

\noindent where $L$ is the luminosity of the clump in the filter in which it is selected, $size$ is the PSF-corrected core diameter, $m$ is the angular coefficient of the correlation, $q$ is the y-axis intercept and G$(\varepsilon)$ is the intrinsic scatter, computed from a Gaussian distribution centered in $m\,\mathrm{Log} (size)\,+\,q$ with standard deviation $\varepsilon$.

\begin{figure*}
\centering
\gridline{\resizebox{\textwidth}{!}{\includegraphics[height=1cm]{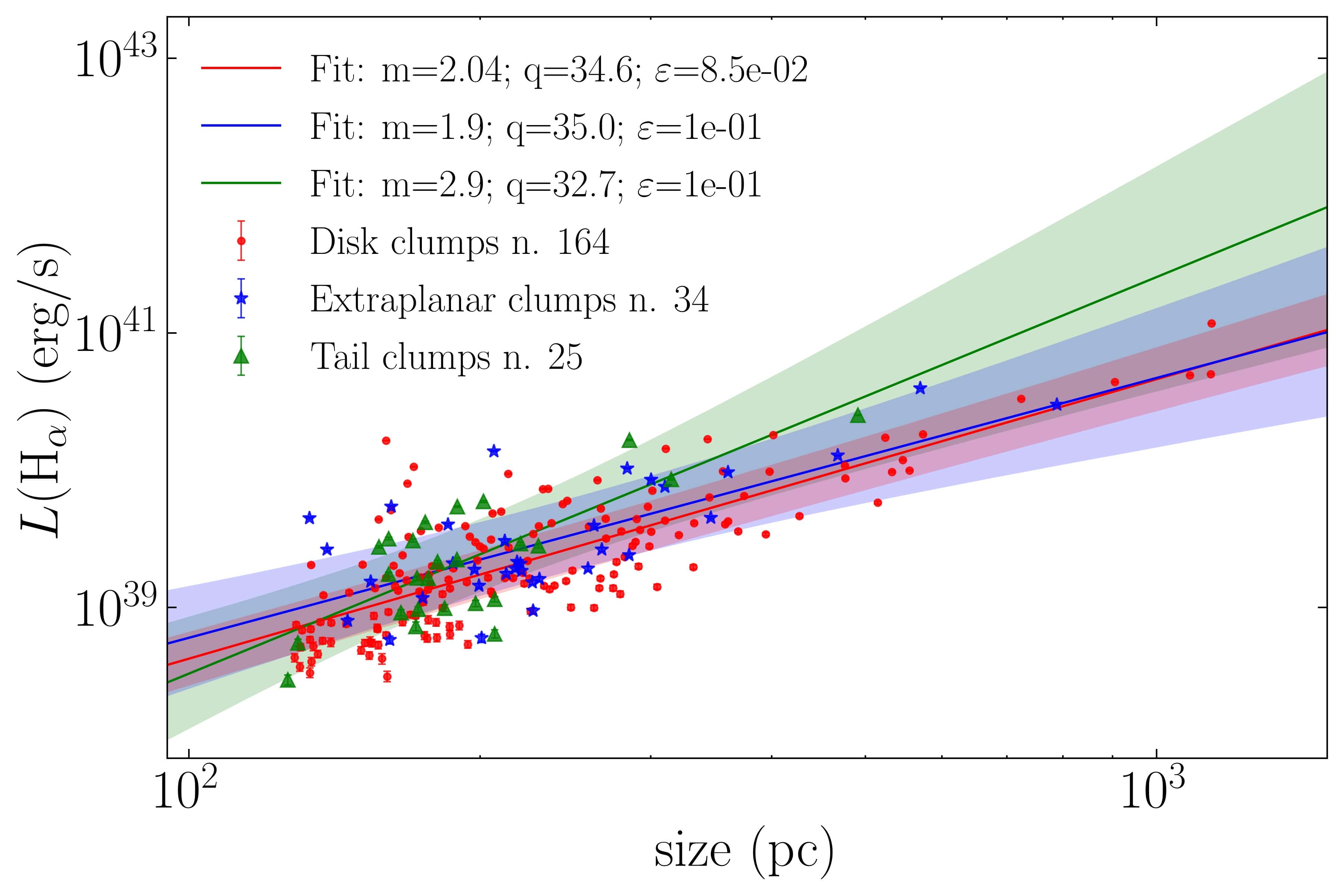}
          \includegraphics[height=1cm]{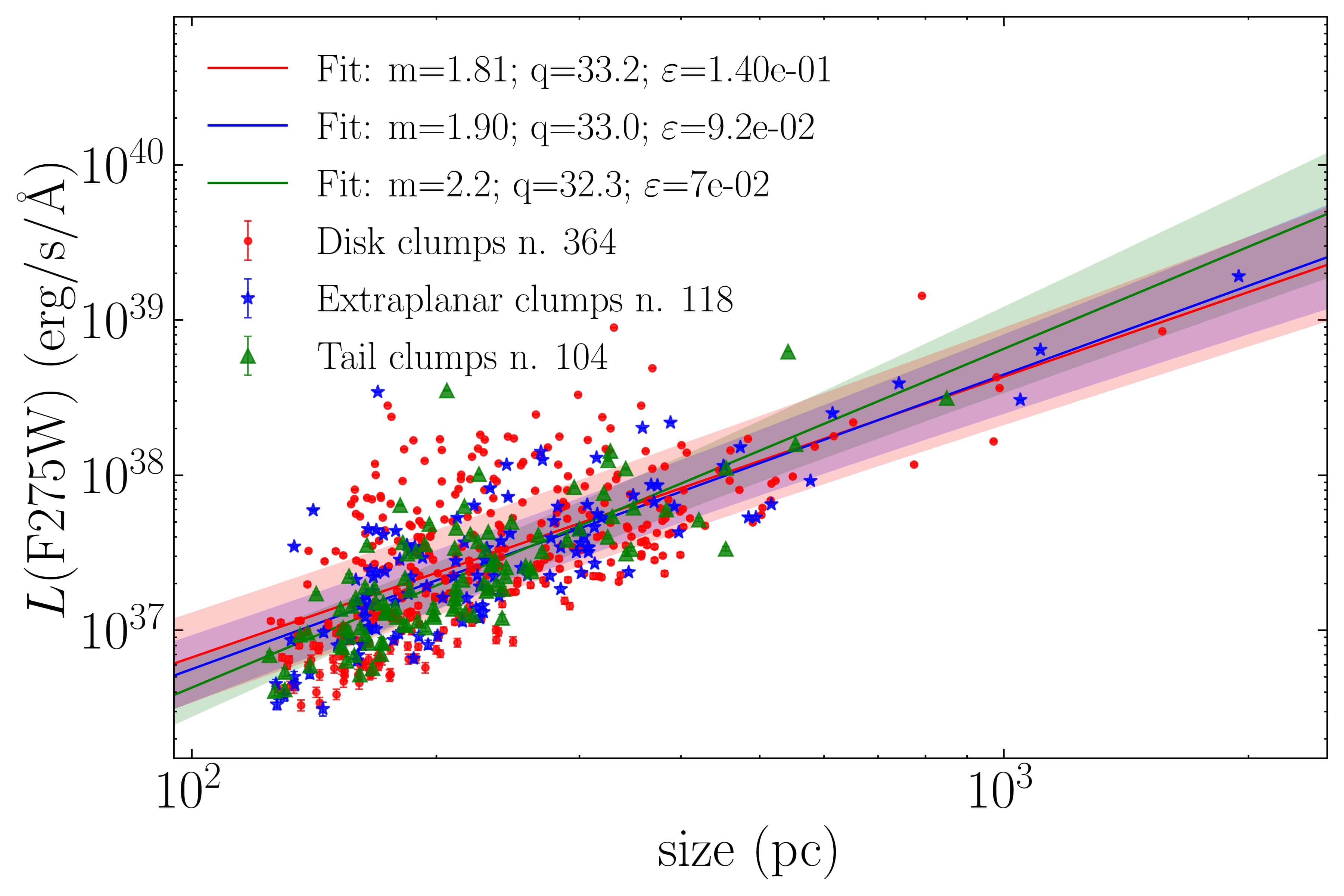}
          }}
\caption{Luminosity-size relations for H$\alpha$-resolved clumps (on the left) and UV-resolved clumps (on the right). The clumps are plotted according to their spatial category: disk (red circles), extraplanar (blue stars), tail (green triangles). The best-fitting lines to the three categories are plotted as solid lines of the corresponding color. The shaded areas are the uncertainties on the fits at 2$\sigma$. Note that H$\alpha$ luminosity is in erg/s, while F275W in erg/s/$\mathrm{\AA}$.
}
\label{lum_size_plot}
\end{figure*}

In Fig. \ref{lum_size_plot} we plot the datapoints in the $(\mathrm{Log} L,\mathrm{Log}(size))$ plane and the best-fitting lines, both for H$\alpha$- and UV-resolved clumps (left and right panel, respectively). Clumps are divided according to their spatial position.

Best-fitting parameters are listed in Table \ref{lum_size_tab}. The average slope is $2.3\pm 0.4$ for H$\alpha$-resolved and $1.97\pm 0.17$ for UV-resolved clumps. The slopes for the disk and extraplanar H$\alpha$-resolved clumps are consistent within 1$\sigma$ and close to 2, while the slope for the tail clumps is steeper. In UV, the slopes of all spatial categories are consistent with each other.

\begin{deluxetable*}{l|ccc|ccc}
\tablecaption{Luminosity-size relations best-fitting parameters.}
\tablewidth{0pt}
\tablehead{
\colhead{} & \colhead{} & \colhead{H$\alpha$} & \colhead{} & \colhead{} & \colhead{UV} & \colhead{}\\\cmidrule(lr){2-4}\cmidrule(lr){5-7}
\colhead{Spat. cat.} & \colhead{$m$} & \colhead{$q$} & \colhead{$\varepsilon$} & \colhead{$m$} & \colhead{$q$} & \colhead{$\varepsilon$}
}
\decimalcolnumbers
\startdata
Disk & $2.04\pm 0.12$ & $34.6\pm 0.3$ & $(8.5\pm 1.0)\times 10^{-2}$ & $1.81\pm 0.12$ & $33.2\pm 0.3$ & $(14.0\pm 1.0)\times 10^{-2}$\\
Extraplanar & $1.9\pm 0.4$ & $35.0\pm 0.8$ & $(10\pm 3)\times 10^{-2}$ & $1.90\pm 0.14$ & $33.0\pm 0.3$ & $(9.2\pm 1.3)\times 10^{-2}$\\
Tail & $2.9\pm 0.5$ & $32.7\pm 1.2$ & $(10\pm 3)\times 10^{-2}$ & $2.2\pm 0.2$ & $32.3\pm 0.4$ & $(7\pm 1)\times 10^{-2}$\\
\enddata
\tablecomments{Best-fitting parameters $(m,q,\varepsilon)$ (with 1$\sigma$ uncertainties) of the luminosity-size linear relations (Eq. \ref{lum_size_eq}) for H$\alpha$-resolved clumps (columns 2, 3, 4) and UV-resolved clumps (columns 5, 6, 7). Each spatial category (column 1) is fitted separately.}
\label{lum_size_tab}
\end{deluxetable*}

The Str\"omgren sphere model predicts the slope to be 3 \citep{Beckman2000}, hinting that disk and extraplanar clumps are not well described by this model as a consequence of additional effects to be taken into account, such as RPS, transition from ionization bound to density bound, dust, metallicity and magnetic fields \citep{Wisnioski2012}. On the other hand, \cite{Nath2020} obtained a slope equal to 2 simulating the expansion of ionized bubbles in a Milky Way-like ISM environment (in this case, H$\alpha$-resolved tail clumps would be the only ones deviating from the prediction of the model).
\cite{Cosens2018} proposed a model explaining why in galactic disks clumps with a low star formation rate surface density ($\Sigma_{\mathrm{SFR}}$) seem to follow a steeper relation (slope closer to 3) than clumps with high $\Sigma_{\mathrm{SFR}}$ (slope closer to 2). According to this model, if the expected radius of a Str\"omgren sphere is larger than the scale-height of the disk ($H$), the ionized bubble can keep growing only across the galactic plane, according to a power law with slope closer to 2 rather than 3. 
The flattening occurs only if the ionized region is brighter than a critical value. The fact that our slopes are consistent with 2 might therefore suggest that our clumps have an enhancement in the H$\alpha$ luminosity, maybe caused by ram-pressure stripping. Such a model would explain also why H$\alpha$-resolved tail clumps are likely to follow a steeper relation than clumps in the disk or extraplanar region. Tail clumps are embedded in the spherically symmetric ICM, in place of the gaseous disk of the galaxy, therefore they are not bound by $H$.

\subsection{Comparison with previous results}
\begin{figure*}
\centering
\includegraphics[height=0.6\textwidth]{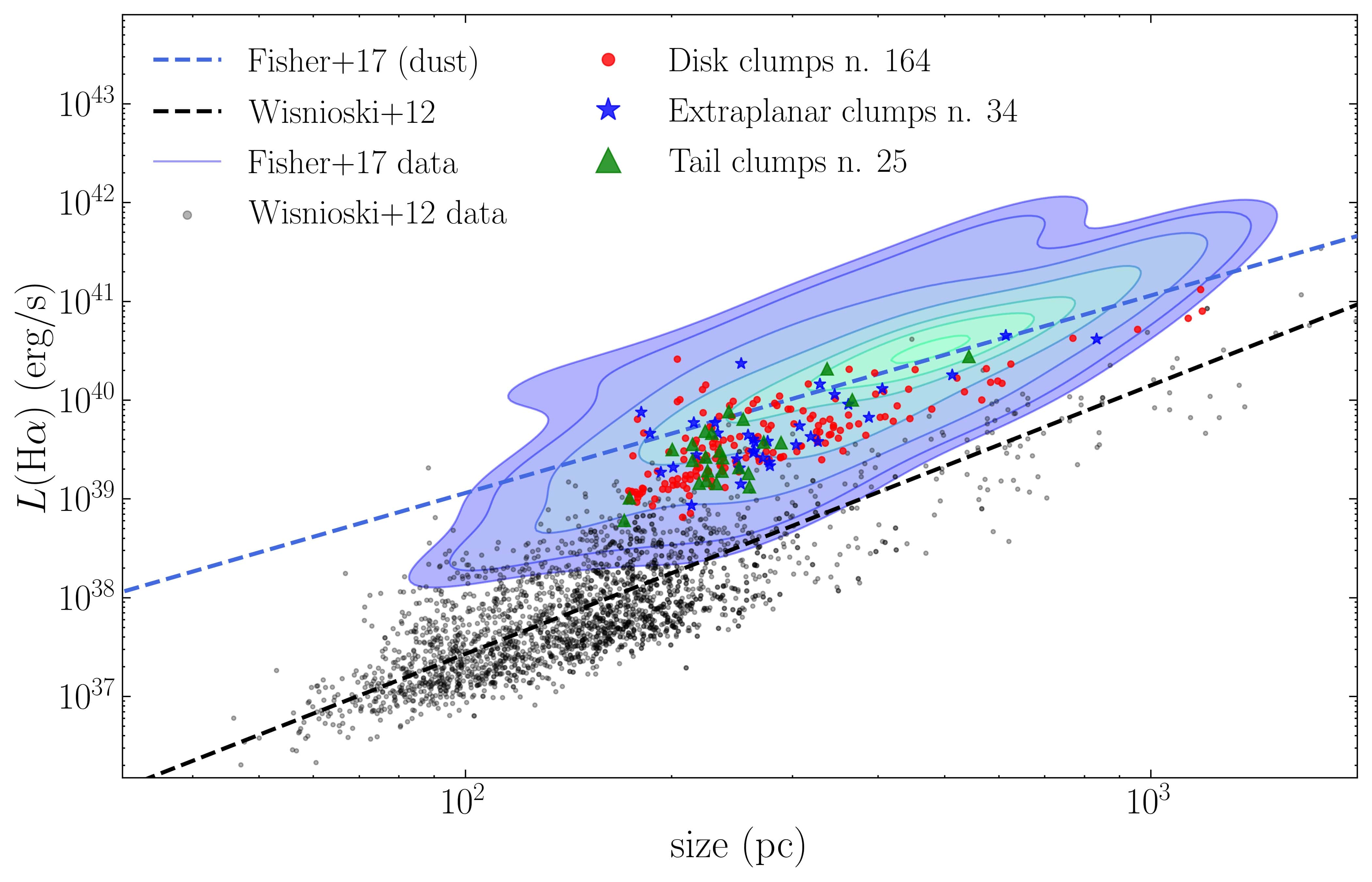}
\caption{$\log(\mathrm{L}(\mathrm{H}\alpha))-\log(\mathrm{size})$ comparing our H$\alpha$-resolved clumps with those in DYNAMO starburst galaxies (\citealt{Fisher2017}, blue contours) and those in local, isolated, star-forming galaxies presented in \citealt{Wisnioski2012} (\citealt{Arsenault1988,Kennicutt2003,Rozas2006,Gallagher1983,Monreal2007}, black dots). Our clumps are plotted according to their spatial category: disk (red circles), extraplanar (blue stars), tail (green triangles). Our clump luminosities and sizes are corrected in order to make the comparison more trustworthy; for the same reason, DYNAMO clump luminosities are corrected re-adding the effects of dust extinction. The black dashed line is the best-fitting relation by \cite{Wisnioski2012}, the blue dashed line is obtained fitting the dust-extincted DYNAMO clumps keeping the slope fixed at 2, as done in \cite{Fisher2017}. Clumps in our sample lie in between the two sample, being close in particular to clumps in starburst galaxies.}
\label{wisn_lum_size_plot}
\end{figure*}

In Fig. \ref{wisn_lum_size_plot} we compare the position of our H$\alpha$-resolved clumps in the $\log L-\log(size)$ with those presented in the literature. We show results from \cite{Fisher2017}, who study clumps belonging to turbulent, extremely H$\alpha$-bright DYNAMO galaxies, and those by \cite{Wisnioski2012}, who studied $z\sim0$, isolated, star-forming galaxies \citep{Arsenault1988,Kennicutt2003,Rozas2006,Gallagher1983,Monreal2007}. We also show the best-fitting relations they present in their works.

The luminosities of the DYNAMO clumps in Fig. \ref{wisn_lum_size_plot} were corrected re-adding extinction caused by dust, since both our luminosities and those computed by \cite{Wisnioski2012} are not dust-corrected. Dust-extincted DYNAMO clumps are then fitted to a power law with slope fixed at 2, as done in \cite{Fisher2017}.

As described in detail in \cite{Fisher2017}, the radii of DYNAMO clumps were found fitting a 2D Gaussian to the light distribution, with the addition of a constant representing the local background level \citep{Fisher2017}. 
To make the comparison with DYNAMO as consistent as possible, we derive new PSF-corrected core radii ($r_{\mathrm{gauss}}$) fitting a 2D Gaussian+constant to our tail H$\alpha$-resolved clumps, which are more isolated and in a fainter local background than disk and extraplanar clumps. We then visually select only clumps for which a  fit is appropriate. For these clumps, we compute $r_{\mathrm{gauss}}-\rcorecorr$, finding that it does not correlate with $\rcorecorr$, it ranges between 0 and 50 pc and it has a median value of 25.5 pc. Assuming that this difference is a good representation of the value of $r_{\mathrm{gauss}}-\rcorecorr$ for all the H$\alpha$-resolved clumps of our sample, we computed a new PSF-corrected core radius $\widetilde{r}_{\mathrm{core,corr}}=\rcorecorr+25.5$ pc. Therefore the new sizes are $2\widetilde{r}_{\mathrm{core,corr}}$.
The luminosities are re-computed integrating the light within a circle of radius $3\widetilde{r}_{\mathrm{core,corr}}$.
The procedure adopted in \cite{Wisnioski2012} to compute luminosity and size is similar to the one applied in \cite{Fisher2017}, though not identical. Therefore we are confident that the corrections we applied to our clumps allow us to make a fair comparison also with the results by \cite{Wisnioski2012}.

Our clumps lie between the Fisher and Wisnioski relations, being closer to the one obtained by \cite{Fisher2017}, even though they have lower luminosities and sizes compared to the peak of their distribution. With respect to the Wisnioski clump distribution, our resolved clumps are on average larger and, at a given size, brighter.

As shown by \cite{Johnson2017} and \cite{Cosens2018}, DYNAMO clumps have both higher SFR and $\Sigma_{\mathrm{SFR}}$ than clumps in isolated spiral galaxies, as a consequence of the starbursty star formation of their hosting galaxies. Being closer to the DYNAMO sample in the luminosity-size relation may suggest our H$\alpha$-resolved clumps to have a high $\Sigma_{\mathrm{SFR}}$, too (hints of that has already been found in \citealt{Vulcani2020}, in which they studied the resolved SFR-stellar mass relation for the MUSE H$\alpha$ clumps).

\section{Catalog}\label{sec:catalog}
We release the catalogs of H$\alpha$- and UV-selected clumps, separately, as online Table. 
Each clump is univocally determined by the name of the galaxy, a letter (referred to the \textsc{Astrodendro} run in which it has been detected, see Sec. \ref{appendix}) and an ID number.
We then list the RA and DEC coordinates, the
luminosity in the selection filter (not corrected for dust, but corrected for $\mathrm{NII}$ in the case of H$\alpha$-selected clumps), the morphological quantities (area, major and minor sigma, position angle, core radius and PSF-corrected core radius), the photometric fluxes and their errors in each band, including F680N continuum-subtracted (H$\alpha+\mathrm{NII}$), a flag for the clump properties in the tree structure (0 trunks which are not leaves, 1 trunks which are also leaves, 2 branches, 4 leaves which are not trunks), a flag for the spatial category (0 tail, 1 extraplanar and 2 disk) and a flag for the BPT classification (0 no BPT diagram available, 1 star-forming, 2 composite, 3 AGN, 4 LINER). Details about how these quantities are computed are given in Secs. \ref{clu:det} and \ref{sf:clumps}.

As an example, in Appendix (Table \ref{clu:cat}) we report the first ten rows of the H$\alpha$-selected clumps catalog. For clarity, for some values not all the significant digits are reported.

\section{Summary}\label{sec:summary}
In this paper we have built a sample of star-forming clumps and complexes in six jellyfish galaxies, using a set of \textit{HST} images in five photometric bands.
Clumps were detected independently in UV ($\sim 2700\,\mathrm{\AA}$) and H$\alpha$, in order to probe star formation on different timescales \citep{Kennicutt1998a}, while the star-forming complexes were detected from optical emission ($\sim6000\,\mathrm{\AA}$) to fully recover the stellar content formed from the stripped material in each clump region. Clumps were also divided in three spatial categories to study separately clumps formed within the disk (disk clumps), clumps likely originated in extraplanar gas but still close to the disk (extraplanar clumps) and clumps formed in the stripped gas out of the galactic disk and embedded in the ICM (tail clumps). Also, clumps in the tail give the unprecedented opportunity to study young stellar populations with no influence or underlying contamination by the stellar disk.

The method we use to detect clumps (Sec. \ref{clu:det}) yields the hierarchical cascade structure of these star-forming regions.
While disk clumps are often characterized by complex structures where many clumps are localized in bigger structures, extraplanar and tail clumps are typically simple structures with no sub-regions. Moreover, tail clumps tend to be aligned in elongated or arched structures along RPS sub-tails. Interestingly, the H$\alpha$-selected clumps, UV-selected clumps and F606W star-forming complexes are often nested into each other, with H$\alpha$-selected clumps being embedded in larger UV-selected clumps, and star-forming complexes containing sometimes several UV- and H$\alpha$-selected clumps. 

The final samples comprise 2406 H$\alpha$-selected clumps (1708 disk clumps, 375 extraplanar clumps and 323 tail clumps), 3745 UV-selected clumps (2021 disk clumps, 825 extraplanar clumps and 899 tail clumps) and 424 star-forming complexes (located only in the tails by construction). The covered luminosity range goes from $\sim 3\times 10^{38}$ to $\sim 4\times 10^{40}$ erg/s in H$\alpha$, and from $\sim 3\times 10^{36}$ to $\sim 2\times 10^{39}$ erg/s/$\mathrm{\AA}$ in UV.
On average, $\sim15\%$ of them are resolved, meaning that their sizes are larger than $\sim140$ pc, up to more than 1 kpc.

We studied the luminosity distribution functions (LDFs), size distribution functions (SDFs) and luminosity-size relations of H$\alpha$- and UV-selected clumps as a function of the spatial category. The LDF slopes averaged on all the spatial categories are $1.84\pm 0.03$ for H$\alpha$-selected clumps and $1.73\pm 0.09$ for UV-selected clumps. The average slopes of SDFs are $3.6\pm 0.6$ for H$\alpha$-resolved clumps and $3.1\pm 0.3$ for UV-resolved clumps. Finally, the average slopes of the luminosity-size relations are $2.3\pm0.4$ for H$\alpha$-resolved clumps and $1.97\pm0.17$ for UV-resolved clumps.
We find no clear difference among disk, extraplanar and tail clumps. The best-fitting slopes of these distributions and relations are consistent among each other, as well as with the results obtained in previous works \citep{Kennicutt1989,Cook2016,Santoro2022} and with theoretical predictions of hierarchical turbulence-driven star formation \citep{Elmegreen1996,Elmegreen2006}. On the other hand, the luminosity-size relation of the H$\alpha$-resolved clumps is more similar to that of clumps in starburst galaxies and therefore it suggests that these clumps, regardless of the spatial category, are experiencing an enhancement in $\Sigma_{\mathrm{SFR}}$. These preliminary results suggest ram pressure to compress the ISM and increase the H$\alpha$ luminosity of the clumps, while neither the presence of a disk and its gravity, nor the gaseous conditions of the surrounding medium have a strong impact on the star formation process, once the cold gas cloud conditions are set.

Future works on the mass, age and star formation of the clumps, on trends and gradients with the distance from the galaxies and on their fate will elucidate how and how much these clumps differ from those in undisturbed galaxies, in order to shed light on the effects of ram pressure on the galactic ISM and of environment on star formation.

\begin{acknowledgments}
EG would like to thank A. Ignesti, A. Wolter, A. Marasco, R. Smith, K. George and the GASP team for the useful discussions and comments. We would like to thank also Emily Wisnioski for providing us the data from her paper.
This paper is based on observations made with the NASA/ESA Hubble Space Telescope obtained from the Space Telescope Science Institute, which is operated by the Association of Universities for Research in Astronomy, Inc., under NASA contract NAS 5-26555. These observations are under the programme GO-16223. All the \textit{HST} data used in this paper can be found in MAST: \dataset[10.17909/tms2-9250]{http://dx.doi.org/10.17909/tms2-9250}.
This paper used also observations collected at the European Organization for Astronomical Research in the Southern Hemisphere associated with the ESO programme 196.B-0578.
This research made use of Astropy, a community developed core Python package for Astronomy by the Astropy Collaboration (\citeyear{Astropy2018}). This project has received funding from the European Research Council (ERC) under the European Union’s Horizon 2020 research and innovation programme (grant agreement No. 833824) and "INAF main-streams" funding programme (PI B. Vulcani).
\end{acknowledgments}

\appendix

\section{\textsc{Astrodendro} parameters setting}\label{appendix}

Three parameters regulate how \textsc{Astrodendro} builds the tree structure:

\begin{itemize}
    \item{\textsc{min\_value}: the algorithm stops when the flux threshold reaches this value, instead of zero;}
    \item{\textsc{min\_npix}: minimum number of pixels for a clump to be included to the tree structure;}
    \item{\textsc{min\_delta}: the threshold is not lowered in a continuum way, but at steps of \textsc{min\_delta}. If no \textsc{min\_delta} is given, the algorithm identifies each local maximum as a new sub-clump. \textsc{min\_delta} should be high enough to avoid the detection of noise peaks in the surface brightness distribution as sub-clumps.}
\end{itemize}

We performed three runs of \textsc{Astrodendro} for the F275W and H$\alpha$ images of each galaxy, adopting the following parameters (\textsc{min\_npix}$=5$ in all the runs):

\begin{itemize}
    \item{RUN A: \textsc{min\_value}$=2.5\sigma$; \textsc{min\_delta}$=5\sigma$;}
    \item{RUN B: \textsc{min\_value}$=2.5\sigma$; no \textsc{min\_delta}. 
    Given that a clump candidate is detected only if its brightest pixel is brighter than about \textsc{min\_value}$+$\textsc{min\_delta}, regions for which each pixel has counts between \textsc{min\_value} and \textsc{min\_value}$+$\textsc{min\_delta} are not detected. Since we want to detect also these fainter clumps, we ran \textsc{Astrodendro} a second time without defining \textsc{min\_delta}.
    This run is executed on an image masked for the clumps detected in run A and only trunk clumps are retained, to avoid including spurious local maxima.
    }
    \item{RUN C: \textsc{min\_value}$=2\sigma$; no \textsc{min\_delta}. This run is performed on an image masked for the clumps detected and kept in runs A and B. For the same reasons explained for run B, we kept only the trunks clumps of run C. Also, as a consequence of removing the high-frequency components of the image, denoising introduces a sort of smoothing, and part of the light of the brightest regions of the image, already detected as clumps, may eventually smooth out of the masks defined from runs A and B. Thus, even masking the image for the clumps already detected, the residual smoothed emission adjacent to these masks may be possibly flagged as a clump in run C. Since such an emission is clearly not due to a real clump, we excluded from the sample generated by run C all the clump candidates adjacent to the clumps found in the previous runs.
    }
\end{itemize}

In the case of F606W images, \textsc{Astrodendro} was run on the denoised F606W images, with \textsc{min\_value}$=3\sigma$ and no \textsc{min\_delta}.

\setcounter{table}{0}
\renewcommand{\thetable}{A\arabic{table}}

\begin{deluxetable*}{ll|ccc|ccc}
\tablecaption{Summary of the number of clumps in each sub-sample used throughout this paper, depending on the galaxy and on the spatial category.}
\tablewidth{0pt}
\tablehead{
\colhead{} & \colhead{} & \multicolumn{3}{c}{LT sample} & \multicolumn{3}{c}{resolved sample} \\\cmidrule(lr){3-5}\cmidrule(lr){6-8}
\colhead{Filter} & \colhead{gal} & \colhead{A} & \colhead{B} & \colhead{C} & \colhead{A} & \colhead{B} & \colhead{C}
}
\decimalcolnumbers
\startdata
    & JO$175$ & 65(65) & 214(214) & 11(11) & 31(31) & 7(7) & 0(0)\\
    & JO$201$ & 206(165) & 438(296) & 19(15) & 97(81) & 18(7) & 0(0)\\
    & JO$204$ & 81(60) & 287(233) & 5(3) & 36(29) & 4(1) & 0(0)\\
H$\alpha$ & JO$206$ & 99(84) & 318(274) & 21(19) & 41(34) & 3(3) & 0(0)\\
    & JW$39$ & 27(20) & 198(143) & 10(5) & 11(9) & 3(3) & 0(0)\\
    & JW$100$ & 94(40) & 299(232) & 14(12) & 29(13) & 6(5) & 0(0)\\\cmidrule(lr){2-8}
    & tot & 572(434) & 1754(1392) & 80(65) & 245(197) & 41(26) & 0(0)\\\cmidrule(lr){1-8}
    & JO$175$ & 74(74) & 204(204) & 9(9) & 35(35) & 7(7) & 0(0)\\
    & JO$201$ & 399(372) & 779(670) & 66(58) & 198(180) & 35(31) & 0(0)\\
    & JO$204$ & 106(102) & 397(348) & 28(25) & 68(65) & 14(13) & 0(0)\\
UV & JO$206$ & 137(135) & 573(568) & 31(30) & 82(80) & 24(24) & 0(0)\\
    & JW$39$ & 55(55) & 282(276) & 18(18) & 28(8) & 3(3) & 0(0)\\
    & JW$100$ & 173(152) & 392(362) & 22(19) & 80(67) & 12(11) & 0(0)\\\cmidrule(lr){2-8}
    & tot & 949(891) & 2627(2428) & 174(159) & 491(455) & 100(94) & 0(0)\\
\enddata
\tablecomments{Table with the number of clumps detected in each galaxy and depending on the \textsc{Astrodendro} run. From left to right: the photometric band in which the clumps were detected (1), the name of the galaxy (2), the number of clumps in the LT sample detected in runs A (3), B (4) and C (5), the number of clumps in the resolved sample detected in runs A (6), B (7) and C (8). In brackets, the number of clumps in the same sample, but selected in order to avoid regions powered by AGN emission (see Sect. \ref{sf:clumps}).}
\label{table:runs}
\end{deluxetable*}

\renewcommand{\thetable}{A\arabic{table}}

\begin{deluxetable*}{c|ccc|ccc}
\tablecaption{\textit{KS}-test P values.}
\tablewidth{0pt}
\tablehead{
\colhead{} & \colhead{} & \colhead{H$\alpha$} & \colhead{} & \colhead{} & \colhead{UV} & \colhead{}\\\cmidrule(lr){2-4}\cmidrule(lr){5-7}
\colhead{} & \colhead{D-E} & \colhead{D-T} & \colhead{E-T} & \colhead{D-E} & \colhead{D-T} & \colhead{E-T}
}
\decimalcolnumbers
\startdata
Lum & $0.792$ & $0.350$ & $0.254$ & $0.004$ & $2\times 10^{-8}$ & $0.152$\\ \hline
size & $3\times 10^{-6}$ & $0.613$ & $0.012$ & $7\times 10^{-5}$ & $9\times 10^{-4}$ & $0.002$\\
\enddata
\tablecomments{P values of \textit{KS}-tests (Sect. \ref{lum:dis}) for luminosity (first row) and size distributions (second row). Columns (2), (3) and (4) list the values for the H$\alpha$-selected clumps, when comparing disk and extraplanar clumps (D-E), disk and tail clumps (D-T) and extraplanar and tail clumps (E-T), respectively. Columns (5), (6) and (7) show the same results, but for UV-selected clumps. H$\alpha$-selected clumps are also selected to avoid AGN-powered regions when performing the \textit{KS}-test on the luminosity distributions.}
\label{ks:test}
\end{deluxetable*}

\renewcommand{\thetable}{A\arabic{table}}

\begin{table*}
\centering
\caption{First ten rows of the catalog of H$\alpha$-selected clumps available online. For clarity, for some values not all the significant digits are reported. Columns from 1 to 11: ID of the clump (ID\_clump), name of the host galaxy (galaxy), \textsc{Astrodendro} run (id\_cat, details in Sec. \ref{appendix}), clump id (\_idx), coordinates of the center (RA and Dec), luminosity and uncertainty in the selection filter (Lum and errL, notice that H$\alpha$ luminosity is in erg/s and corrected to remove NII emission, while UV luminosity is in erg/s/$\mathring{A}$), area $A$, semi-major axis (major\_sigma, $\sigma_{\mathrm{M}}$), semi-minor axis (minor\_sigma, $\sigma_{\mathrm{m}}$). }
\begin{tabular}{ccccccccccc}
\toprule\toprule
ID\_clump & galaxy & id\_cat & \_idx & RA & Dec & Lum & errL & $A$ & $\sigma_{\mathrm{M}}$ & $\sigma_{\mathrm{m}}$ \\
 &  &  &  & $\mathrm{{}^{\circ}}$ & $\mathrm{{}^{\circ}}$ & $\mathrm{erg\,s^{-1}}$ & $\mathrm{erg\,s^{-1}}$ & $\mathrm{arcsec^{2}}$ & $\mathrm{{}^{\prime\prime}}$ & $\mathrm{{}^{\prime\prime}}$ \\\hline
JO175\_A6\_halpha & JO175 & A & 6 & 312.851 & -52.834 & 3.264e+38 & 2.236e+37 & 0.040 & 0.071 & 0.041 \\
JO175\_A10\_halpha & JO175 & A & 10 & 312.829 & -52.827 & 6.320e+38 & 2.865e+37 & 0.061 & 0.091 & 0.046 \\
JO175\_A11\_halpha & JO175 & A & 11 & 312.829 & -52.827 & 1.979e+39 & 4.976e+37 & 0.173 & 0.137 & 0.080 \\
JO175\_A15\_halpha & JO175 & A & 15 & 312.823 & -52.824 & 1.068e+38 & 1.339e+37 & 0.016 & 0.060 & 0.026 \\
JO175\_A16\_halpha & JO175 & A & 16 & 312.824 & -52.823 & 3.202e+38 & 2.193e+37 & 0.038 & 0.055 & 0.047 \\
JO175\_A17\_halpha & JO175 & A & 17 & 312.823 & -52.823 & 8.186e+38 & 3.692e+37 & 0.106 & 0.111 & 0.088 \\
JO175\_A18\_halpha & JO175 & A & 18 & 312.823 & -52.823 & 2.851e+38 & 1.742e+37 & 0.022 & 0.050 & 0.025 \\
JO175\_A19\_halpha & JO175 & A & 19 & 312.823 & -52.823 & 3.118e+39 & 6.811e+37 & 0.336 & 0.267 & 0.123 \\
JO175\_A20\_halpha & JO175 & A & 20 & 312.824 & -52.823 & 1.183e+39 & 3.940e+37 & 0.112 & 0.094 & 0.080 \\
JO175\_A21\_halpha & JO175 & A & 21 & 312.819 & -52.823 & 2.433e+38 & 2.055e+37 & 0.035 & 0.059 & 0.055 \\
\bottomrule
\end{tabular}
\label{clu:cat}
\end{table*}

\addtocounter{table}{-1}
\renewcommand{\thetable}{A\arabic{table} (continued)}

\begin{table*}
\centering
\caption{First ten rows of the H$\alpha$-selected clumps catalog. Columns from 12 to 17: position angle ($\theta$), core radius ($\rcore$), PSF-corrected core raius ($\rcorecorr$), density flux and uncertainty in the filter F275W (F275W and errF275W), density flux in the filter F336W (F336W).}
\begin{tabular}{cccccc}
\toprule\toprule
$\theta$ & $r_{\mathrm{core}}$ & $r_{\mathrm{core,corr}}$ & F275W & errF275W & F336W \\
$\mathrm{{}^{\circ}}$ & $\mathrm{{}^{\prime\prime}}$ & $\mathrm{kpc}$ & $\mathrm{erg\,\mathring{A}^{-1}\,s^{-1}\,cm^{-2}}$ & $\mathrm{erg\,\mathring{A}^{-1}\,s^{-1}\,cm^{-2}}$ & $\mathrm{erg\,\mathring{A}^{-1}\,s^{-1}\,cm^{-2}}$ \\\hline
174.007 & 5.383e-02 & 4.027e-02 & 6.931e-19 & 4.989e-20 & 4.697e-19 \\
-155.162 & 6.429e-02 & 5.116e-02 & 1.321e-18 & 6.366e-20 & 8.863e-19 \\
-163.439 & 1.050e-01 & 9.040e-02 & 3.742e-18 & 1.092e-19 & 2.674e-18 \\
120.172 & 3.928e-02 & 2.300e-02 & 5.896e-20 & 2.790e-20 & 2.779e-20 \\
149.070 & 5.085e-02 & 3.701e-02 & 1.767e-19 & 4.521e-20 & 2.139e-19 \\
117.811 & 9.889e-02 & 8.468e-02 & 2.876e-18 & 8.714e-20 & 2.590e-18 \\
114.278 & 3.533e-02 & 1.706e-02 & 1.080e-19 & 3.385e-20 & 1.221e-19 \\
-149.026 & 1.810e-01 & 1.603e-01 & 5.979e-18 & 1.504e-19 & 5.986e-18 \\
118.371 & 8.677e-02 & 7.318e-02 & 2.055e-18 & 8.589e-20 & 2.233e-18 \\
106.350 & 5.689e-02 & 4.352e-02 & 1.504e-19 & 4.306e-20 & 1.526e-19 \\
\bottomrule
\end{tabular}
\end{table*}

\addtocounter{table}{-1}
\renewcommand{\thetable}{A\arabic{table} (continued)}

\begin{table*}
\centering
\caption{First ten rows of the H$\alpha$-selected clumps catalog. Columns from 18 to 23: uncertainty on the density flux in the filter F336W (errF336W), density flux and uncertainty in the filter F606W (F606W and errF606W), density flux and uncertainty in the filter F680N (F680N and errF680N), density flux in the filter F814W (F814W).}
\begin{tabular}{cccccc}
\toprule\toprule
errF336W & F606W & errF606W & F680N & errF680N & F814W \\
$\mathrm{erg\,\mathring{A}^{-1}\,s^{-1}\,cm^{-2}}$ & $\mathrm{erg\,\mathring{A}^{-1}\,s^{-1}\,cm^{-2}}$ & $\mathrm{erg\,\mathring{A}^{-1}\,s^{-1}\,cm^{-2}}$ & $\mathrm{erg\,\mathring{A}^{-1}\,s^{-1}\,cm^{-2}}$ & $\mathrm{erg\,\mathring{A}^{-1}\,s^{-1}\,cm^{-2}}$ & $\mathrm{erg\,\mathring{A}^{-1}\,s^{-1}\,cm^{-2}}$ \\\hline
3.332e-20 & 1.636e-19 & 7.199e-21 & 3.663e-19 & 1.761e-20 & 6.024e-20 \\
4.264e-20 & 4.281e-19 & 1.003e-20 & 7.618e-19 & 2.318e-20 & 1.010e-19 \\
7.370e-20 & 1.191e-18 & 1.703e-20 & 2.308e-18 & 4.012e-20 & 3.208e-19 \\
1.809e-20 & 3.245e-20 & 3.987e-21 & 1.104e-19 & 1.035e-20 & 2.372e-20 \\
3.051e-20 & 3.086e-19 & 8.181e-21 & 5.045e-19 & 1.843e-20 & 2.006e-19 \\
6.171e-20 & 2.520e-18 & 1.948e-20 & 2.837e-18 & 3.703e-20 & 1.847e-18 \\
2.281e-20 & 3.534e-19 & 7.613e-21 & 5.500e-19 & 1.626e-20 & 3.052e-19 \\
1.056e-19 & 8.298e-18 & 3.536e-20 & 9.999e-18 & 6.858e-20 & 6.803e-18 \\
6.131e-20 & 2.869e-18 & 2.062e-20 & 3.449e-18 & 3.973e-20 & 2.161e-18 \\
2.871e-20 & 1.442e-19 & 6.735e-21 & 3.060e-19 & 1.633e-20 & 9.034e-20 \\
\bottomrule
\end{tabular}
\end{table*}

\addtocounter{table}{-1}
\renewcommand{\thetable}{A\arabic{table} (continued)}

\begin{table*}
\centering
\caption{First ten rows of the H$\alpha$-selected clumps catalog. Columns from 24 to 29: uncertainty on the filter F814W (errF814W), flux and uncertainty in continuum-subtracted F680N (F680N\_line\_flux and errF680N\_line\_flux, note the units are $\mathrm{erg\,s^{-1}\,cm^{-2}}$ in this case), position of the clump in the tree hierarchy (dendro\_flag), spatial category of the clump (tail\_gal\_flag), BPT category (BPT\_flag).}
\begin{tabular}{cccccc}
\toprule\toprule
errF814W & F680N\_line\_flux & errF680N\_line\_flux & dendro\_flag & tail\_gal\_flag & BPT\_flag \\
$\mathrm{erg\,\mathring{A}^{-1}\,s^{-1}\,cm^{-2}}$ & $\mathrm{erg\,s^{-1}\,cm^{-2}}$ & $\mathrm{erg\,s^{-1}\,cm^{-2}}$ &  &  &  \\\hline
8.368e-21 & 1.023e-16 & 7.008e-18 & 1 & 0 & 0 \\
1.050e-20 & 1.981e-16 & 8.981e-18 & 1 & 0 & 0 \\
1.820e-20 & 6.204e-16 & 1.560e-17 & 1 & 0 & 1 \\
5.095e-21 & 3.347e-17 & 4.198e-18 & 1 & 2 & 1 \\
9.345e-21 & 1.004e-16 & 6.875e-18 & 1 & 2 & 1 \\
2.083e-20 & 2.566e-16 & 1.157e-17 & 1 & 2 & 1 \\
8.726e-21 & 8.937e-17 & 5.459e-18 & 1 & 2 & 1 \\
3.914e-20 & 9.773e-16 & 2.135e-17 & 1 & 2 & 1 \\
2.211e-20 & 3.707e-16 & 1.235e-17 & 1 & 2 & 1 \\
8.150e-21 & 7.624e-17 & 6.440e-18 & 1 & 0 & 1 \\
\bottomrule
\end{tabular}
\end{table*}

\setcounter{figure}{0}
\renewcommand\thefigure{A\arabic{figure}} 
\begin{figure*}
\centering
\includegraphics[width=\textwidth]{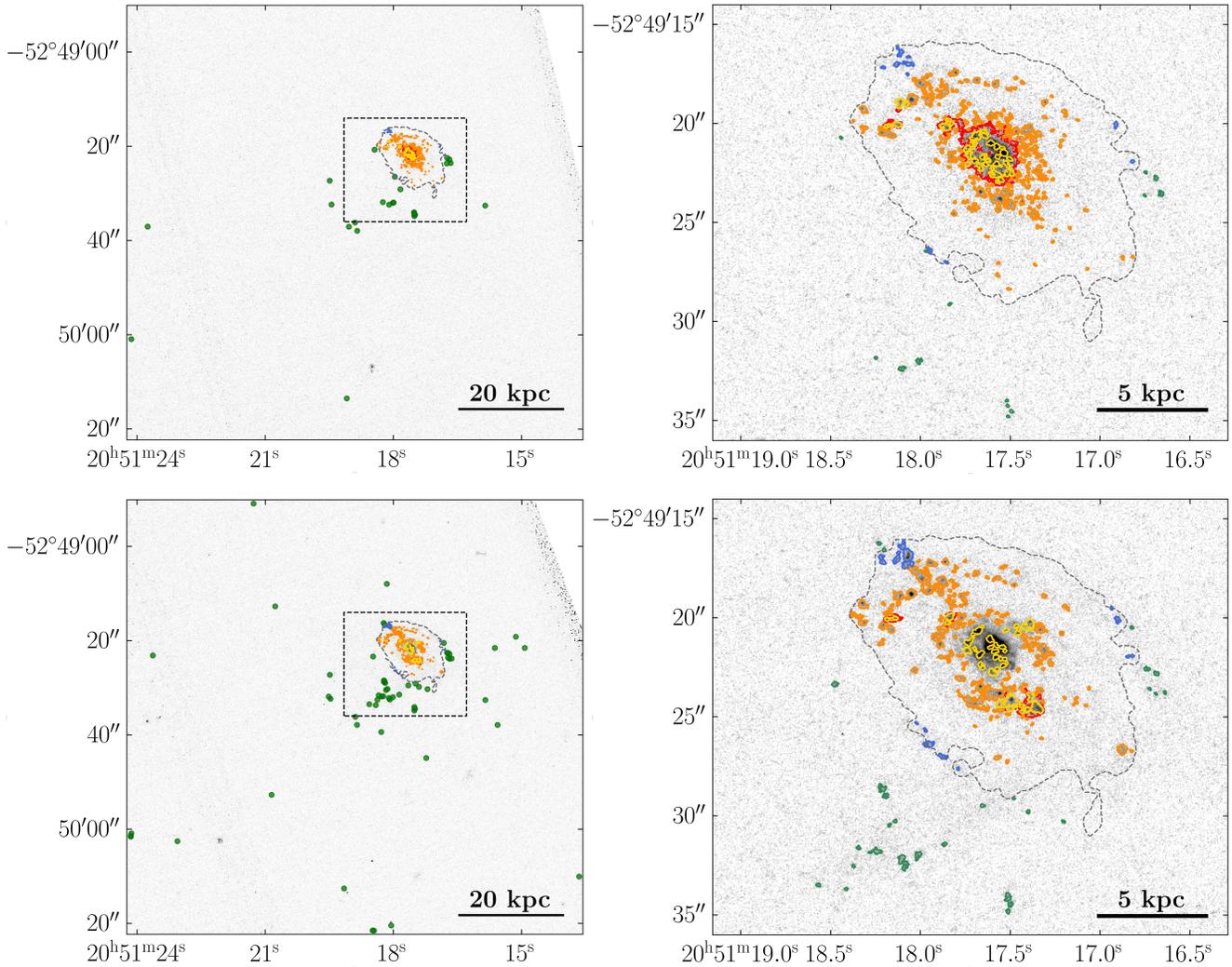}
\caption{Same as figure \ref{fig:detected_clumps}, but for JO175.}
\label{JO175_clumps}
\end{figure*}

\addtocounter{figure}{-1}
\renewcommand\thefigure{A\arabic{figure} (continued)} 
\begin{figure*}
\centering
\includegraphics[width=\textwidth]{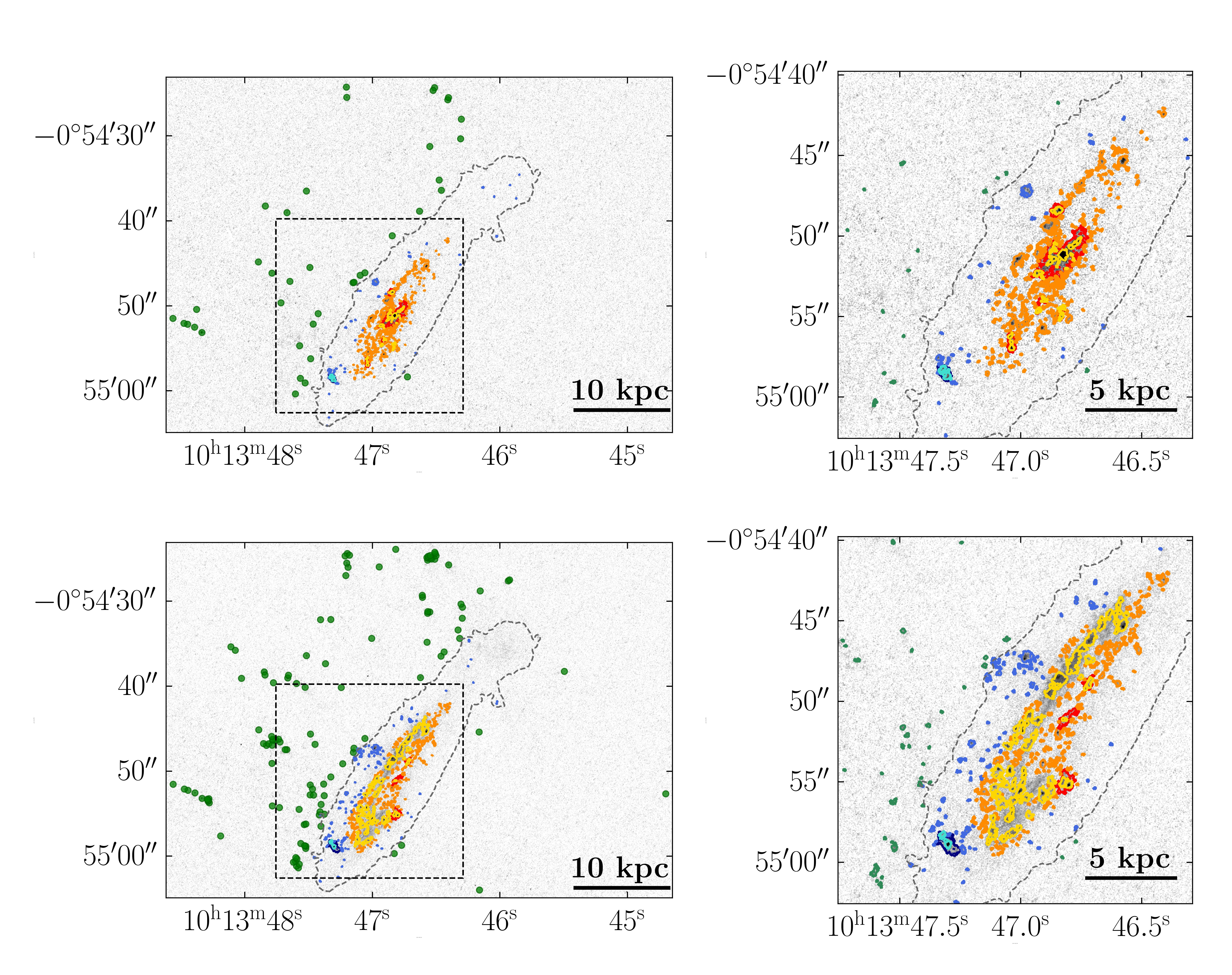}
\caption{Same as figure \ref{fig:detected_clumps}, but for JO204.}
\label{JO204_clumps}
\end{figure*}

\addtocounter{figure}{-1}
\renewcommand\thefigure{A\arabic{figure} (continued)}
\begin{figure*}
\centering
\includegraphics[width=\textwidth]{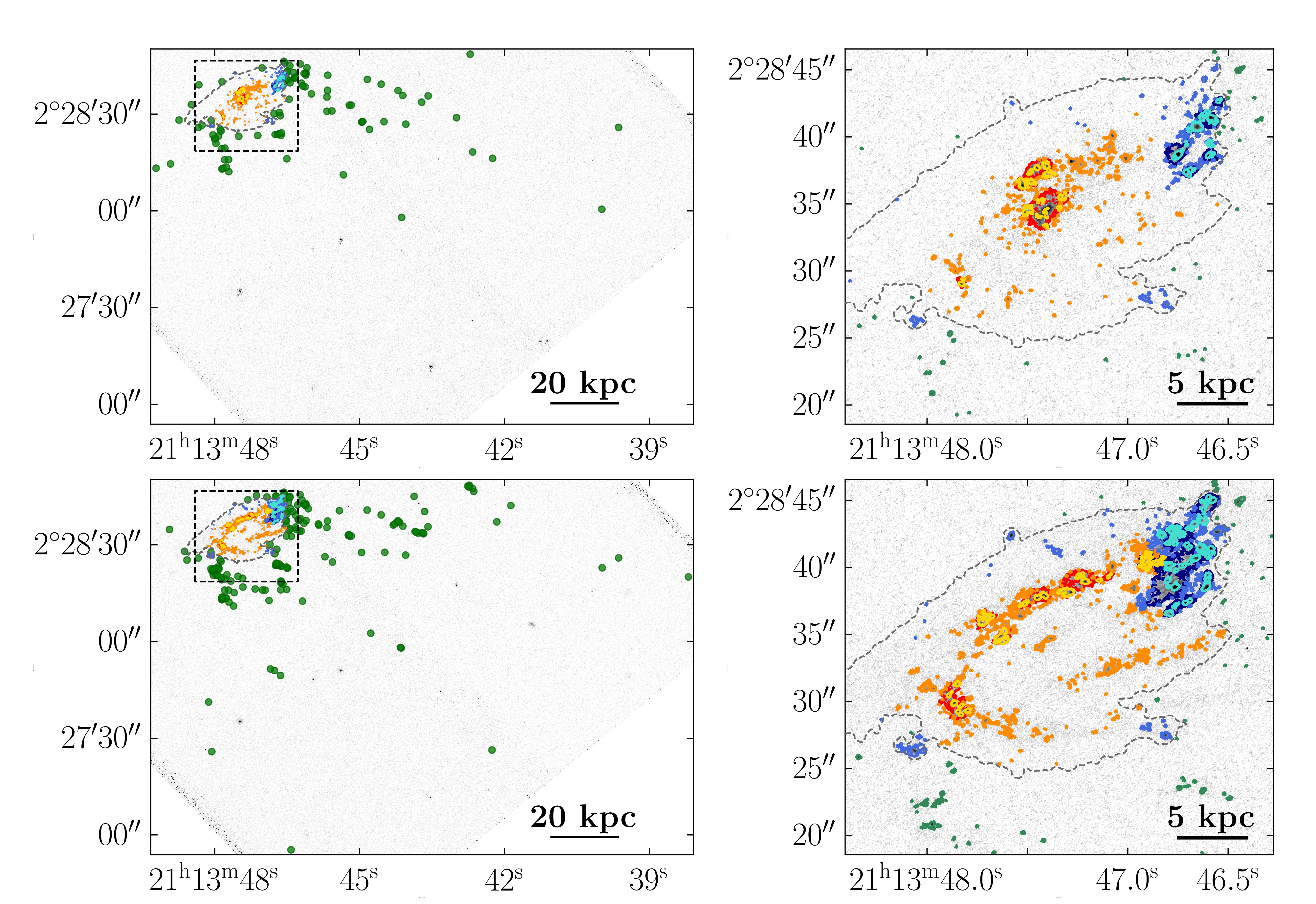}
\caption{Same as figure \ref{fig:detected_clumps}, but for JO206.}
\label{JO206_clumps}
\end{figure*}

\addtocounter{figure}{-1}
\renewcommand\thefigure{A\arabic{figure} (continued)}
\begin{figure*}
\centering
\includegraphics[width=\textwidth]{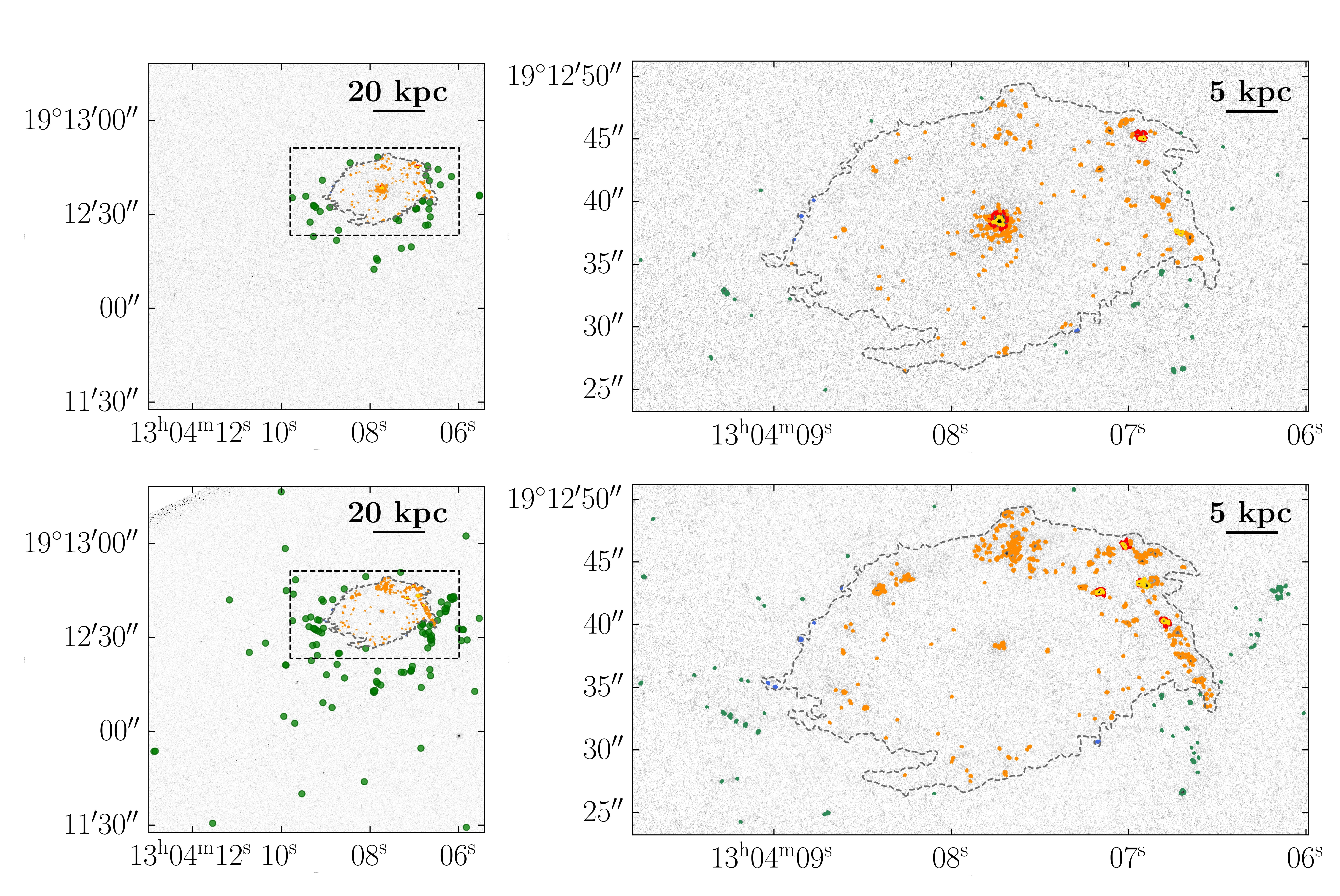}
\caption{Same as figure \ref{fig:detected_clumps}, but for JW39.}
\label{JW39_clumps}
\end{figure*}

\addtocounter{figure}{-1}
\renewcommand\thefigure{A\arabic{figure} (continued)}
\begin{figure*}
\centering
\includegraphics[width=0.8\textwidth]{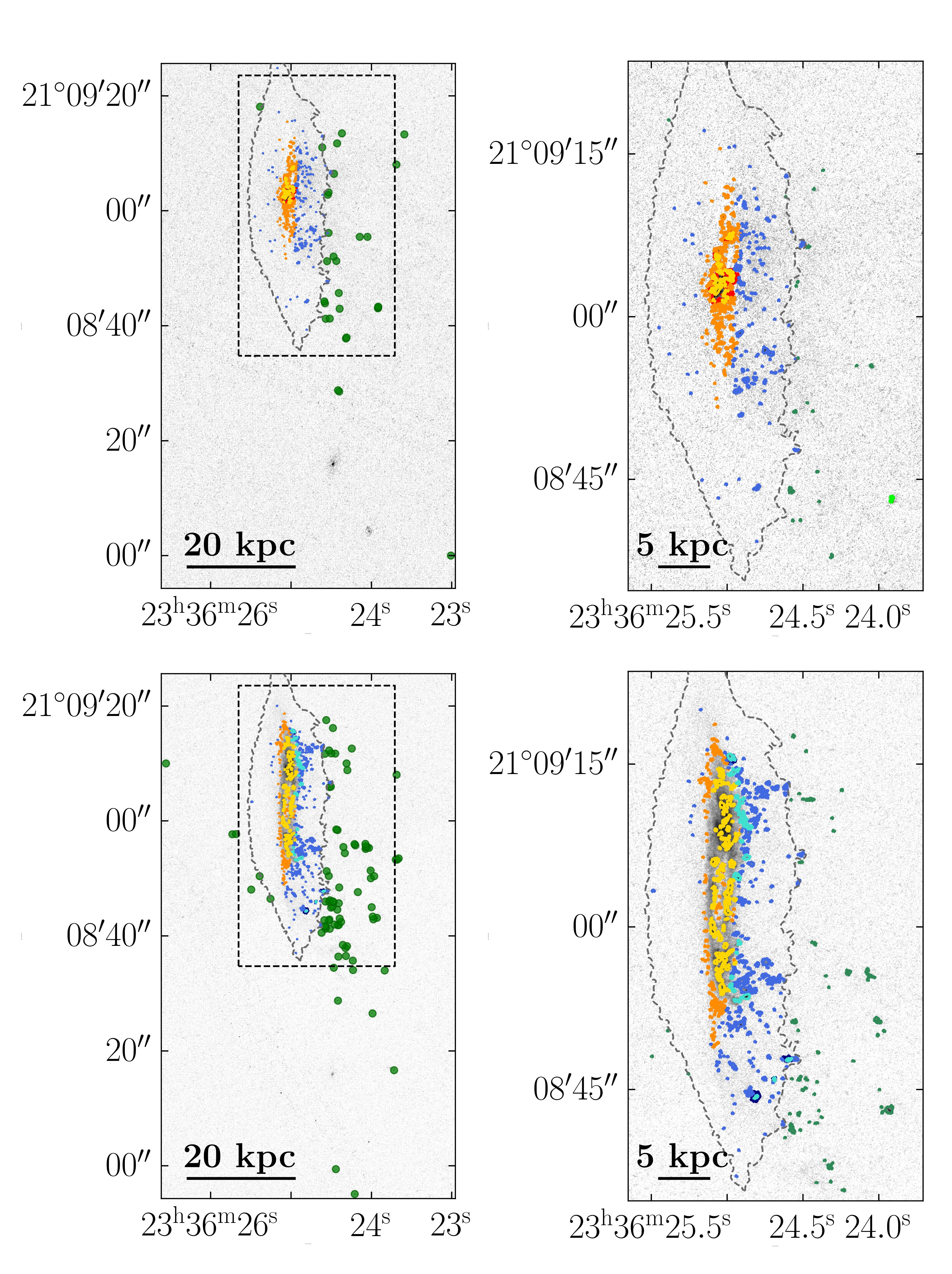}
\caption{Same as figure \ref{fig:detected_clumps}, but for JW100.}
\label{JW100_clumps}
\end{figure*}

\bibliography{biblio}{}
\bibliographystyle{aasjournal}

\end{document}